\pdfoutput=1
\makeatletter\@ifundefined{AddToHook}{}{\AddToHook{class/revtex4-2/after}{\def\switch@array{}}}\makeatother
\documentclass[%
reprint,
superscriptaddress,
aps,
pra,
]{revtex4-2}
\usepackage{amsmath,amssymb,mathtools}
\usepackage{xcolor}
\definecolor{mplblue}{HTML}{1F77B4}
\definecolor{mplgreen}{HTML}{2CA02C}
\definecolor{mplred}{HTML}{D62728}
\definecolor{mplpurple}{HTML}{9467BD}
\definecolor{mplolive}{HTML}{BCBD22}
\usepackage[colorlinks=true, allcolors=mplblue]{hyperref}
\usepackage{graphicx}
\usepackage{dcolumn}
\usepackage{bm}
\usepackage{dsfont}
\usepackage{physics}
\usepackage{url}
\usepackage{mathrsfs}
\usepackage{amsthm}
\usepackage{booktabs}
\usepackage{multirow}
\usepackage{array}
\usepackage{tabularx}
\usepackage{enumitem}
\usepackage{overpic}

\graphicspath{{img/}}

\newtheorem{theorem}{Theorem}
\newtheorem{corollary}{Corollary}
\newtheorem{proposition}{Proposition}
\newtheorem{lemma}{Lemma}
\theoremstyle{remark}

\newcommand{\N}{\mathds{N}}

\newcommand{\R}{\mathds{R}}
\newcommand{\C}{\mathds{C}}
\renewcommand\norm[1]{\left\Vert#1\right\Vert}
\renewcommand\abs[1]{\left\vert#1\right\vert}
\renewcommand\trace{\operatorname{Tr}}
\renewcommand{\d}{\mathrm{d}}

\newcommand{\ad}{\hat{a}^{\dagger}}
\newcommand{\nhat}{\hat{n}}
\newcommand{\Dop}{\hat\Lambda}
\newcommand{\Ex}[1]{\langle #1 \rangle}
\newcommand{\KL}{D_{\mathrm{KL}}}
\newcommand{\bs}{\bm}
\newcommand{\smax}{\sigma_{\max}}
\newcommand{\Om}{\Omega}
\newcommand{\pS}{p_S}
\newcommand{\Cov}{\operatorname{Cov}}
\newcommand{\Var}{\operatorname{Var}}
\newcommand{\supp}{\operatorname{supp}}
\newcommand{\conv}{\operatorname{conv}}
\newcommand{\diag}{\operatorname{diag}}
\newcommand{\Barr}{\Psi}
\newcommand{\Prob}{\operatorname{Pr}}

\newcommand{\papertitle}{Efficient maximization of the heralding probability of non-Gaussian state preparation}
\newcommand{\QST}{\cite{fiurasekHeraldingProbabilityOptimization2026}}
\newcommand{\PRA}{\cite{fiurasekMaximumHeraldingProbabilities2026}}
\newcommand{\affDTU}{Center for Macroscopic Quantum States (bigQ), Department of Physics, Technical University of Denmark, Building 307, Fysikvej, 2800 Kgs.\ Lyngby, Denmark}
\newcommand{\affUPOL}{Department of Optics, Palack\'y University, 17. listopadu 1192/12, 779 00 Olomouc, Czech Republic}

\begin{document}

\title{\papertitle}

\author{Vojt\v{e}ch Kucha\v{r}}
\affiliation{\affUPOL}
\affiliation{\affDTU}
\author{Jarom\'{\i}r Fiur\'a\v{s}ek}
\affiliation{\affUPOL}

\date{\today}
\begin{abstract}
	The heralding probability sets the rate of non-Gaussian state preparation with Gaussian sources and photon-number-resolving detectors. Damping the heralding modes with a noiseless filter changes this probability but not the heralded state. We show that, for ideal detection, the damping that maximizes the heralding probability can be found efficiently, even for many heralding modes. The damped photon statistics form an exponential family, in which the heralding probability is the likelihood of the damping given the registered pattern. This likelihood is log-concave in the logarithmic damping parameters, so every local maximum is global. When every heralding mode is coupled directly to the signal mode and registers at least one photon, there exists a unique maximum, where mean photon numbers match the pattern, and Newton's method finds it within tens of milliseconds.
	
\end{abstract}

\maketitle

\vspace{-1.25em}
\section{Introduction}\label{sec:intro}
\vspace{-1.25em}

Universal quantum computation with continuous variables requires non-Gaussian resources~\cite{lloydQuantumComputationContinuous1999, bartlettEfficientClassicalSimulation2002a, weedbrookGaussianQuantumInformation2012, walschaersNongaussianQuantumStates2021}. In trapped ions and superconducting circuits, non-Gaussian states such as Gottesman--Kitaev--Preskill (GKP) states are prepared by coupling a harmonic oscillator, such as a motional mode of a trapped ion or a microwave resonator, to a qubit~\cite{fluhmannEncodingQubitTrappedion2019, campagneibarcqQuantumErrorCorrection2020}. In propagating optical fields, bulk nonlinearities are too weak to create non-Gaussian states deterministically, so they are instead prepared conditionally~\cite{daknaGeneratingSchrodingercatlikeStates1997, lundConditionalProductionSuperpositions2004, suConversionGaussianStates2019, quesadaSimulatingRealisticNonGaussian2019}. Typically, squeezed light is sent through an interferometer, and some modes of the resulting multimode Gaussian state, the \emph{heralding modes}, are measured with photon-number-resolving (PNR) detectors [Fig.~\ref{fig:scheme}(a)]. The \emph{pattern} of registered photon~numbers then heralds a non-Gaussian state in the remaining \emph{signal modes}. This approach leads to optical cat states~\cite{gerritsGenerationOpticalCoherentstate2010, endoNongaussianQuantumState2023, takaseGenerationOpticalSchrodinger2021, endoHighRateFourPhoton2025} and GKP states~\cite{eatonNonGaussianGottesmanKitaev2019, tzitrinProgressPracticalQubit2020, takaseGottesmanKitaevPreskillQubitSynthesizer2023, zhengGaussianConversionProtocol2023, pizzimentiOpticalGottesmanKitaevPreskillQubit2024, konnoLogicalStatesFaulttolerant2024, larsenIntegratedPhotonicSource2025}, which are fundamental in fault-tolerant photonic architectures~\cite{menicucciFaultTolerantMeasurementBasedQuantum2014, bourassaBlueprintScalablePhotonic2021, aghaeeradScalingNetworkingModular2025}.

Designing an optimal state-generation scheme means choosing the squeezing, the interferometer and the detection pattern for a given target state. This is done by numerical optimization~\cite{sabapathyProductionPhotonicUniversal2019, arrazolaMachineLearningMethod2019, fukuiEfficientBackcastingSearch2022, yaoRiemannianOptimizationPhotonic2024, khanpourRapidGaussianBoson2026} based on efficient and accurate simulation of the heralded states~\cite{gagatsosEfficientRepresentationGaussian2019, provaznikTamingNumericalErrors2022}. The states are then characterized, for example, by their stellar rank~\cite{chabaudStellarRepresentationNongaussian2020, fiurasekEfficientConstructionWitnesses2022b, fiurasekCertificationStellarRanks2025, provaznikWitnessesNonGaussianFeatures2026}, by nonlinear squeezing~\cite{marekGroundStateNature2024,brauerGeneralizedSqueezingWitness2025, kucharNonlinearSqueezingSuperpositions2025}, and by other measures of non-Gaussianity and quality~\cite{genoniQuantifyingNonGaussianCharacter2008, brauerCatabilityMetricEvaluating2025, kucharEfficientCharacterizationGeneral2026}.

The rate of preparation is set by the \emph{heralding probability} $\pS$, the probability that the detectors register the chosen pattern, which generally decreases with the number of detected photons~\PRA. Hanamura \emph{et al.}~\cite{hanamuraStellarRankControl2025a} noticed that $\pS$ can be raised without changing the target. The heralding modes can be \emph{damped} by a noiseless filter, which multiplies the amplitude of $k$ photons in a heralding mode by the $k$th power of a factor smaller than one. Absorbed into the Gaussian input, the damping changes $\pS$ but leaves the heralded state unchanged, and Hanamura \emph{et al.} optimized it numerically.

For one heralding mode, the optimal damping has a closed form~\PRA. For $m$ heralding modes, the conditions for a stationary point of $\pS$ are $m$ polynomial equations in the $m$ damping parameters $X_j$ (one per heralding mode), which Ref.~\QST\ solved for up to four heralding modes and observed that $\pS$ has a single maximum on the set of physical damping parameters and left open whether this property holds in general~\cite[Sec.~4]{fiurasekHeraldingProbabilityOptimization2026}.

Note that the filter does not need an experimental implementation. It is instead intended as a tool to efficiently parametrize and find the optimal Gaussian source state, which is then prepared directly, with different squeezing and a different interferometer. Generally, non-Gaussian state preparation then consists of two nested stages: \emph{target-state design} over the control parameters that fix the heralded state, and, for a given candidate, \emph{damping optimization} over the damping parameters to maximize the heralding probability.

Here we show that the damping optimization problem can be solved efficiently. We prove that, with ideal PNR detection of a single pattern and at fixed control parameters, every stationary point of $\pS$, treated as function of damping parameters, is a global maximum, with or without displacement of the source. The optimal damping can therefore be found efficiently, even for many heralding modes. Damping forms an exponential family of probability distributions~\cite{barndorffnielsenInformationExponentialFamilies1978, brownFundamentalsStatisticalExponential1986}, and the heralding probability is the probability that a member of this family assigns to the registered pattern $\bs n$. As a function of the logarithmic damping parameters $\theta_j=\ln X_j$, this probability is the likelihood of $\bs\theta$. For an exponential family its logarithm is concave, so every local maximum is global.

The same structure has appeared in other contexts. Banchi, Quesada, and Arrazola~\cite{banchiTrainingGaussianBoson2020} trained Gaussian boson samplers~\cite{hamiltonGaussianBosonSampling2017} and found that the derivative of a pattern probability with respect to a scaling factor is proportional to $n_j-\Ex{\nhat_j}$. Their distribution of patterns is an exponential family in the logarithms of the scaling factors, so its log-likelihood is concave in these parameters. Kulesza and Taskar~\cite{kuleszaLearningDeterminantalPoint2012} proved that the log-likelihood of a determinantal point process with kernel $\diag(\bs q)\,S\,\diag(\bs q)$ is concave when $\ln\bs q$ depends linearly on the parameters, which is the form of the damped sources below.\newpage

Our main contributions and results are as follows. We show that, for any supported pattern ($W_{\bs n}>0$), $\ln\pS$ is concave on the convex set of admissible $\bs\theta$ for every pure Gaussian source, and strictly concave unless some subset of heralding modes is in a pure state (Thm.~\ref{thm:main}). Every stationary point of $\pS$ is therefore a global maximum. When every heralding mode is coupled directly to the signal mode and registers at least one photon, the maximum exists and is unique (Cor.~\ref{cor:hill}). At the maximum, the mean photon number of every heralding mode equals its registered photon number, and we call this condition \emph{mean matching}. Newton's method finds the maximum in a few steps of $\mathcal{O}(m^3)$ arithmetic operations each, independently of the number of detected photons. Optimality can then be checked using mean photon numbers of the heralding modes.

A bound on the squeezing keeps the problem convex. For $m$ heralding modes with $n$ photons each, the maximal $\pS$ decays as $n^{-(m+1)/2}$ for a heralded Fock state, and a squeezing bound makes this decay exponential. With lossy detectors, damping changes the fidelity of the heralded state, but $\pS$ still has at most one maximum for one heralding mode. For several heralding modes several maxima can appear as the efficiency decreases.

The rest of the paper is organized as follows. Section~\ref{sec:model} introduces the Gaussian source parametrization, the damping operator, and our standing assumptions. Section~\ref{sec:ideal} establishes the exponential-family structure for ideal detection, proves concavity and uniqueness, and describes the Newton solver. Section~\ref{sec:bounded} treats optimization under a squeezing bound, Sec.~\ref{sec:trials} analyzes photon-number scaling, and Sec.~\ref{sec:detection} studies imperfect detection. Section~\ref{sec:conclusion} concludes, while proofs and numerical details are given in the Appendices and Supplemental Material. Our implementation is available in Ref.~\cite{heraldingDampingRepository}.

\vspace{-0.5em}
\section{Sources and damping parameters}\label{sec:model}
\vspace{-0.25em}
\vspace{-0.25em}
\subsection{Gaussian sources and heralding}
\vspace{-0.25em}\vspace{-0.25em}

\begin{figure}[t]
	\centering
	\includegraphics[width=\columnwidth]{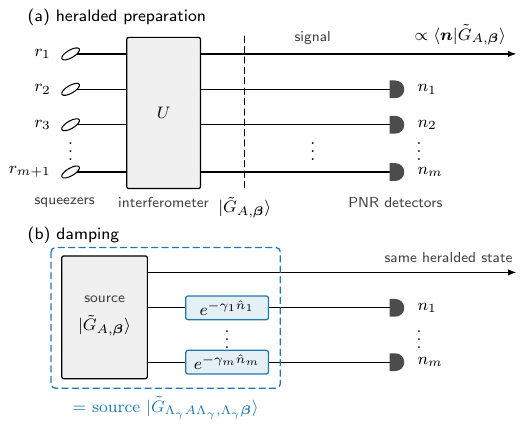}
	\caption{Heralded preparation and damping. \textbf{(a)} Squeezed vacua (optionally displaced) with squeezing parameters $r_i$ pass a passive interferometer $U$ and form the source $|\tilde G_{A,\bs\beta}\rangle$ of one signal mode and $m$ heralding modes. Photon-number-resolving (PNR) detectors register $\bs n=(n_1,\dots,n_m)$ and leave the signal mode in the state proportional to $\langle\bs n|\tilde G_{A,\bs\beta}\rangle$. \textbf{(b)} The source $|\tilde G_{A,\bs\beta}\rangle$ followed by noiseless filters $e^{-\gamma_j\nhat_j}$ on the heralding modes equals the source $|\tilde G_{\Lambda_\gamma A\Lambda_\gamma,\Lambda_\gamma\bs\beta}\rangle$ with $\Lambda_\gamma=\diag(1,e^{-\gamma_1},\dots,e^{-\gamma_m})$, with $\theta_j=-2\gamma_j$ in Eq.~\eqref{eq:damp}. The heralded state is the same as without the filters, and the filters need not be built, because squeezers and an interferometer prepare the damped source directly (dashed blue box).
	\vspace{-0.5em}
	}
	\label{fig:scheme}
\end{figure}

We consider a pure Gaussian state of one signal mode, labeled $0$, and $m$ heralding modes, labeled $1,\dots,m$, and call it the \emph{source}. It can be written as~\QST\vspace{-0.25em}
\begin{equation}\label{eq:state}
	|\tilde G_{A,\bs\beta}\rangle=\exp\!\Big(\tfrac12\textstyle\sum_{j,k=0}^{m}\ad_jA_{jk}\ad_k+\sum_{j=0}^{m}\beta_j\ad_j\Big)\ket{\bs 0},\vspace{-0.25em}
\end{equation}
where $\ad_j$ are creation operators, $\ket{\bs 0}$ the vacuum of all $m+1$ modes, $A$ is a complex symmetric $(m+1)\times(m+1)$ matrix, and the vector $\bs\beta\in\C^{m+1}$ determines the displacement. The tilde indicates that the state is not normalized. It can be normalized iff the largest singular value $\smax(A)$ of $A$ is smaller than one (App.~\ref{app:gauss}), then
\begin{equation}\label{eq:norm}
	\langle\tilde G_{A,\bs\beta}|\tilde G_{A,\bs\beta}\rangle=\frac{\exp\mathrm{Re}\big[\bs\beta^{\dagger}(I-AA^{\dagger})^{-1}(\bs\beta+A\bs\beta^{*})\big]}{\big[\det(I-AA^{\dagger})\big]^{1/2}}.
\end{equation}

This state can be obtained by sending single-mode squeezed vacua with squeezing $r_i$ through a passive interferometer $U$ and displacing its outputs [Fig.~\ref{fig:scheme}(a)]. This follows from the \emph{Takagi factorization}~\cite{takagiAlgebraicProblemRelated1924}, \cite[Cor.~4.4.4(c)]{hornMatrixAnalysis2017} $A=U\diag(\sigma_1,\dots,\sigma_{m+1})U^{\mathsf T}$, where $U$ is unitary and the singular values of $A$ are $\sigma_i=\tanh r_i$. We call the squeezed vacua the \emph{normal modes} of the source and quote their squeezing $10\log_{10}e^{2r_i}\,\mathrm{dB}$. The limit $\smax(A)\to1$ corresponds to infinite squeezing of at least one normal mode.

The heralding modes are measured with ideal PNR detectors. If they register the Fock-state pattern $\bs n=(n_1,\dots,n_m)$, the signal mode is left in the state proportional to\vspace{-0.25em}
\begin{equation*}
	\langle\bs n|\tilde G_{A,\bs\beta}\rangle\equiv\big(\mathds 1_0\otimes\langle n_1|\otimes\dots\otimes\langle n_m|\big)|\tilde G_{A,\bs\beta}\rangle .\vspace{-0.25em}
\end{equation*}
The heralding probability is the probability of this event,\vspace{-0.25em}
\begin{equation}\label{eq:pSdef}
	\pS=\frac{\big\|\langle\bs n|\tilde G_{A,\bs\beta}\rangle\big\|^2}{\langle\tilde G_{A,\bs\beta}|\tilde G_{A,\bs\beta}\rangle}.\vspace{-0.25em}\vspace{-0.25em}\vspace{-0.5em}
\end{equation}

\vspace{-1em}
\subsection{Damping and control parameters}\label{sec:Dop}\vspace{-1em}

Damping multiplies every creation operator $\ad_j$ of a heralding mode by $e^{\theta_j/2}$, where $\theta_j$ is the \emph{logarithmic damping parameter} and $X_j=e^{\theta_j}$ the \emph{damping parameter}. Applied to some \emph{reference source} $|\tilde G_{B,\bs b}\rangle$ with the matrix $B$ and the displacement vector $\bs b$, it gives the \emph{damped source}~\eqref{eq:state} with~\cite{hanamuraStellarRankControl2025a, fiurasekHeraldingProbabilityOptimization2026}\vspace{-0.5em}
\begin{equation}\label{eq:damp}
	A=\Lambda B\Lambda,\ \ \bs\beta=\Lambda\bs b,\ \ \Lambda=\diag\big(1,e^{\theta_1/2},\dots,e^{\theta_m/2}\big).\vspace{-0.5em}
\end{equation}

Damping a damped source by $\bs\theta'$ gives damping $\bs\theta+\bs\theta'$ of the reference, so all damped sources form one family, and any member of it can serve as the reference. The reference itself need not be normalizable. Equivalently, $|\tilde G_{A,\bs\beta}\rangle=\Dop(\bs\theta)|\tilde G_{B,\bs b}\rangle$ with the \emph{damping operator}\vspace{-0.75em}
\begin{equation}\label{eq:Dop}
	\Dop(\bs\theta)=\prod_{j=1}^me^{\theta_j\nhat_j/2},\vspace{-0.75em}
\end{equation}
where $\nhat_j$ is the photon-number operator of heralding mode $j$. Every Fock amplitude then acquires a factor that depends only on $\bs k=(k_1,\dots,k_m)$, the photon numbers of the heralding modes,\vspace{-0.5em}
\begin{equation}\label{eq:amp}
	\langle k_0,\bs k|\tilde G_{A,\bs\beta}\rangle=e^{\bs k\cdot\bs\theta/2}\,\langle k_0,\bs k|\tilde G_{B,\bs b}\rangle,\vspace{-0.25em}\vspace{-0.25em}
\end{equation}
where $k_0$ is the photon number of the signal mode and $\bs k\cdot\bs\theta=\sum_jk_j\theta_j$. Importantly, when registering $\bs n$ in the heralding modes, $e^{\bs n\cdot\bs\theta/2}$ does not depend on $k_0$. The heralded state $\langle\bs n|\tilde G_{A,\bs\beta}\rangle$ is thus the same for every $\bs\theta$, up to normalization,\vspace{-0.25em}
\begin{equation*}
	\langle\bs n|\tilde G_{A,\bs\beta}\rangle=e^{\bs n\cdot\bs\theta/2}\,\langle\bs n|\tilde G_{B,\bs b}\rangle,\vspace{-0.25em}\vspace{-0.25em}
\end{equation*}
and only the heralding probability depends on $\bs\theta$.

Taking a physical source as the reference, $\theta_j<0$ applies the purity-preserving filter $e^{\theta_j\nhat_j/2}$, also known as noiseless attenuation~\cite{micudaNoiselessLossSuppression2012,hanamuraStellarRankControl2025a} [Fig.~\ref{fig:scheme}(b)], while $\theta_j>0$ corresponds to noiseless amplification~\cite{ralphNondeterministicNoiselessLinear2009, micudaNoiselessLossSuppression2012}. These filters need not be built, because the damped source is again~\eqref{eq:state} and is prepared directly by squeezers, an interferometer and displacements. Changing $\bs\theta$ generally changes the squeezing and the interferometer, so a fixed interferometer can reach only some of the damped sources.

We call a source \emph{directly coupled} if $B_{0j}\neq0$ for every heralding mode, i.e., if for every heralding mode $j$ the exponent of Eq.~\eqref{eq:state} contains the term $\ad_0\ad_j$. A heralding mode that is coupled to the signal mode only through other heralding modes still changes the heralded state, so it cannot be left out.

For a directly coupled source we can set $B_{0j}=1$ without loss of generality: a phase rotation of mode $j$ makes $B_{0j}$ positive, and damping multiplies it by $e^{\theta_j/2}$, so $\abs{B_{0j}}$ only fixes the origin of $\theta_j$. The reference matrix is then\vspace{-0.25em}
\begin{equation}\label{eq:B}
	B_{00}=s_0,\ \ B_{0j}=1,\ \ B_{jj}=s_j,\ \ B_{jk}=\nu_{jk}\ \ (j\neq k),\vspace{-0.25em}
\end{equation}
and the damped source has signal--herald couplings $A_{0j}=\sqrt{X_j}$. The first column of $A$ has squared norm $\abs{s_0}^2+\sum_jX_j$, which must not exceed $\smax(A)^2<1$, so every normalizable damped source has $X_j<1$ ($\theta_j<0$). Relative to this reference, damping always attenuates, hence its name. We call $s_0$, $s_j$, $\nu_{jk}=\nu_{kj}$ and the displacement $\bs b$ the \emph{control parameters}; they fix the heralded state, while $\bs\theta$ affects only $\pS$.

The heralded state itself is a squeezed and displaced superposition of Fock states up to $N=\sum_j n_j$ photons, so its stellar rank~\cite{chabaudStellarRepresentationNongaussian2020} is at most $N$ and equals~$N$ under direct coupling (App.~\ref{app:gauss}). Its squeezing and displacement are set by $B_{00}=A_{00}$ and $b_0=\beta_0$, which damping leaves unchanged; we assume $\abs{B_{00}}<1$ throughout so that normalizable damped sources exist. Setting $B_{00}=\beta_0=0$ gives finite Fock superpositions known as \emph{core states}~\cite{menziesGaussianoptimizedPreparationNongaussian2009, chabaudStellarRepresentationNongaussian2020, fiurasekHeraldingProbabilityOptimization2026}, and we choose $s_0=0$ in examples.

\vspace{-1.5em}
\subsection{Heralding probability}\label{sec:pS}
\vspace{-0.75em}\vspace{-0.25em}\vspace{-0.25em}

The \emph{weights}\vspace{-1em}\vspace{-0.25em}
\begin{equation}\label{eq:W}
	W_{\bs k}=\sum_{k_0}\abs{\langle k_0,\bs k|\tilde G_{B,\bs b}\rangle}^2\vspace{-0.5em}\vspace{-0.25em}
\end{equation}
are finite even if $|\tilde G_{B,\bs b}\rangle$ itself is not normalizable, because $W_{\bs k}e^{\bs k\cdot\bs\theta}\le\langle\tilde G_{A,\bs\beta}|\tilde G_{A,\bs\beta}\rangle$ whenever the damped source is normalizable. Equations~\eqref{eq:norm}, \eqref{eq:pSdef} and~\eqref{eq:amp} imply\vspace{-0.25em}\vspace{-0.25em}
\begin{equation}\label{eq:pS}
	\pS(\bs\theta)=\frac{W_{\bs n}\,e^{\bs n\cdot\bs\theta}}{\langle\tilde G_{A,\bs\beta}|\tilde G_{A,\bs\beta}\rangle},\vspace{-0.25em}
\end{equation}
for heralding probability defined on the \emph{admissible set}\vspace{-0.25em}\vspace{-0.25em}
\begin{equation}\label{eq:Omega}
	\Om=\big\{\bs\theta\in\R^m:\ \smax(\Lambda B\Lambda)<1\big\},\vspace{-0.25em}\vspace{-0.25em}
\end{equation}
where the damped source is normalizable.

\vspace{-1em}\vspace{-0.25em}\vspace{-0.25em}
\subsection{Assumptions}\label{sec:assumptions}
\vspace{-1em}

We generally use three assumptions on the source and the pattern, though not every result needs all three.
\begin{enumerate}[label=(H\arabic*), ref=H\arabic*, leftmargin=2.4em, labelsep=0.5em, topsep=0.25em, itemsep=0.25em, parsep=0pt, partopsep=0pt]
	\item\label{as:photon} Every heralding mode registers at least one photon, $n_j\ge1$. Projecting a mode onto the vacuum leaves a pure Gaussian state of the other modes, which heralds the same state with at least the same probability and at most the same squeezing (App.~\ref{app:gauss}), as a Gaussian measurement does not increase the squeezing of the remaining modes~\cite{krausEntanglementGenerationHamiltonian2003,aghaeeradScalingNetworkingModular2025, fiurasekSimplifiedSchemeContinuousvariable2026}.
	\item\label{as:pure} No set of heralding modes is in a pure state. Such a set does not affect the heralded state, and leaving it undetected raises $\pS$, excluding wasteful schemes.
	\item\label{as:direct} The source is directly coupled, $B_{0j}\neq0$ for every $j$. This is stronger than~\ref{as:pure}.
\end{enumerate}
Assumption~\ref{as:pure} makes the optimal damping unique (Thm.~\ref{thm:main}), and the stronger~\ref{as:direct} also guarantees that it exists (Cor.~\ref{cor:hill}). Both are needed only for these proofs. A local method maximizes $\pS$ also without them, and Sec.~\ref{sec:unique} describes what it then finds. With imperfect detection, the argument for \ref{as:photon} is invalid as a detected $0$ is not a projection onto the vacuum, see Sec.~\ref{sec:detection}.

Targets with several signal modes, such as the two-signal-mode states of Ref.~\QST, can be treated in the same way. Results carry over with~\ref{as:pure} unchanged and~\ref{as:direct} requiring each heralding mode to couple directly to at least one signal mode (Suppl., Sec.~\ref{supp:signal}).\vspace{-0.5em}\vspace{-0.25em}
\section{Ideal detection}\label{sec:ideal}
\vspace{-0.25em}

\vspace{-0.5em}\vspace{-0.25em}
\subsection{Exponential family}\vspace{-0.5em}\vspace{-0.25em}\vspace{-0.25em}

Summing the squared amplitudes~\eqref{eq:amp} over $k_0$ and normalizing gives the photon-number distribution of the heralding modes in the damped state, which we call the \emph{herald statistics},\vspace{-0.25em}
\begin{equation}\label{eq:tilt}
	\begin{aligned}
		p_{\bs\theta}(\bs k) & =W_{\bs k}\,e^{\bs k\cdot\bs\theta-\Phi(\bs\theta)},                                                                   \\
		\Phi(\bs\theta)      & =\ln\sum_{\bs k}W_{\bs k}e^{\bs k\cdot\bs\theta}=\ln\langle\tilde G_{A,\bs\beta}|\tilde G_{A,\bs\beta}\rangle,\vspace{-0.5em}\vspace{-0.75em}\vspace{-0.25em}
	\end{aligned}
\end{equation}
where the second form follows from Eqs.~\eqref{eq:amp} and~\eqref{eq:W}, and reduces to $\Phi=-\tfrac12\ln\det(I-AA^{\dagger})$ without displacement. The heralding probability~\eqref{eq:pS} is the value of this distribution at the registered pattern, $\pS(\bs\theta)=p_{\bs\theta}(\bs n)$.

\begin{figure*}[t]
	\centering
	\includegraphics[width=0.95\textwidth]{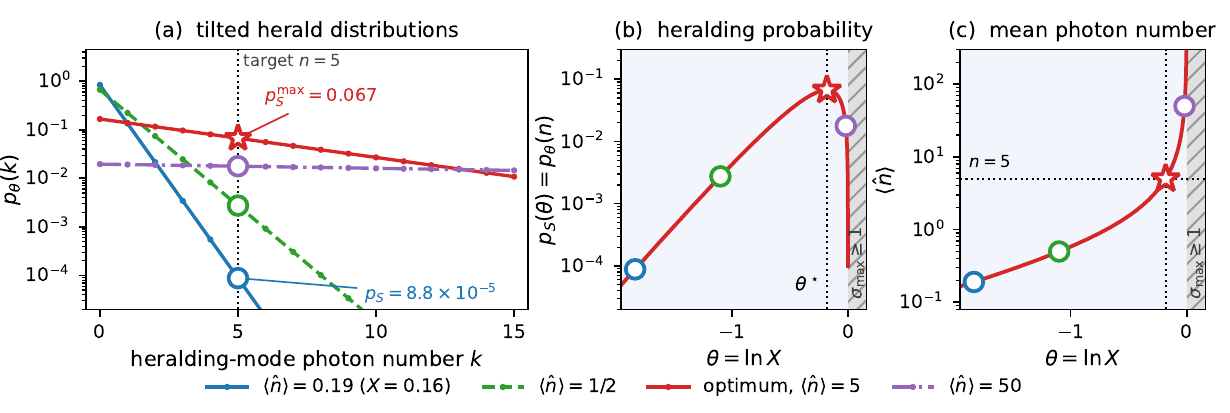}
	\vspace{-0.8em}
	\caption{Damping as exponential tilting for one heralding mode with $s_1=0$, postselected on $n=5$, which heralds the Fock state $\ket5$. The heralding mode is thermal with mean photon number $X/(1-X)$, and the admissible set is $\theta<0$. Four members of the family share colors and markers in all panels: $\Ex{\nhat}=0.19$ ($X=0.16$), $\Ex{\nhat}=1/2$, the optimum ($\Ex{\nhat}=n=5$, star) and $\Ex{\nhat}=50$. \textbf{(a)} Herald distributions $p_\theta(k)$ of Eq.~\eqref{eq:tilt}. On the logarithmic axis, tilting is a shear and a shift. The heralding probability is $8.8\times10^{-5}$ at $X=0.16$ and $0.067$ at the optimum, a gain of $7.6\times10^{2}=\exp[\KL(p_{\theta^\star}\|p_{\theta})]$, with $\KL$ the Kullback--Leibler divergence [Eq.~\eqref{eq:S-gain}]. \textbf{(b)} $\pS(\theta)=p_\theta(n)$ is log-concave, with a single maximum at $\theta^\star=\ln(5/6)$, and vanishes at the boundary of the admissible set (hatched region: not normalizable). \textbf{(c)} The mean photon number grows monotonically, $\d\Ex{\nhat}/\d\theta=\Sigma>0$, and diverges at the boundary, so it equals $n$ exactly once [Eq.~\eqref{eq:match}].\vspace{-1em}}
	\label{fig:tilting}
\end{figure*}

Equation~\eqref{eq:tilt} multiplies the fixed weights $W_{\bs k}$ by $e^{\bs k\cdot\bs\theta}$, a reweighting known as \emph{exponential tilting}. Increasing $\theta_j$ shifts probability towards larger photon numbers in mode $j$, as shown in Fig.~\ref{fig:tilting} for one heralding mode.

In statistics, a set of distributions of the form~\eqref{eq:tilt} is called an \emph{exponential family}~\cite{barndorffnielsenInformationExponentialFamilies1978, brownFundamentalsStatisticalExponential1986}, and $\Phi$, the logarithm of the normalization constant, is called its \emph{log-partition function}. Table~\ref{tab:glossary} in App.~\ref{app:expfam} translates other terms of the theory into the heralding problem.

Differentiating under the sum, $\partial\Phi/\partial\theta_j=\sum_{\bs k}k_jp_{\bs\theta}(\bs k)$, gives the mean photon numbers and the photon-number covariance matrix $\Sigma$ of the damped state\vspace{-0.5em}
\begin{equation}\label{eq:derivs}
	\frac{\partial\Phi}{\partial\theta_j}=\Ex{\nhat_j},\qquad
	\frac{\partial^2\Phi}{\partial\theta_j\partial\theta_k}=\Cov(\nhat_j,\nhat_k)\equiv\Sigma_{jk}.\vspace{-0.5em}
\end{equation}

Regarded as a function of $\bs\theta$ at fixed $\bs n$, the probability $p_{\bs\theta}(\bs n)$ is the \emph{likelihood} of the parameter $\bs\theta$ given the observation $\bs n$. Maximizing $\pS$ is therefore formally a maximum-likelihood estimation of $\bs\theta$ in which the chosen pattern $\bs n$ plays the role of a sole observation.

\vspace{-0.5em}\vspace{-0.25em}
\subsection{Concavity, uniqueness and existence}\label{sec:unique}
\vspace{-0.5em}\vspace{-0.25em}

The sum in Eq.~\eqref{eq:tilt} converges if and only if $\bs\theta$ lies in the admissible set $\Om$ of Eq.~\eqref{eq:Omega} (Prop.~\ref{prop:natural} in App.~\ref{app:expfam}), because it is the squared norm of $|\tilde G_{A,\bs\beta}\rangle$.

\begin{theorem}\label{thm:main}
	Let $A=\Lambda B\Lambda$ with complex symmetric $B$ and displacement $\bs\beta=\Lambda\bs b$ be such that no set of heralding modes is in a pure state (\ref{as:pure}), and let $\bs n$ be the registered pattern. Then $\Om$ is convex, and if $W_{\bs n}>0$, $\ln\pS$ is strictly concave on $\Om$. Hence $\pS$ has at most one stationary point, and a local maximum of $\pS$ on $\Om$ is its unique global maximum.\vspace{-0.5em}
\end{theorem}
\begin{proof}
	Expanding on~\cite[Thm.~1.13]{brownFundamentalsStatisticalExponential1986}, we first show $\Om$ is convex: Let $\bs\theta,\bs\theta'\in\Om$ and $u\in(0,1)$. H\"older's inequality states $\sum_{\bs k}a_{\bs k}b_{\bs k}\le(\sum_{\bs k}a_{\bs k}^{1/u})^u(\sum_{\bs k}b_{\bs k}^{1/(1-u)})^{1-u}$ for $a_{\bs k}$, $b_{\bs k}\geq 0$. With $a_{\bs k}=(W_{\bs k}e^{\bs k\cdot\bs\theta})^u$ and $b_{\bs k}=(W_{\bs k}e^{\bs k\cdot\bs\theta'})^{1-u}$ it gives\vspace{-1.5em}
	\begin{equation*}
		\qquad\sum_{\bs k}W_{\bs k}e^{\bs k\cdot[u\bs\theta+(1-u)\bs\theta']}\le e^{u\Phi(\bs\theta)+(1-u)\Phi(\bs\theta')}<\infty,\vspace{-0.5em}\vspace{-0.25em}
	\end{equation*}
	so the point $u\bs\theta+(1-u)\bs\theta'$ lies in $\Om$.

	Next, by Eq.~\eqref{eq:tilt}, $\ln\pS(\bs\theta)=\ln W_{\bs n}+\bs n\cdot\bs\theta-\Phi(\bs\theta)$. The first term does not depend on $\bs\theta$, the second is linear in $\bs\theta$, so $\nabla^2\ln\pS\!=\!(\partial^2\ln\pS/\partial\theta_j\partial\theta_k)=-\Sigma$ with the photon-number covariance matrix $\Sigma$ of Eq.~\eqref{eq:derivs}. For any real vector $\bs v$, the quadratic form $\bs v^{\mathsf T}\Sigma\bs v=\Var\big(\sum_jv_j\nhat_j\big)$ is never negative, so $\ln\pS$ is concave.

	Concavity already makes every local maximum global. For uniqueness we prove strict concavity, $ \bs v\neq\bs0 \implies \bs v^{\mathsf T}\Sigma\bs v>0$. Suppose instead that $\bs v^{\mathsf T}\Sigma\bs v=\Var(\hat L)=0$ for some $\bs v\neq\bs0$ and $\hat L=\sum_jv_j\nhat_j$. We then show that a set of heralding modes is in a pure state, contradicting~\ref{as:pure}.

	First, the damped source is invariant under a phase rotation. A state in which $\hat L$ has zero variance is an eigenvector of $\hat L$; this includes $|\tilde G_{A,\bs\beta}\rangle$. Its vacuum amplitude is one and $\hat L\ket{\bs0}=0$, so the eigenvalue is $0$, and the phase rotation $e^{-it\hat L}$, which multiplies every $\ad_j$ by $e^{-iv_jt}$, leaves the source unchanged for every real $t$.

	Second, with $v_0=0$ for the signal mode, the rotated source is again of the form~\eqref{eq:state}, with $A_{jk}$ multiplied by $e^{-i(v_j+v_k)t}$ and $\beta_j$ by $e^{-iv_jt}$. A source of this form determines $A$ and $\bs\beta$ through its amplitudes with one and two photons, so $(v_j+v_k)A_{jk}=0$ for all $j$ and $k$.

	Third, the modes with $v_j\neq0$ decouple from the others. Let $\mathcal P$ be the set of these modes. It is not empty because $\bs v\neq\bs0$, and it contains only heralding modes because $v_0=0$. For $j\in\mathcal P$ and $k\notin\mathcal P$, $(v_j+v_k)A_{jk}=v_jA_{jk}=0$ gives $A_{jk}=0$, so no term of the exponent in Eq.~\eqref{eq:state} couples the modes in $\mathcal P$ to the others. The damped source is then a product of a state of the modes in $\mathcal P$ and a state of the others, and the modes in $\mathcal P$ are in a pure state.

	Hence $\Sigma$ is positive definite on $\Om$, and $\ln\pS$ is strictly concave. A strictly concave function on a convex set has at most one stationary point, and a local maximum is its unique global maximum~\cite[Sec.~4.2]{boydConvexOptimization2004}.
\end{proof}

If a set of heralding modes is in a pure state, so that \ref{as:pure} fails, $\ln\pS$ is still concave, so every local maximum is still global, but the maximizers can form a convex set instead of a single point. All of them give the same damped source (App.~\ref{app:expfam}), so the optimal source is still unique.

By Eqs.~\eqref{eq:tilt} and~\eqref{eq:derivs}, the gradient of $\ln\pS$ is $\bs n-\Ex{\hat{\bs n}}$. Setting it to zero gives the optimality condition\vspace{-0.5em}
\begin{equation}\label{eq:match}
	\Ex{\nhat_j}=n_j,\qquad j=1,\dots,m .\vspace{-0.5em}
\end{equation}
We call Eq.~\eqref{eq:match} the \emph{mean-matching condition}. It states that a source is optimally damped if and only if the mean photon number of every heralding mode equals the photon number $n_j$ in the chosen pattern.

Theorem~\ref{thm:main} does not state that a maximum exists. For a minimal exponential family (\ref{as:pure}), the likelihood attains its maximum if and only if~\cite[Cor.~9.6]{barndorffnielsenInformationExponentialFamilies1978}, \cite[Thm.~5.5]{brownFundamentalsStatisticalExponential1986}\vspace{-0.5em}
\begin{equation}\label{eq:hull}
	\bs n\in\operatorname{int}\overline{\conv}\,\{\bs k:\ W_{\bs k}>0\},\vspace{-0.5em}
\end{equation}
where $\overline{\conv}$ is the closed convex hull and $\operatorname{int}$ the interior. For a directly coupled source (\ref{as:direct}), $W_{\bs k}>0$ for every pattern (App.~\ref{app:gauss}), so the hull is the closed orthant $[0,\infty)^m$, and condition~\eqref{eq:hull} holds whenever $n_j\ge1, \forall j$ (\ref{as:photon}).

\begin{corollary}\label{cor:hill}
	For a directly coupled source (\ref{as:direct}) with $n_j\ge1, \forall j$(\ref{as:photon}), $\pS$ attains its maximum on $\Om$ at a unique maximizer $\bs\theta^\star$, and $\bs\theta^\star$ is the only solution of Eq.~\eqref{eq:match}.
\end{corollary}
\begin{proof}
	The damping parameters are bounded from above on $\Om$. $A_{j0}=B_{0j}e^{\theta_j/2}$ implies $\smax(A)^2\ge\sum_j\abs{B_{0j}}^2e^{\theta_j}$, and $\smax(A)<1$ implies $\theta_j<-2\ln\abs{B_{0j}}$.

	Since $\langle\bs0|\tilde G_{A,\bs\beta}\rangle=1$, $e^{\Phi(\bs\theta)}=\langle\tilde G_{A,\bs\beta}|\tilde G_{A,\bs\beta}\rangle\ge1$, and Eq.~\eqref{eq:pS} gives $\pS\le W_{\bs n}e^{\bs n\cdot\bs\theta}$. Since $n_j>0$ and $\theta_k$ are bounded from above, this bound tends to zero as any $\theta_j\to-\infty$. At the boundary of $\Om$, $e^{\Phi(\bs\theta)}$ diverges while $W_{\bs n}e^{\bs n\cdot\bs\theta}$ stays finite [Eq.~\eqref{eq:norm}; App.~\ref{app:gauss}], so $\pS\to0$ there.

	Take $\varepsilon=\pS(\bs\theta_0)$ for some $\bs\theta_0\in\Om$. The set of points of $\Om$ with $\pS\ge\varepsilon$ is then nonempty. It is bounded, because $\pS<\varepsilon$ once some $\theta_j$ is sufficiently negative, and closed, because it stays away from the boundary of $\Om$. The continuous function $\pS$ attains its maximum on this set, and this is its maximum on $\Om$. By Thm.~\ref{thm:main}, the maximizer is unique and is the only stationary point.
\end{proof}
The proof uses only $n_j>0$. For real patterns with $n_j>0$, it shows in the same way that $\bs n\cdot\bs\theta-\Phi(\bs\theta)$, which equals $\ln\pS-\ln W_{\bs n}$ for integer patterns, has a unique maximizer, at which $\Ex{\hat{\bs n}}=\bs n$. The continuation of Sec.~\ref{sec:newton} uses such patterns.

The proof also gives intuition for locating the maximum. A heralding mode with $\theta_j\to-\infty$ approaches the vacuum and cannot supply its $n_j\ge1$ photons, and at the boundary of $\Om$ the source is infinitely squeezed and every fixed pattern becomes improbable, i.e.\vspace{-0.33em}
\begin{equation*}
	\pS(\bs\theta)\to0\quad\text{for}\quad\theta_j\to-\infty\quad\text{and for}\quad\smax(A)\to1. \vspace{-0.33em}
\end{equation*}
The maximum lies between these limits and direct coupling (\ref{as:direct}) bounds every $\theta_j$ from above. Without it, $\theta_j\to+\infty$ can appear, along which $\pS$ increases towards a supremum with heralding modes in a pure state. Leaving these modes out gives a smaller scheme that heralds the same state with a higher probability (Suppl., Sec.~\ref{supp:nodirect}). Without direct coupling, the maximum can therefore fail to exist only for schemes that can be outperformed by a smaller scheme.

\vspace{-0.75em}
\subsection{Finding the optimum}\label{sec:newton}
\vspace{-0.75em}

For one heralding mode, the mean-matching condition~\eqref{eq:match} is a quadratic equation for $X$, or a quartic one if the source is displaced and $s_1\neq0$~\PRA. For $m\ge2$ and without displacement, Ref.~\QST\ writes the stationarity conditions as polynomial equations $Q_j=0$. With $K=\diag(1,X_1,\dots,X_m)$ and $J(\bs X)=I-BKB^{\dagger}K$, for which $\det J=\det(I-AA^{\dagger})$, its extremal equations are\vspace{-0.33em}
\begin{equation}\label{eq:Qdef}
		Q_j(\bs X)\!\equiv\!\big(X_j\partial_{X_j}+2n_j\big)\det J = 2\det J\,\big(n_j-\Ex{\nhat_j}\big),\vspace{-0.33em}
\end{equation}
where the second form holds on $\Om$ (App.~\ref{app:gauss}). The extremal equations are therefore the mean-matching condition multiplied by $2\det J$, which is positive on $\Om$.

The roots of $Q_j(\bs X)$ in $\Om$ are stationary points of~$\pS$. Since $X_j\partial_{X_j}(\bs X^{2\bs n}\det J)=\bs X^{2\bs n}Q_j$ with $\bs X^{2\bs n}=\prod_jX_j^{2n_j}$, all roots with nonzero $X_j$ are stationary points of the polynomial $\bs X^{2\bs n}\det J$, which equals $(\pS/W_{\bs n})^2$ on $\Om$. Its other stationary points have zero, negative, or complex $X_j$, or positive $X_j$ outside $\Om$, where the source is not normalizable. Computing and discarding them is what makes the system hard to solve for many heralding modes.

By Thm.~\ref{thm:main} and Cor.~\ref{cor:hill}, these equations need not be solved. It suffices to find a local maximum, which can be done efficiently using values and derivatives of $\ln\pS$ at the iterates of local optimization methods~\cite[Sec.~1.4.1]{boydConvexOptimization2004}.

By Eqs.~\eqref{eq:tilt} and~\eqref{eq:derivs} and Wick's theorem (App.~\ref{app:gauss}), the gradient and the Hessian of $\ln\pS$ are\vspace{-0.5em}
\begin{equation}\label{eq:gradhess}
	\nabla\ln\pS=\bs n-\Ex{\hat{\bs n}},\qquad \nabla^2\ln\pS=-\Sigma .
\end{equation}
Defining $\mathcal C=(I-A^{\dagger}A)^{-1}-I$ and $\mathcal M=A(I-A^{\dagger}A)^{-1}$, their entries are the moments $\Ex{\ad_j\hat a_k}$ and $\Ex{\hat a_j\hat a_k}$ of an undisplaced source. For $\bs b=\bs0$,
\begin{equation}\label{eq:cov}
	\Ex{\nhat_j}=\mathcal C_{jj},\qquad \Sigma_{jk}=\abs{\mathcal M_{jk}}^2+\abs{\mathcal C_{jk}}^2+\delta_{jk}\,\mathcal C_{jj},
\end{equation}
so one iterate requires inverting an $(m+1)\times(m+1)$ matrix. Displacements need no further inversion [Eq.~\eqref{eq:S-displaced-cov}].

We maximize $\ln\pS$ by Newton's method with backtracking~\cite[Sec.~9.5]{boydConvexOptimization2004}. Its step $\Sigma^{-1}(\bs n-\Ex{\hat{\bs n}})$ is also the Newton step for solving Eq.~\eqref{eq:match} as a system of equations, but the value of $\ln\pS$ also measures progress and steps that leave $\Om$ or decrease $\pS$ are shortened. The iterates therefore stay in the set where $\pS$ is at least its initial value, which is closed and bounded (proof of Cor.~\ref{cor:hill}). On this set $\Sigma$ is Lipschitz continuous and its smallest eigenvalue is nonzero, which allows for a standard convergence proof~\cite[Sec.~9.5.3]{boydConvexOptimization2004}.

\begin{figure}[tb]
	\centering
	\includegraphics[width=\columnwidth]{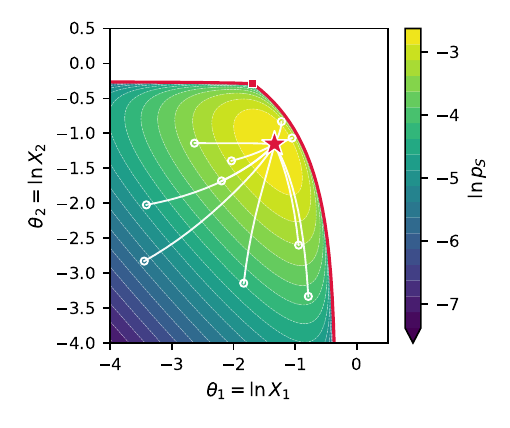}\vspace{-1em}
	\caption{An example with two heralding modes, real $B$, control parameters $s_1=2/5$, $s_2=-3/10$, $\nu_{12}=3/5$ and the pattern $\bs n=(1,1)$: $\ln\pS$ plotted over $\theta_j=\ln X_j$. The crimson curve $\smax(A)=1$ bounds the admissible set $\Om$, and the square marks the corner where two normal modes are infinitely squeezed at once. White lines track the flow $\d\bs\theta/\d t=\Sigma^{-1}(\bs n-\Ex{\hat{\bs n}})$, the continuous-time limit of Newton's method, from admissible starting points (circles). Trajectories of this flow cannot cross, and along each of them $\Ex{\hat{\bs n}}$ moves on a straight line to $\bs n$. All reach the maximum (star) at $\bs X^\star=(0.2627,0.3158)$, where $\Ex{\hat{\bs n}}=(1,1)$.\vspace{-1.25em}}%
	\label{fig:landscape}
\end{figure}

The method thus converges to the maximum from every admissible starting point, quadratically near the maximum (Fig.~\ref{fig:landscape}). For random sources with up to $50$ heralding modes and up to nine photons per mode, it needed a median of $6$ to $10$ steps and took milliseconds to tens of milliseconds on one CPU core.

Patterns with thousands of photons per mode can need several hundred Newton steps, because the iterates take many short steps near the boundary of $\Om$. We can instead use a continuation in the pattern whenever some $n_j$ exceeds $16$. The method first maximizes $\bs n'\cdot\bs\theta-\Phi(\bs\theta)$ for $\bs n'=\bs n/4^k$, where $k$ is the smallest integer for which no entry of $\bs n'$ exceeds $16$, and then for $\bs n'=\bs n/4^{k-1},\dots,\bs n/4$ and finally $\bs n$, each time starting from the previous optimum. The intermediate patterns are generally not integers, but the maximizer still exists and is unique (remark after Cor.~\ref{cor:hill}). The factor $4$ is an arbitrary choice that worked well in our tests. With this continuation, at most $51$ steps sufficed for patterns with up to $9\times10^4$ photons per mode (Suppl., Sec.~\ref{supp:newton}).

\vspace{-0.5em}
\section{Bounded squeezing}\label{sec:bounded}
\vspace{-0.75em}
\subsection{Squeezing bound}
\vspace{-0.75em}
So far, every source with $\smax(A)<1$ has been allowed, implying arbitrarily large squeezing. The optima without a bound often require strong squeezing. For the undisplaced random sources of Sec.~\ref{sec:newton} ($s_0=0$, independent standard complex Gaussian $s_j,\nu_{jk}$, and $1\le n_j\le 9$; Suppl., Table~\ref{tab:newton}), the median squeezing of the most squeezed normal mode at the optimum is $15$\,dB for $m=1$ and $32$\,dB for $m=50$. For the heralded Fock state $\ket n$ from a two-mode squeezed vacuum, the optimal squeezing is $20\log_{10}(\sqrt n+\sqrt{n+1})$\,dB~\cite{christLimitsDeterministicCreation2012, tiedauScalabilityParametricDownconversion2019, fiurasekMaximumHeraldingProbabilities2026}, which is $15.3$\,dB at $n=8$.
Squeezing of $15$\,dB has been detected directly~\cite{vahlbruchDetection15DB2016}, but any given experiment is limited by the squeezing that its resources provide.

The example sources of Fig.~\ref{fig:bounded} need between $12.8$ and $20$\,dB at their unbounded optima, so a bound of $10$ or $12$\,dB is active for every one of them, as it is for essentially every random source at $10$\,dB (Suppl., Sec.~\ref{supp:bounded}). With the squeezing available in experiments, the optimum under a bound on the maximum available squeezing is therefore the relevant one. Following Ref.~\QST, we impose a bound on the squeezing of every normal mode,\vspace{-0.25em}
\begin{equation}\label{eq:bound}
	\smax(A)\le\mu,\qquad \mu=\tanh r_{\max}<1 .\vspace{-0.25em}
\end{equation}
In decibels the bound is $10\log_{10}e^{2r_{\max}}$, and a bound of $10$\,dB corresponds to $\mu=9/11$. We write $\pS^{\max}(\mu)$ for the maximum of $\pS$ under the bound and $\pS^{\max}$ for the maximum without a bound.

Damping does not change $B_{00}=A_{00}$. The norm of the first column of $A$ cannot exceed $\smax(A)$, so\vspace{-0.5em}
\begin{equation*}
	\smax(A)^2\ge\abs{B_{00}}^2+\sum_j\abs{A_{0j}}^2>\abs{B_{00}}^2\vspace{-0.33em}\vspace{-0.33em}
\end{equation*}
for every $\bs\theta$. The last inequality is strict because some $A_{0j}\neq0$, as otherwise all heralding modes together would be in a pure state, which \ref{as:pure} excludes. As all $\theta_j\to-\infty$, the heralding modes approach the vacuum and $\smax(A)\to\abs{B_{00}}$. Some damping therefore satisfies the bound~\eqref{eq:bound} if and only if $\mu>\abs{B_{00}}$, and we assume this from now. With $\abs{B_{00}}=\tanh r_0$, the condition reads $r_{\max}>r_0$, so the bound has to allow more squeezing than the squeezing $r_0$ of the heralded state.

The bound changes neither the objective $\ln\pS$ nor the heralded state. It only shrinks the feasible set, and the question is whether this set is still convex. The bound~\eqref{eq:bound} replaces the admissible set $\Om$ by\vspace{-0.33em}
\begin{equation}\label{eq:Omegamu}
	\Om_\mu=\big\{\bs\theta\in\R^m:\ \smax(\Lambda B\Lambda)\le\mu\big\}.\vspace{-0.33em}
\end{equation}

The set $\Om_\mu$ is convex, because $\ln\smax(\Lambda B\Lambda)$ is a convex function of $\bs\theta$ (Prop.~\ref{prop:logconvex} in App.~\ref{app:capped}) and $\Om_\mu$ is the set where this function is at most $\ln\mu$.

The bounded problem therefore again maximizes a concave function over a convex set, which is a convex optimization problem. For directly coupled sources (\ref{as:direct}), the maximum exists by the argument of Cor.~\ref{cor:hill}, it is unique, and every local maximum is global, so a local method again finds the global maximum.

Since the bounded problem is convex, its maximum can be found with standard interior-point methods~\cite[Ch.~11]{boydConvexOptimization2004}. We use the barrier method~\cite[Sec.~11.3]{boydConvexOptimization2004}, which maximizes\vspace{-0.33em}\vspace{-0.33em}\vspace{-0.33em}
\begin{equation}\label{eq:barrier}
	\tau\ln\pS(\bs\theta)+\ln\det\big(\mu^2I-AA^{\dagger}\big)\vspace{-0.33em}
\end{equation}
by Newton's method for an increasing sequence of $\tau>0$. The second term diverges at the boundary of $\Om_\mu$ and keeps the iterates inside (Suppl., Sec.~\ref{supp:bounded}).
\vspace{-1em}
\subsection{Optimality conditions}\label{sec:kkt}\vspace{-0.33em}
\vspace{-0.5em}
If $\smax<\mu$ at the optimum, the bound is \emph{inactive} and mean matching~\eqref{eq:match} still holds. If $\smax=\mu$, the bound is \emph{active} and the optimum lies on the boundary of $\Om_\mu$. There the gradient $\bs n-\Ex{\hat{\bs n}}$ of $\ln\pS$ need not vanish, but it must point out of $\Om_\mu$, because no direction into $\Om_\mu$ may increase $\pS$~\cite[Sec.~4.2.3]{boydConvexOptimization2004}.

To make this explicit, we use the Takagi factorization $A=U\diag(\sigma_1,\sigma_2,\dots)U^{\mathsf T}$ of Sec.~\ref{sec:model}, with $\sigma_1=\smax$. The columns of $U$ describe the normal modes (App.~\ref{app:gauss}), and $U_{ji}$ is the amplitude of mode $j$ in the normal mode with singular value $\sigma_i$. Hence $\abs{U_{j1}}^2$ is the weight of heralding mode $j$ in the most squeezed normal mode. If $\sigma_1$ is nondegenerate, the outward normal of the boundary is the gradient of $\ln\smax$, whose components are $\abs{U_{j1}}^2$ (App.~\ref{app:capped}), and the optimality condition takes the Karush--Kuhn--Tucker form\vspace{-0.25em}
\begin{equation}\label{eq:kkt}
	\Ex{\nhat_j}=n_j-\xi\abs{U_{j1}}^2,\qquad\xi=\frac{\d\ln\pS^{\max}(\mu)}{\d\ln\mu}\ge0.\vspace{-0.25em}
\end{equation}
By the sensitivity interpretation of Lagrange multipliers~\cite[Sec.~5.6]{boydConvexOptimization2004}, $\xi$ is the relative gain of the maximal heralding probability per unit increase of $\ln\mu$. When $\xi>0$, a source with bounded squeezing thus delivers fewer photons in total than the pattern prescribes, mostly in the heralding modes that constitute the most squeezed normal mode.

The optimum can also be degenerate where the two largest singular values both equal $\mu$. There the boundary of $\Om_\mu$ can have an edge (a corner for $m=2$), where the outward normal is not unique and a positive semidefinite $2\times2$ matrix takes the place of $\xi$ (Suppl., Sec.~\ref{supp:bounded}). Such optima occur often---for real control parameters from $m=2$ and for complex ones from $m=3$---and the barrier method finds them (Suppl., Sec.~\ref{supp:bounded}).

\vspace{-0.5em}\vspace{-0.25em}
\subsection{Dependence on the bound}\label{sec:dB}
\vspace{-0.5em}

\begin{figure}[t]
	\centering
	\includegraphics[width=\columnwidth]{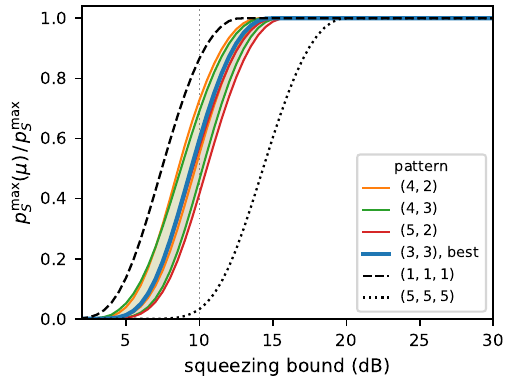}
	\caption{Effect of a squeezing bound with ideal detection: maximal heralding probability with the bound, $\pS^{\max}(\mu)$, relative to the unbounded maximum. Two heralding modes: sources that herald the balanced superpositions $(\ket0+\ket2+\ket4+\ket6)/2$ with the patterns $(3,3)$ and $(4,2)$, and $(\ket1+\ket3+\ket5+\ket7)/2$ with $(4,3)$ and $(5,2)$. For each pattern, six choices of $s_1$, $s_2$ and $\nu_{12}$ herald the target~\cite{fiurasekHeraldingProbabilityOptimization2026}. For $(4,2)$, $(4,3)$ and $(5,2)$, two thin lines bound the range over these six sources. For $(3,3)$, the thick line shows the source with the largest $\pS^{\max}$, $s_1=s_2^*=-0.9462+0.6972i$, $\nu_{12}=1.2394$, and the curves of the other five sources with this pattern lie within $0.015$ of it. Three heralding modes: black lines show the source $s_j=i$, $\nu_{12}=6/7$, $\nu_{13}=-5/11$, $\nu_{23}=2/3$ with the patterns $(1,1,1)$ (dashed) and $(5,5,5)$ (dotted). The dotted vertical line marks $10$\,dB.\vspace{-1em}}
	\label{fig:bounded}
\end{figure}

With the inequality~\eqref{eq:bound}, $\pS^{\max}(\mu)$ grows with $\mu$ and saturates at $\pS^{\max}$ once the bound exceeds the squeezing of the unbounded optimum. Fixing $\smax=\mu$ instead, as in Ref.~\cite[Fig.~5]{fiurasekHeraldingProbabilityOptimization2026}, restricts $\bs\theta$ to the boundary of $\Om_\mu$, which is not convex, so the maximum decreases again beyond this squeezing and $\pS$ can have several local maxima on the boundary (Suppl., Sec.~\ref{supp:bounded}). If the bound~\eqref{eq:bound} is active, its unique maximum lies on this boundary and is also the global maximum there.

Figure~\ref{fig:bounded} illustrates the cost of the bound for specific target states and patterns from Ref.~\QST. For these four-component Fock-superposition sources with two heralding modes, a $10$\,dB bound keeps $40$ to $75\,\%$ of their unbounded maximum, whereas for the three-mode example with the pattern $(5,5,5)$, whose unbounded optimum needs about $20$\,dB, it keeps only a few percent.

Sources whose unbounded optima need more squeezing lose more. For the benchmarking random sources of Sec.~\ref{sec:newton}, the median of $\pS^{\max}(\mu)/\pS^{\max}$ at $10$\,dB falls from $0.15$ for $m=2$ to $1.4\times10^{-8}$ for $m=12$.

\vspace{-0.5em}
\section{Photon-number scaling}\label{sec:trials}
\vspace{-0.25em}
\subsection{Without a bound}\label{sec:trials-free}
\vspace{-0.5em}

A heralded preparation needs on average $1/\pS^{\max}$ trials. Without a bound, $\pS^{\max}$ has been found to decay as a power of the number of detected photons $N=\sum_jn_j$ at a fixed number of heralding modes~\cite{fiurasekMaximumHeraldingProbabilities2026, fiurasekHeraldingProbabilityOptimization2026}. Mean matching gives a heuristic estimate of this decay.

At the optimum the pattern $\bs n$ is the mean of the herald statistics, so $\pS^{\max}$ is the probability that these statistics assign to their own mean. If they vary smoothly on the scale of one photon, this probability is close to the peak of a Gaussian with the same covariance matrix $\Sigma^\star$,\vspace{-0.25em}
\begin{equation}\label{eq:clt}
	\pS^{\max}\approx\big[(2\pi)^m\det\Sigma^\star\big]^{-1/2}.\vspace{-0.25em}
\end{equation}
Without displacement, Eq.~\eqref{eq:cov} and the Cauchy--Schwarz inequality $\abs{\Ex{\hat a_j^2}}^2\le\Ex{\nhat_j}(\Ex{\nhat_j}+1)$ give $\Sigma^\star_{jj}\le2n_j(n_j+1)$, and $\det\Sigma^\star\le\prod_j\Sigma^\star_{jj}$, so that the estimate~\eqref{eq:clt} decays at most as a power of $N$.

\begin{figure}[t]
	\centering
	\includegraphics[width=0.95\columnwidth]{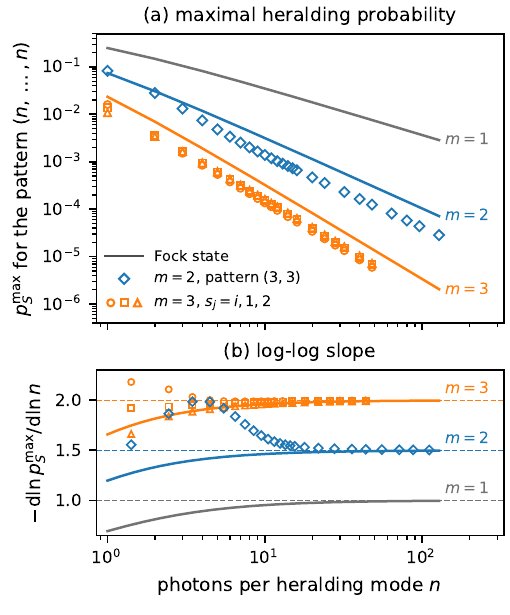}\vspace{-0.5em}
	\caption{Maximal heralding probability for the patterns $(n,\dots,n)$ at fixed control parameters. Lines: the Fock state of Eq.~\eqref{eq:fockm} with $m=1$, $2$ and $3$ heralding modes. Markers, from Newton's method and the exact herald statistics: three sources with three heralding modes~\cite[Fig.~6]{fiurasekHeraldingProbabilityOptimization2026}, $s_j=i$ (circles) and $s_j=1$ (squares) with $\nu_{12}=6/7$, $\nu_{13}=-5/11$ and $\nu_{23}=2/3$, and $s_j=2$ with $\nu_{jk}=1$ (triangles), and the source $s_1=s_2^*=-0.9462+0.6972i$, $\nu_{12}=1.2394$, which with the pattern $(3,3)$ heralds $(\ket0+\ket2+\ket4+\ket6)/2$ (diamonds). \textbf{(a)} $\pS^{\max}$. \textbf{(b)} The log-log slope $-\d\ln\pS^{\max}/\d\ln n$ of (a), which equals $p$ where $\pS^{\max}\propto n^{-p}$, for the markers from consecutive values of $n$. Dashed lines mark $(m+1)/2$.\vspace{-1em}}
	\label{fig:scaling}
\end{figure}

The source with $s_j\!=\!\nu_{jk}\!=\!0$ in the form~\eqref{eq:B}, two-mode squeezed vacuum split into $m$ heralding modes, heralds the Fock state $\ket N$ with $\pS=(1-\sum_jX_j)\,N!\prod_jX_j^{n_j}/n_j!$. Mean matching gives $X_j=n_j/(N+1)$ and\vspace{-0.25em}
\begin{equation}\label{eq:fockm}
	\pS^{\max}=\frac{N!}{(N+1)^{N+1}}\prod_{j=1}^m\frac{n_j^{n_j}}{n_j!}.\vspace{-0.25em}
\end{equation}

Here $\Sigma^\star=\diag(\bs n)+\bs n\bs n^{\mathsf T}$ and $\det\Sigma^\star=(N+1)\prod_jn_j$, so for the pattern $(n,\dots,n)$, $\pS^{\max}$ decays as $n^{-(m+1)/2}$, and by Stirling's formula it lies below the estimate~\eqref{eq:clt} only by the factor $e/\sqrt{2\pi}=1.08$ for large $n_j$.

Figure~\ref{fig:scaling} shows the same power for the other example sources, which accounts for the power close to $2$ that Ref.~\QST\ found for three heralding modes. As $n$ grows, the optimum approaches the boundary of $\Om$, where one normal mode (or a degenerate pair for the Fock source) becomes infinitely squeezed and supplies almost all photons. The fluctuations of their total number give $\Sigma^\star$ one eigenvalue of order $n^2$, and their distribution over the heralding modes gives it $m-1$ eigenvalues of order $n$, so $\det\Sigma^\star\propto n^{m+1}$. At moderate $n$, the slope can still differ from $(m+1)/2$ by up to $0.5$ [Fig.~\ref{fig:scaling}(b)].

Not every source follows this power law. A displacement can lower the power, for example to the decay $\pS^{\max}\propto n^{-3/4}$ of photon-added coherent states~\PRA, which the estimate~\eqref{eq:clt} reproduces. The estimate fails if the optimal herald statistics vary strongly between neighboring patterns (Suppl., Sec.~\ref{supp:onemode}).

\vspace{-0.5em}\vspace{-0.25em}
\subsection{With a bound}\label{sec:trials-bound}
\vspace{-1em}\vspace{-0.25em}

A squeezing bound makes $\pS^{\max}$ decay exponentially with the number of detected photons, because the photons must come from normal modes with bounded squeezing, and a fixed displacement supplies only a bounded mean number of them. At a fixed number of heralding modes and fixed control parameters, $\ln\pS^{\max}$ decreases asymptotically by at least $\ln(1/\mu)$ per detected photon, without displacement for every source and with a fixed displacement $\bs b$ for directly coupled sources (\ref{as:direct}; Prop.~\ref{prop:expbound} in App.~\ref{app:trials}). For the heralded Fock state, $\smax(A)=\sqrt X$ and $\pS=(1-X)X^n$, so for $\mu^2<n/(n+1)$ the bound is active, $\pS^{\max}(\mu)=(1-\mu^2)\mu^{2n}$, and $\ln\pS^{\max}$ decreases by exactly $2\ln(1/\mu)$ per photon.

For random sources with two and three heralding modes, fits give a decrease of $1.4$ to $2$ times $\ln(1/\mu)$ per photon (Suppl., Sec.~\ref{supp:onemode}). At $10$\,dB, $\pS^{\max}$ then falls by a factor of about $1.3$ to $1.5$ per detected photon. A displacement in the heralding mode of the Fock-state source, which then heralds photon-added coherent states, keeps this decrease per photon but raises $\pS^{\max}$ by a factor that grows as $n^{-1/4}e^{2\abs{b_1}\sqrt n}$ for large $n$ (App.~\ref{app:trials}).

\vspace{-0.75em}\vspace{-0.25em}\vspace{-0.25em}\vspace{-0.25em}\vspace{-0.25em}
\section{Imperfect detection}\label{sec:detection}
\vspace{-0.25em}\vspace{-0.5em}\vspace{-0.25em}

The results of Secs.~\ref{sec:ideal} to~\ref{sec:trials} rely on two properties:\vspace{-0.25em}
\begin{enumerate}[label=(P\arabic*), ref=P\arabic*, leftmargin=2.4em, labelsep=0.5em, topsep=0.25em, itemsep=0.25em, parsep=0pt, partopsep=0pt]
	\item\label{pr:project} The detector projects the heralding modes onto a single photon-number pattern, so damping changes the herald statistics but not the heralded state.
	\item\label{pr:prepare} For every $\bs\theta$, the damped state is again in the form~\eqref{eq:state}, which squeezers, an interferometer and displacements can prepare, so optimization over~$\bs\theta$ is an optimization over sources of the same kind.
\end{enumerate}
Either property can fail:
\begin{itemize}[leftmargin=1.2em, labelsep=0.5em, topsep=0.25em, itemsep=0.25em, parsep=0pt, partopsep=0pt]
	\item For a mixed source $\rho$, the states $\Dop\rho\Dop/\trace(\Dop\rho\Dop)$ still form an exponential family, but the same noisy source cannot in general prepare them, so~\ref{pr:prepare} fails.
	\item If several detection patterns are accepted~\cite{ismailzadehMultioutcomeCircuitOptimization2026}, \ref{pr:project} fails and the heralded state is then a mixture that depends on $\bs\theta$ (unless all accepted patterns herald the same state) and $\pS$ is a sum of log-concave functions, which can have several maxima (App.~\ref{app:detection}).
	\item With imperfect detection, \ref{pr:project} fails as well. The registered pattern no longer fixes the photon numbers that arrive at the detectors, so $\pS$ is an average of likelihoods, and the heralded state is in general mixed and depends on $\bs\theta$ (Prop.~\ref{prop:projector} in App.~\ref{app:detection}).
\end{itemize}
Real PNR detectors lose photons and register dark counts~\cite{litaCountingNearinfraredSinglephotons2008, dingPhotonnumberresolvingSinglephotonDetector2026}. We treat loss, the simplest detector model, and App.~\ref{app:detection} treats dark counts and other models. Since the damping now also changes the heralded state, maximizing $\pS$ compares sources that herald different states. A high $\pS$ requires strong squeezing, which makes the heralded state more sensitive to loss, so the damping trades $\pS$ against the fidelity (Sec.~\ref{sec:fidelity}). The maximum of $\pS$ alone therefore does not determine the best damping, but its landscape at least shows which results of Sec.~\ref{sec:ideal} are applicable under imperfect detection.

\begin{figure*}[t]
	\centering
	\includegraphics[width=0.9\textwidth]{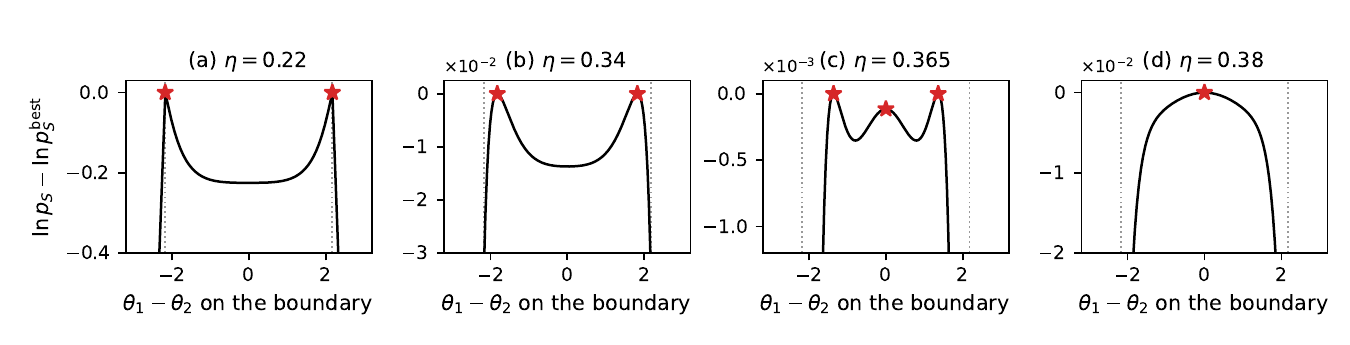}\vspace{-1em}
	\caption{Several maxima under loss for the symmetric source $s_1=s_2=-0.5$, $\nu_{12}=5$, the pattern $(4,4)$, a bound of $16$\,dB and detectors of efficiency $\eta$: $\ln\pS$ along the boundary of $\Om_\mu$, which contains all local maxima at these efficiencies, against $\theta_1-\theta_2$, relative to its largest value. Stars mark the local maxima, and dotted lines the corners of $\Om_\mu$, where the two largest singular values of $A$ both equal $\mu$. The best point on the symmetry axis is at $\theta_1-\theta_2=0$.\vspace{-1.5em}\vspace{-0.25em}\vspace{-0.25em}}
	\label{fig:detection}
\end{figure*}
\vspace{-0.5em}\vspace{-0.25em}
\subsection{Detector model}
\vspace{-0.5em}\vspace{-0.25em}\vspace{-0.25em}\vspace{-0.25em}

We write $\bs n$ for the registered pattern, $\bs k$ for the photon numbers that arrive at the detectors, and $g_{\bs n}(\bs k)$ for the probability that the detectors register $\bs n$ when $\bs k$ arrive. A detector of \emph{efficiency} $\eta_j$ registers each arriving photon independently with probability $\eta_j$, so that $g_{\bs n}(\bs k)\!=\!\prod_j\binom{k_j}{n_j}\eta_j^{n_j}(1-\eta_j)^{k_j-n_j}$. Losses between interferometer and detectors and losses equal in all modes after squeezers act as lower detector efficiency (App.~\ref{app:detection}).
\vspace{-1.5em}\vspace{-0.5em}\vspace{-0.25em}\vspace{-0.25em}
\subsection{Gradient and Hessian}
\vspace{-0.75em}\vspace{-0.25em}

The heralding probability is $\pS(\bs\theta)=\sum_{\bs k}g_{\bs n}(\bs k)p_{\bs\theta}(\bs k)$, and since $\nabla p_{\bs\theta}(\bs k)=(\bs k-\Ex{\hat{\bs n}})\,p_{\bs\theta}(\bs k)$ by Eqs.~\eqref{eq:tilt} and~\eqref{eq:derivs},\vspace{-0.25em}
\begin{equation}\label{eq:louis}
	\nabla\ln\pS=\Ex{\bs k}_q-\Ex{\hat{\bs n}},\qquad\nabla^2\ln\pS=\Cov_q(\bs k)-\Sigma,\vspace{-0.25em}
\end{equation}
where $q(\bs k)=g_{\bs n}(\bs k)p_{\bs\theta}(\bs k)/\pS$ is the distribution of the photon numbers given the registered pattern (App.~\ref{app:detection}).

At a stationary point, $\Ex{\hat{\bs n}}=\Ex{\bs k}_q$, so the source delivers on average as many photons as arrive at the detectors given the registered pattern. The term $-\Sigma$ bends $\ln\pS$ downward, while the term $\Cov_q(\bs k)$ bends it upward and vanishes for ideal detection. This term can make $\ln\pS$ nonconcave, but that alone does not give several maxima, and for one heralding mode $\pS$ still has at most one local maximum (Sec.~\ref{sec:onemode}). In statistics, Eq.~\eqref{eq:louis} is the \emph{missing-information principle}~\cite{orchardMissingInformationPrinciple1972, louisFindingObservedInformation1982}. The information carried by the registered pattern equals the information of the arrived photon numbers minus the information lost because they are not observed.

\vspace{-0.5em}\vspace{-0.25em}\vspace{-0.25em}\vspace{-0.25em}\vspace{-0.25em}
\subsection{Uniqueness}\label{sec:onemode}
\vspace{-0.5em}\vspace{-0.25em}\vspace{-0.25em}\vspace{-0.25em}
\begin{theorem}\label{thm:onemode}
	Let $m=1$ with the heralding mode not in a pure state (\ref{as:pure}), and let the detector register $n$ with probability $g_n(k)$ when $k$ photons arrive. Assume that $g_n$ is unimodal and nonconstant. Then $\pS(\theta)=\sum_kg_n(k)p_\theta(k)$ has at most one local maximum on $\Om$, which is global on~$\Om$ if it exists. The same holds on $\Om_\mu$ for every bound $\mu<1$.\vspace{-0.25em}
\end{theorem}
Unimodality means that the detector response~$g_n(k)$ first rises and then falls with $k$ (including monotone responses, such as threshold detectors), while nonconstancy excludes trivial responses independent of~$k$. Loss with any nonzero efficiency satisfies these assumptions, and so does loss followed by Poissonian dark counts, whose $g_n$ is log-concave in $k$ (Lemma~\ref{lem:detector} in App.~\ref{app:detection}).

For loss and $n\ge1$ the maximum exists, with dark counts it need not. The proof (App.~\ref{app:detection}) applies Descartes' rule of signs~\cite{jamesonCountingZerosGeneralised2006} to the identity\vspace{-0.25em}
\begin{equation}\label{eq:descartes}
	e^{\Phi(\theta)}\big[\pS(\theta)-a\big]=\sum_k\big[g_n(k)-a\big]W_kX^k,\quad X=e^{\theta},\vspace{-0.25em}\vspace{-0.25em}
\end{equation}
for every $a>0$. Since $g_n$ is unimodal, the coefficients change sign at most twice, and then in the order $-,+,-$. So does $\pS-a$ as a function of $\theta$, whereas two separate local maxima would give the order $+,-,+$ for some $a$.

With one heralding mode, damping makes all photon numbers larger or smaller together, and $\pS$ inherits the single peak of $g_n$. With several heralding modes, damping can also move photons between them. Along such a direction a power of $e^t$ collects patterns for which $g_{\bs n}(\bs k)-a$ has both signs, and the sign count fails. Uniqueness then holds only close to ideal detection. For directly coupled sources (\ref{as:direct}), patterns with $n_j\ge1$ (\ref{as:photon}) and no squeezing bound, the maximum is unique if all efficiencies exceed a threshold $\eta_0<1$ (Prop.~\ref{prop:higheta} in App.~\ref{app:detection}). Above it, $\Cov_q(\bs k)$ is smaller than $\Sigma$ on a compact convex set that contains all stationary points, so $\ln\pS$ is strictly concave there by Eq.~\eqref{eq:louis}. At lower efficiencies, and with a bound even where the unbounded problem has a single maximum, several maxima can occur.

\vspace{-0.5em}\vspace{-0.25em}\vspace{-0.25em}
\subsection{Several maxima}\label{sec:multimax}
\vspace{-0.5em}\vspace{-0.25em}\vspace{-0.25em}\vspace{-0.25em}

Several maxima occur mainly for sources that are symmetric under exchange of heralding modes. Figure~\ref{fig:detection} shows an example with a bound of $16$\,dB, which has a single maximum on the symmetry axis for ideal detection. Under loss, a stationary point requires the source to deliver more photons than are registered [Eq.~\eqref{eq:louis}], and on the axis the bound limits how many it can deliver. At low efficiency two mirror-image maxima near the corners of $\Om_\mu$ are therefore higher, and above $\eta_c=0.367$ the maximum on the axis is again the only one (Fig.~\ref{fig:detection}).

In our searches (Suppl., Sec.~\ref{supp:detection}), several maxima occurred for symmetric sources at efficiencies up to $0.95$ (and up to $\eta_c=0.91$ for the sources of Suppl., Fig.~\ref{fig:etac}), but at efficiencies of $0.9$ and above the best maximum exceeded the best symmetric point by less than $1\,\%$. The $24$ two-mode sources of Fig.~\ref{fig:bounded} had a single maximum each, at most $5.7\,\%$ above $\pS$ at their optimum for ideal detection.

Since no criterion identifies the global maximum, a search over the local maxima, including the boundary of $\Om_\mu$, is needed. A good starting point is the optimum for ideal detection with the non-integer pattern $n_j/\eta_j$, because a detector of efficiency $\eta_j$ registers on average the fraction $\eta_j$ of the arriving photons. For random two-mode sources it came within about $1\,\%$ of the largest $\pS$ (Suppl., Fig.~\ref{fig:heuristic}), but for symmetric sources it lies on the symmetry axis and misses the maxima away from it.

\vspace{-0.5em}\vspace{-0.5em}\vspace{-0.25em}
\subsection{Heralding probability and fidelity}\label{sec:fidelity}\vspace{-0.5em}\vspace{-0.25em}
\begin{figure}[t]
	\centering
	\includegraphics[width=0.85\columnwidth]{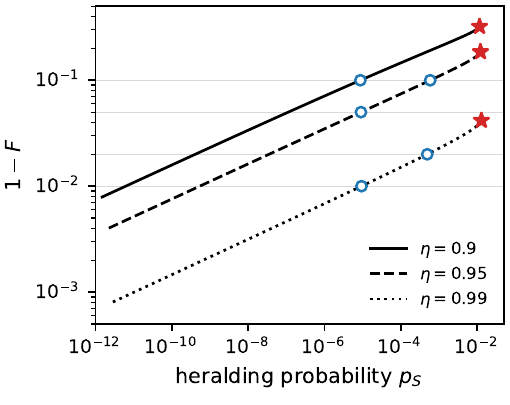}\vspace{-0.5em}\vspace{-0.25em}
	\caption{Heralding probability against fidelity under loss for the source $s_1=s_2^*=-0.9462+0.6972i$, $\nu_{12}=1.2394$ with the pattern $(3,3)$, which heralds $(\ket0+\ket2+\ket4+\ket6)/2$, at $\eta=0.9$, $0.95$ and $0.99$, without a squeezing bound: infidelity $1-F$ of the heralded state with the state heralded by ideal detection, against $\pS$, along the line of uniform damping that starts at the maximum of $\pS$ (star). Circles mark the largest $\pS$ with $F\ge0.9$ at $\eta=0.9$, with $F\ge0.9$ and $0.95$ at $\eta=0.95$, and with $F\ge0.98$ and $0.99$ at $\eta=0.99$.\vspace{-1em}\vspace{-0.5em}\vspace{-0.25em}\vspace{-0.25em}}
	\label{fig:fidelity}
\end{figure}
\vspace{-0.5em}

We keep the control parameters chosen for ideal detection and vary only the damping. Optimizing the control parameters themselves for imperfect detection is part of target-state design (Sec.~\ref{sec:Dop}). With imperfect detection the heralded state $\rho$ is mixed, and the damping changes its fidelity $F=\langle\psi|\rho|\psi\rangle$ with the target $\ket\psi$ heralded by ideal detection. Figure~\ref{fig:fidelity} shows this for the two-mode source of Fig.~\ref{fig:scaling} that heralds $(\ket0+\ket2+\ket4+\ket6)/2$.

Starting from the maximum of $\pS$, attenuating both heralding modes more strongly lowers $\pS$ and raises $F$, so the damping trades $\pS$ against $F$. At $\eta=0.95$, $F=0.815$ at the maximum of $\pS$, and requiring $F\ge0.9$ or $F\ge0.95$ lowers $\pS$ by factors of $21$ and $1350$.

\vspace{-0.5em}\vspace{-0.5em}
\section{Conclusions}\label{sec:conclusion}
\vspace{-0.5em}\vspace{-0.5em}

We have shown that, with ideal detection of one pattern, the damping optimization problem, maximizing the heralding probability at fixed control parameters, can be solved efficiently. The damped photon statistics form an exponential family, in which the heralding probability is the likelihood of the damping and is log-concave in the logarithmic damping parameters. For every pure Gaussian source, with or without displacement, every local maximum of the heralding probability is global.

For directly coupled sources with at least one photon per heralding mode, the maximum exists, is unique and is characterized by mean matching, answering the question of uniqueness raised in Ref.~\QST. Newton's method finds it within tens of milliseconds and the optimization stays practical for many heralding modes and large patterns.

A squeezing bound keeps the problem convex, so its maximum is again unique. At an active bound, which is the typical case for current squeezers, the optimal source delivers fewer photons than the pattern prescribes [Eq.~\eqref{eq:kkt}]. Mean matching also gives a heuristic Gaussian estimate of the maximal heralding probability. Without a squeezing bound and without displacement, this estimate decays as a power of the number of detected photons, specifically giving the exponent $(m+1)/2$ for $m$ heralding modes with $n$ photons each for a heralded Fock state and the other undisplaced examples we computed. Coherent displacements can alter this power~\PRA. A squeezing bound makes this decay exponential, also for directly coupled sources with displacement.

Imperfect detection breaks property~\ref{pr:project}, that the detectors project onto a single pattern. With detector loss, the heralding probability for one heralding mode still has at most one local maximum, with or without a bound. For more heralding modes the maximum is unique at efficiencies close to one without a bound, and otherwise several maxima are possible, mainly for symmetric sources. We know of no general criterion for the global maximum under imperfect detection, so a search over the local maxima is needed, for which the optimum for ideal detection with the pattern $n_j/\eta_j$ is a good starting point. Under loss the damping also changes the heralded state and trades the heralding probability against the fidelity.

Target-state design over the control parameters is not convex in general and remains open. As discussed in Ref.~\QST, the target state may fix the control parameters up to a discrete set of candidates to compare after damping optimization, or leave some control parameters undetermined, allowing further optimization of $\pS$.\vspace{-0.5em}\vspace{-0.5em}\vspace{-0.25em}\vspace{-0.25em}

\begin{acknowledgments}\vspace{-0.5em}\vspace{-0.5em}\vspace{-0.25em}
	We thank Petr Marek for carefully reading through an early version of this work. The authors used OpenAI's ChatGPT-6 Astra for assistance with selected reasoning, and help in simplifying the proof of Thm. 1. The authors used Anthropic's Claude Opus 5.5 for assistance with visualizations, grammar, and general presentation, as well as to perform validity checks and identify gaps and unstated assumptions. The authors acknowledge support from IGA-PrF-2026-005. VK acknowledges support from the Danish National Research Foundation, Center for Macroscopic Quantum States (bigQ, DNRF0142) and the EU project CLUSTEC (grant agreement no. 101080173), and a research grant (VIL78622) from Villum Fonden.
\end{acknowledgments}\vspace{-0.5em}\vspace{-0.25em}

\section*{Data availability}\vspace{-0.5em}\vspace{-0.5em}\vspace{-0.25em}
The code and data that reproduce the figures, tables and numbers of this article are available in Ref.~\cite{heraldingDampingRepository}.\vspace{-0.5em}
\clearpage
\appendix
\vspace{-0.5em}\section{Exponential families}\label{app:expfam}
\vspace{-0.5em}\vspace{-0.25em}
Let the outcomes be the vectors $\bs k\in\N_0^m$ of photon numbers registered by the $m$ heralding detectors. A family of distributions\vspace{-0.25em}\vspace{-0.25em}
\begin{equation}\label{eq:S-family}
	p_{\bs\theta}(\bs k)=W_{\bs k}\,e^{\bs\theta\cdot\bs k-\Phi(\bs\theta)},\qquad \Phi(\bs\theta)=\ln\sum_{\bs k}W_{\bs k}e^{\bs\theta\cdot\bs k},\vspace{-0.25em}\vspace{-0.25em}\vspace{-0.25em}
\end{equation}
is an exponential family, and Table~\ref{tab:glossary} lists its ingredients. The weights $W_{\bs k}\ge0$, called the \emph{carrier}, need not be normalizable~\cite[Ch.~1]{brownFundamentalsStatisticalExponential1986}. The set of parameters for which the sum~\eqref{eq:S-family} converges is the \emph{natural parameter space}.

By Eq.~\eqref{eq:tilt}, the herald statistics of the damped sources form such a family, with the carrier~\eqref{eq:W}. In the interior of the natural parameter space, $\Phi$ may be differentiated term by term~\cite[Prop.~3.1]{wainwrightGraphicalModelsExponential2008}, which gives Eq.~\eqref{eq:derivs}.

\begin{proposition}\label{prop:natural}
	For the sources of Eq.~\eqref{eq:damp}, the natural parameter space is $\Om$, and $\Om$ is open.
\end{proposition}
\begin{proof}
	By Eq.~\eqref{eq:amp}, $e^{\Phi(\bs\theta)}=\langle\tilde G_{A,\bs\beta}|\tilde G_{A,\bs\beta}\rangle$, a sum of nonnegative terms that may diverge. App.~\ref{app:gauss} shows that it is finite if and only if $\smax(\Lambda B\Lambda)<1$, with or without displacement. The set $\Om$ is open because $\smax$ depends continuously on $\bs\theta$.
\end{proof}

The family is \emph{regular} if its natural parameter space is open, which holds here by Prop.~\ref{prop:natural}. It is \emph{minimal} if no combination $\sum_jv_jk_j$ with $\bs v\neq\bs0$ takes the same value for every $\bs k$ in $\supp W=\{\bs k:W_{\bs k}>0\}$. Minimality is equivalent to $\Sigma\succ 0$ on $\Om$, and hence to strict concavity of $\ln\pS$. Sources in which no set of heralding modes is in a pure state (\ref{as:pure}) are minimal by the proof of Thm.~\ref{thm:main}.

For regular minimal exponential families, the mean-value parametrization theorem~\cite[Thm.~3.6]{brownFundamentalsStatisticalExponential1986},~\cite[Prop.~3.2, Thm.~3.3]{wainwrightGraphicalModelsExponential2008} states that the map $\bs\theta\mapsto\Ex{\hat{\bs n}}$ is a one-to-one map of $\Om$ onto $\operatorname{int}\overline{\conv}\supp W$. A directly coupled source (\ref{as:direct}) has $W_{\bs k}>0$ for every pattern $\bs k$ (App.~\ref{app:gauss}), so this hull is the closed nonnegative orthant, with the interior $\{\bs n:n_j>0\ \forall j\}$.

\vspace{-0.5em}\vspace{-0.25em}
\subsection{Heralding modes in a pure state}
\vspace{-0.5em}\vspace{-0.25em}\vspace{-0.25em}

If a set of heralding modes is in a pure state, the family can fail to be minimal. The directions with $\bs v^{\mathsf T}\Sigma\bs v=\Var(\bs v\cdot\hat{\bs n})=0$ form the subspace\vspace{-0.25em}\vspace{-0.25em}
\begin{equation}\label{eq:S-V}
	\mathcal V=\big\{\bs v:\ \bs v\cdot\bs k=0\ \text{for all}\ \bs k\in\supp W\big\},\vspace{-0.25em}\vspace{-0.25em}
\end{equation}
which is identical at every $\bs\theta$, because $\bs v\cdot\bs k$ must be constant on $\supp W$ and $\bs 0\in\supp W$. By Eq.~\eqref{eq:amp}, damping along $\bs v\in\mathcal V$ multiplies every nonzero Fock amplitude by $e^{\bs v\cdot\bs k/2}=1$, so it does not change the source. If $\bs\theta^\star$ and $\bs\theta^\star+\bs v$ are both maximizers, the concave function $\ln\pS$ is constant between them, so $\bs v\in\mathcal V$. The maximizers, if they exist, therefore form the set $\bs\theta^\star+\mathcal V$, and all of them give the same damped source. By Thm.~\ref{thm:main}, $\mathcal V\neq\{\bs0\}$ only if a set of heralding modes is in a pure state.\vspace{-0.25em}\vspace{-0.25em}\vspace{-0.25em}

\begin{table}[b]
	\caption{Terms of the theory of exponential families~\cite{barndorffnielsenInformationExponentialFamilies1978, brownFundamentalsStatisticalExponential1986, wainwrightGraphicalModelsExponential2008} and their meaning for the herald statistics.\vspace{-0.25em}}
	\label{tab:glossary}
	\small
	\begin{tabular*}{\columnwidth}{@{\extracolsep{\fill}}ll@{}}
		\toprule
		Exponential family               & Heralding                              \\
		\midrule
		sufficient statistic $\bs k$     & herald photon numbers                  \\
		carrier $W_{\bs k}$              & undamped weights, Eq.~\eqref{eq:W}     \\
		natural parameter $\bs\theta$    & damping, $\theta_j=\ln X_j$            \\
		log-partition function $\Phi$    & $\ln\langle\tilde G_{A,\bs\beta}|\tilde G_{A,\bs\beta}\rangle$ \\
		natural parameter space          & admissible set $\Om$                   \\
		mean $\nabla\Phi$                & mean photon numbers $\Ex{\hat{\bs n}}$ \\
		Fisher information $\nabla^2\Phi$ & photon-number covariance $\Sigma$     \\
		likelihood of $\bs\theta$        & heralding probability $\pS$            \\
		likelihood equation              & mean matching, Eq.~\eqref{eq:match}    \\
		regular                          & $\Om$ is open                          \\
		minimal                          & $\Sigma$ is positive definite          \\
		\bottomrule
	\end{tabular*}
	\vspace{-0.5em}\vspace{-0.25em}
\end{table}
\vspace{-0.5em}\vspace{-0.25em}\vspace{-0.25em}
\subsection{Gain at the optimum}
\vspace{-0.5em}\vspace{-0.25em}\vspace{-0.25em}

Under~\ref{as:pure}, for admissible $\bs\theta$ and $n_j\ge1$, the log-ratio\vspace{-0.25em}\vspace{-0.25em}
\begin{equation}\label{eq:S-logratio}
	\ln\frac{p_{\bs\theta^\star}(\bs k)}{p_{\bs\theta}(\bs k)}=(\bs\theta^\star-\bs\theta)\cdot\bs k-\Phi(\bs\theta^\star)+\Phi(\bs\theta)\vspace{-0.25em}
\end{equation}
is linear in $\bs k$. Its average over $p_{\bs\theta^\star}$, whose mean is $\bs n$, therefore equals its value at $\bs k=\bs n$,\vspace{-0.25em}
\begin{equation}\label{eq:S-gain}
	\begin{aligned}
		\!\!\ln\pS(\bs\theta^\star)-\ln\pS(\bs\theta) & =\KL\big(p_{\bs\theta^\star}\,\|\,p_{\bs\theta}\big)                                                   \\
		                                          & \simeq\tfrac12\big(\bs n-\Ex{\hat{\bs n}}\big)^{\mathsf T}\Sigma^{-1}\big(\bs n-\Ex{\hat{\bs n}}\big),
	\end{aligned}\vspace{-0.25em}
\end{equation}
with $\KL(p\|p')=\sum_{\bs k}p(\bs k)\ln[p(\bs k)/p'(\bs k)]$ the Kullback--Leibler divergence. The second line follows from two expansions around $\bs\theta$ in $\bs\Delta=\bs\theta^\star-\bs\theta$.

Averaging Eq.~\eqref{eq:S-logratio} gives $\KL=\bs\Delta\cdot\nabla\Phi(\bs\theta^\star)-\Phi(\bs\theta^\star)+\Phi(\bs\theta)\simeq\tfrac12\bs\Delta^{\mathsf T}\Sigma\bs\Delta$, and since $\nabla\Ex{\hat{\bs n}}=\Sigma$ by Eq.~\eqref{eq:derivs}, the mean photon numbers change by $\bs n-\Ex{\hat{\bs n}}\simeq\Sigma\bs\Delta$. Eliminating $\bs\Delta$ gives the approximation, with $\Ex{\hat{\bs n}}$ and $\Sigma$ taken at $\bs\theta$ and a third-order error in $\bs\Delta$. The right-hand side is half the squared Newton decrement, so the stopping rule of Suppl., Sec.~\ref{supp:newton}, approximates the remaining gain in $\ln\pS$.

\vspace{-0.5em}\vspace{-0.25em}\vspace{-0.25em}
\section{Gaussian formulae}\label{app:gauss}
\vspace{-0.5em}\vspace{-0.25em}\vspace{-0.25em}

The basic objects of the optimization are the matrix $A$ and the displacement $\bs\beta$ of the source~\eqref{eq:state}. Written in its normal modes, the source is a product of squeezed coherent states, which gives the norm Eq.~\eqref{eq:norm}, the admissible set $\Om$ and the photon-number moments. The Bargmann representation then allows us to find the heralded state. Finally, we relate mean matching to the polynomial equations of Ref.~\QST\ and show that heralding modes that register no photon can be removed (\ref{as:photon}). Displacement is included unless stated otherwise.

\vspace{-0.5em}\vspace{-0.25em}\vspace{-0.25em}
\subsection{Normal modes and normalization}
\vspace{-0.5em}\vspace{-0.25em}

The Takagi factorization $A=U\diag(\sigma_1,\dots,\sigma_{m+1})U^{\mathsf T}$ of the complex symmetric matrix $A$, with a unitary $U$ and the singular values $\sigma_1\ge\dots\ge\sigma_{m+1}\ge0$~\cite[Cor.~4.4.4(c)]{hornMatrixAnalysis2017}, \cite{takagiAlgebraicProblemRelated1924}, defines the normal-mode operators $\hat c^{\dagger}_i=\sum_{j=0}^mU_{ji}\ad_j$. The columns $\bs U_i$ of $U$, the \emph{Takagi vectors}, satisfy $A\bs U_i^*=\sigma_i\bs U_i$. In the normal modes, the source is a product of displaced single-mode squeezed vacua,\vspace{-0.25em}\vspace{-0.25em}\vspace{-0.25em}\vspace{-0.25em}\vspace{-0.25em}
\begin{equation}\label{eq:S-product}
	|\tilde G_{A,\bs\beta}\rangle=\bigotimes_i\exp\!\Big(\tfrac{\sigma_i}{2}\hat c_i^{\dagger2}+\gamma_i\hat c_i^{\dagger}\Big)\ket{0}_i,\vspace{-0.25em}\vspace{-0.25em}
\end{equation}
with $\sigma_i=\tanh r_i$ and $\bs\gamma=U^{\dagger}\bs\beta$.

To find when the source can be normalized, we use the Bargmann representation. A state with the Fock amplitudes $c_{\bs k}$ is there the function $f(\bs z)=\sum_{\bs k}c_{\bs k}\prod_jz_j^{k_j}/\sqrt{k_j!}$ of $\bs z\in\C^{m+1}$, with the squared norm\vspace{-0.25em}
\begin{equation*}
	\norm{f}^2=\pi^{-m-1}\!\int\!\d^{2(m+1)}z\,\abs{f(\bs z)}^2e^{-\abs{\bs z}^2}.\vspace{-0.25em}
\end{equation*}
The source~\eqref{eq:state} is $f(\bs z)=\exp(\tfrac12\bs z^{\mathsf T}A\bs z+\bs\beta^{\mathsf T}\bs z)$, so its squared norm is the Gaussian integral\vspace{-0.25em}
\begin{equation*}
		\langle\tilde G_{A,\bs\beta}|\tilde G_{A,\bs\beta}\rangle =\pi^{-m-1}\!\!\int\!\d^{2(m+1)}z                                                   \times e^{-\abs{\bs z}^2+\mathrm{Re}(\bs z^{\mathsf T}A\bs z + 2\bs\beta^{\mathsf T}\bs z)}.\vspace{-0.25em}
\end{equation*}
In the coordinates $\bs w=U^{\mathsf T}\bs z$, the quadratic part of the exponent is $\sum_i[(\sigma_i-1)(\mathrm{Re}\,w_i)^2-(1+\sigma_i)(\mathrm{Im}\,w_i)^2]$. The integral therefore converges iff $\smax(A)<1$ and then it equals Eq.~\eqref{eq:norm}. Therefore the sum in Eq.~\eqref{eq:tilt} converges iff $\bs\theta\in\Om$, and $\Om$ does not depend on $\bs b$.

In terms of $\bs\gamma$, the exponent of Eq.~\eqref{eq:norm} is\vspace{-0.25em}
\begin{equation}\label{eq:S-expbound}
	\sum_i\frac{\abs{\gamma_i}^2+\sigma_i\,\mathrm{Re}(\gamma_i^{*2})}{1-\sigma_i^2}\ge\frac{\norm{\bs\beta}^2}{1+\smax(A)}\ge0 .\vspace{-0.25em}
\end{equation}
Hence $e^{\Phi(\bs\theta)}\ge[\det(I-AA^{\dagger})]^{-1/2}$, which diverges at the boundary of $\Om$ (Cor.~\ref{cor:hill}), $\Phi\ge\norm{\bs\beta}^2/2$ (Suppl., Sec.~\ref{supp:nodirect}).

\vspace{-0.5em}\vspace{-0.25em}\vspace{-0.25em}\vspace{-0.25em}
\subsection{Moments}
\vspace{-0.5em}\vspace{-0.25em}

Without displacement, the single-mode moments $\Ex{\hat c^{\dagger}\hat c}=\sigma^2/(1-\sigma^2)=\sinh^2r$ and $\Ex{\hat c\hat c}=\sigma/(1-\sigma^2)$ of the factors of Eq.~\eqref{eq:S-product}, transformed back with $U$, give
\begin{equation}\label{eq:S-moments}
	\begin{aligned}
		\Ex{\ad_j\hat a_k}    & =\big[(I-A^{\dagger}A)^{-1}-I\big]_{jk}\equiv\mathcal C_{jk}, \\
		\Ex{\hat a_j\hat a_k} & =\big[A(I-A^{\dagger}A)^{-1}\big]_{jk}\equiv\mathcal M_{jk}.
	\end{aligned}\vspace{-0.25em}
\end{equation}
In particular, $\Ex{\nhat_j}=\mathcal C_{jj}=\sum_i\abs{U_{ji}}^2\sinh^2r_i$. Wick's theorem with $\Ex{\ad_j\ad_k}=\mathcal M_{jk}^*$ and $\Ex{\hat a_j\ad_k}=\delta_{jk}+\mathcal C_{kj}$ gives\vspace{-0.25em}
\begin{equation*}
	\begin{aligned}
		\Cov(\nhat_j,\nhat_k) & =\Ex{\ad_j\ad_k}\Ex{\hat a_j\hat a_k}+\Ex{\ad_j\hat a_k}\Ex{\hat a_j\ad_k}       \\
		                      & =\abs{\mathcal M_{jk}}^2+\abs{\mathcal C_{jk}}^2+\delta_{jk}\,\mathcal C_{jj},
	\end{aligned}\vspace{-0.25em}
\end{equation*}
which is Eq.~\eqref{eq:cov}. A displacement adds the coherent amplitudes $\bs\alpha=\Ex{\hat{\bs a}}=(I-AA^{\dagger})^{-1}(\bs\beta+A\bs\beta^*)$, and then\vspace{-0.25em}
\begin{align}
	\Ex{\nhat_j} & =\mathcal C_{jj}+\abs{\alpha_j}^2,\nonumber                                                                                    \\
	\Sigma_{jk}  & =\abs{\mathcal M_{jk}}^2+\abs{\mathcal C_{jk}}^2+\delta_{jk}\big(\mathcal C_{jj}+\abs{\alpha_j}^2\big)\nonumber                \\
	             & \quad+2\,\mathrm{Re}\big[\alpha_j\alpha_k^*\mathcal C_{jk}+\alpha_j^*\alpha_k^*\mathcal M_{jk}\big].\label{eq:S-displaced-cov}
\end{align}

\vspace{-0.5em}\vspace{-0.25em}\vspace{-0.25em}\vspace{-0.25em}\vspace{-0.25em}
\subsection{The heralded state}
\vspace{-0.5em}\vspace{-0.25em}

In the Bargmann representation, the signal variable $z$ enters the exponent of the source only through\vspace{-0.25em}
\begin{equation*}
	\tfrac12A_{00}z^2+z\Big(\beta_0+\sum_jA_{0j}z_j\Big).\vspace{-0.25em}
\end{equation*}
The state heralded by the pattern $\bs n$ is, up to a constant, the coefficient of $\prod_jz_j^{n_j}$, so its Bargmann function is\vspace{-0.25em}\vspace{-0.25em}
\begin{equation}\label{eq:S-heralded}
	f_{\bs n}(z)=e^{A_{00}z^2/2+\beta_0z}\,P_{\bs n}(z),\vspace{-0.25em}
\end{equation}
where $P_{\bs n}$ is a polynomial of degree at most $N=\sum_jn_j$.

The coefficient of $z^N$ comes only from the terms $A_{0j}zz_j$ of the exponent and is proportional to $\prod_jA_{0j}^{n_j}$. A directly coupled source therefore heralds a nonzero state for every pattern, so $W_{\bs n}>0$ for every $\bs n$.

\vspace{-0.5em}\vspace{-0.5em}\vspace{-0.25em}
\subsection{Polynomial equations}
\vspace{-0.5em}\vspace{-0.5em}

For $\bs b=\bs0$, the matrices $K=\diag(1,X_1,\dots,X_m)=\Lambda^2$ and $J=I-BKB^{\dagger}K$ of Sec.~\ref{sec:newton} satisfy
\begin{equation*}
	\Lambda J\Lambda^{-1}=I-(\Lambda B\Lambda)(\Lambda B^{\dagger}\Lambda)=I-AA^{\dagger}.
\end{equation*}
Hence $\det J=\det(I-AA^{\dagger})=e^{-2\Phi}$, polynomial in~$\bs X$. Since $X_j\partial_{X_j}=\partial_{\theta_j}$, Eq.~\eqref{eq:derivs} implies $X_j\partial_{X_j}\det J=-2\det J\,\Ex{\nhat_j}$ on $\Om$, which is the second form of Eq.~\eqref{eq:Qdef}.

\vspace{-0.5em}\vspace{-0.5em}
\subsection{Heralding modes that register no photon}
\vspace{-0.5em}\vspace{-0.5em}

Projecting heralding mode $j$ onto the vacuum keeps exactly the terms of Eq.~\eqref{eq:state} without $\ad_j$. It maps the source to the pure Gaussian state of the other modes whose matrix $A'$ is $A$ without row and column $j$ and whose displacement vector is $\bs\beta$ without $\beta_j$. This reduced source heralds the same state as the full source, with the probability\vspace{-0.25em}\vspace{-0.25em}\vspace{-0.25em}\vspace{-0.25em}\vspace{-0.25em}
\begin{equation*}
	\pS'=\frac{\pS}{\Prob(k_j=0)}\ge\pS,\vspace{-0.25em}
\end{equation*}
which is also the limit of $\pS$ as $\theta_j\to-\infty$. Since $\smax(A')\le\smax(A)$, the reduced source needs no stronger squeezing and satisfies every squeezing bound that the full source satisfies. It can be prepared directly, without mode $j$ and its detector.
\vspace{-0.25em}
\vspace{-0.5em}\vspace{-0.25em}
\section{Bounded squeezing}\label{app:capped}
\vspace{-0.5em}\vspace{-0.25em}

Section~\ref{sec:bounded} replaces the admissible set $\Om$ by the constrained set $\Om_\mu=\{\bs\theta:\smax(\Lambda B\Lambda)\le\mu\}$. This appendix shows that $\Om_\mu$ is convex, so that the bounded problem is still a convex optimization problem, and gives the optimality condition that replaces mean matching when the bound is active.\vspace{-0.25em}\vspace{-0.25em}

\vspace{-0.5em}\vspace{-0.5em}
\subsection{The constrained set is convex}
\vspace{-0.5em}\vspace{-0.5em}

Convexity of $\Om_\mu$ is not obvious, because $\smax(\Lambda B\Lambda)$ is the largest value of $\abs{\bs u^{\dagger}\Lambda B\Lambda\bs v}$ over unit vectors, and each of these is the absolute value of a sum of exponentials of $\bs\theta$ with complex coefficients, which can cancel. Proposition~\ref{prop:logconvex} shows that $\ln\smax$ is a convex function of $\bs\theta$, and hence so is $\smax$.

\begin{proposition}\label{prop:logconvex}
	Let $\sigma_1\ge\sigma_2\ge\dots$ be the singular values of $\Lambda B\Lambda$. For all $k \leq \rank(B)$, $\ell_k(\bs\theta)=\sum_{i\le k}\ln\sigma_i(\Lambda B\Lambda)$ is convex on $\R^m$. In particular, $\ln\smax$ is convex.
\end{proposition}
\begin{proof}
	Fix real vectors $\bs\theta,\bs\theta'\in\R^m$ and complex unit vectors $\bs u,\bs v\in\C^{m+1}$. For complex $z$ let
	\begin{equation}\label{eq:S-Lambdaz}
		\begin{aligned}
			\Lambda(z)  & =\diag\big(1,e^{\theta_1(z)/2},\dots,e^{\theta_m(z)/2}\big), \\
			\theta_j(z) & =\theta_j+z(\theta'_j-\theta_j),
		\end{aligned}
	\end{equation}
	so that $\Lambda(0)$ and $\Lambda(1)$ are damping matrices at $\bs\theta$ and $\bs\theta'$.\newpage
	
	The function\vspace{-0.5em}\vspace{-0.25em}
	\begin{equation}\label{eq:S-fz}
		f(z)=\bs u^{\dagger}\Lambda(z)B\Lambda(z)\bs v\vspace{-0.25em}
	\end{equation}
	is a finite sum of exponentials of $z$ and hence analytic. It is also bounded on the strip $0\le\mathrm{Re}\,z\le1$, which the three-lines theorem below requires. The entries of $\Lambda(z)$ have the absolute values $e^{\mathrm{Re}\,\theta_j(z)/2}$, which depend only on $\mathrm{Re}\,z$ and on the strip lie between $e^{\theta_j/2}$ and $e^{\theta'_j/2}$.

	On the line $\mathrm{Re}\,z=0$, $\Lambda(z)$ is $\Lambda$ at $\bs\theta$ times a diagonal unitary matrix $V$, so $\Lambda(z)B\Lambda(z)=V\Lambda B\Lambda V$ has the singular values of $\Lambda B\Lambda$ at $\bs\theta$, and $\abs{f(z)}\le\smax(\bs\theta)$. Likewise $\abs{f(z)}\le\smax(\bs\theta')$ on the line $\mathrm{Re}\,z=1$.
	
	Hadamard's three-lines theorem~\cite[Lemma in Appendix to Sec.~IX.4]{reedMethodsModernMathematical1975} bounds $\abs f$ on an intermediate line by the weighted geometric mean of the two bounds,
	\begin{equation}\label{eq:S-threelines}
		\abs{f(t)}\le\smax(\bs\theta)^{1-t}\,\smax(\bs\theta')^{t},\qquad t\in[0,1].
	\end{equation}

	For real $z=t$, $\Lambda(t)$ is the damping matrix at the interpolated point $\bs\theta_t=\bs\theta+t(\bs\theta'-\bs\theta)$. The bound~\eqref{eq:S-threelines} holds for every pair of unit vectors $\bs u$, $\bs v$, and its right-hand side does not depend on them, so it also holds for the largest value of $\abs{f(t)}$ over all such pairs. That largest value is $\smax(\bs\theta_t)$, because $\abs{\bs u^{\dagger}M\bs v}\le\norm{M\bs v}\le\smax(M)$ for every matrix $M$, with equality when $\bs u$ and $\bs v$ are the leading left and right singular vectors of $M$. Hence
	\begin{equation*}
		\ln\smax(\bs\theta_t)\le(1-t)\ln\smax(\bs\theta)+t\ln\smax(\bs\theta'),
	\end{equation*}
	so $\ln\smax$ is convex for $k=1$.

	For $k>1$ we reduce to the case $k=1$. The $k$th \emph{compound} $C_k(M)$ of a matrix $M$ has its rows and columns labeled by the $k$-element index sets $\mathcal I$ and $\mathcal J$, and its entries are the $k\times k$ minors of $M$ with the rows $\mathcal I$ and the columns $\mathcal J$. It is the matrix of $M$ acting on antisymmetric states of $k$ particles. Three of its properties are needed~\cite{hornMatrixAnalysis2017}. First, the singular values of $C_k(M)$ are the products $\sigma_{i_1}\cdots\sigma_{i_k}$ with $i_1<\dots<i_k$, so $\ln\smax[C_k(\Lambda B\Lambda)]=\ell_k$. Second, the Cauchy--Binet formula gives $C_k(\Lambda B\Lambda)=C_k(\Lambda)\,C_k(B)\,C_k(\Lambda)$. Third, $C_k(\Lambda(z))$ is diagonal with the entries $\exp[\tfrac12\sum_{j\in\mathcal J}\theta_j(z)]$, where $\theta_0\equiv0$, again exponentials of functions affine in $z$. The argument for $k=1$ therefore applies to $C_k(B)$ with $C_k(\Lambda(z))$ in place of $\Lambda(z)$, and it shows that $\ell_k$ is convex.
\end{proof}

Hence $\Om_\mu$, the set where the convex function $\ln\smax$ is at most $\ln\mu$, is convex. Since $\smax\to\abs{B_{00}}<\mu$ as all $\theta_j\to-\infty$, $\Om_\mu$ has points with $\smax<\mu$. On the segment from such a point to any $\bs\theta$ with $\smax(\bs\theta)=\mu$, the convex function $\ln\smax$ lies below $\ln\mu$ except at the endpoint $\bs\theta$, so every boundary point is a limit of points with $\smax<\mu$. Therefore $\Om_\mu$ is the closure of $\{\smax<\mu\}$, the admissible set~\eqref{eq:Omega} with $B$ replaced by $B/\mu$. For directly coupled sources (\ref{as:direct}), a unique maximum of $\pS$ over $\Om_\mu$ exists by the argument of Cor.~\ref{cor:hill}, with $\Om_\mu$ in place of $\Om$.

\vspace{-1em}
\subsection{The optimality condition at an active bound}
\vspace{-0.5em}

When the bound is active, the gradient of $\ln\pS$ need not vanish at the optimum. It is balanced instead by the constraint $\ln\smax\le\ln\mu$. If the largest singular value $\sigma_1$ of $A$ is nondegenerate, first-order perturbation theory with $\partial_{\theta_j}A=\tfrac12(\Pi_jA+A\Pi_j)$, where $\Pi_j$ projects onto mode $j$, gives\vspace{-0.25em}\vspace{-0.25em}\vspace{-0.25em}
\begin{equation}\label{eq:S-gradsmax}
	\frac{\partial\ln\smax}{\partial\theta_j}=\abs{U_{j1}}^2,\vspace{-0.25em}\vspace{-0.25em}
\end{equation}
because $A\bs U_1^*=\sigma_1\bs U_1$ and $\bs U_1^{\dagger}A=\sigma_1\bs U_1^{\mathsf T}$. This vector is the outward normal of the boundary of $\Om_\mu$. The bounded problem is convex and has points with $\smax<\mu$, so the Karush--Kuhn--Tucker conditions are necessary and sufficient for the optimum~\cite[Sec.~5.5.3]{boydConvexOptimization2004}. At an active bound they read $\bs n-\Ex{\hat{\bs n}}=\xi\,(\abs{U_{j1}}^2)_j$ with $\xi\ge0$, which is Eq.~\eqref{eq:kkt}. Where the largest singular value is degenerate, $\smax$ need not be differentiable and the boundary of $\Om_\mu$ may have an edge. Suppl., Sec.~\ref{supp:bounded}, derives the condition that then replaces Eq.~\eqref{eq:kkt}.

\vspace{-0.75em}
\section{An upper bound on the heralding probability}\label{app:trials}
\vspace{-0.5em}

Section~\ref{sec:trials-bound} uses the following bound. Under a squeezing bound, every photon that the detectors register has to come from a normal mode whose squeezing is limited, and a squeezed vacuum at the bound delivers many photons only rarely. The proposition considers $m+1$ squeezed vacua at the bound, which supply the most photons a bounded source can, and asks for the probability that they carry all photons the pattern requires.

\begin{proposition}\label{prop:expbound}
	Take a pure Gaussian source of~$m{+}1$ modes with $\smax(A)\!\le\!\mu\!<\!1$, first without displacement. Let~$P_\psi(k)$ be the probability of $k$ photons in the heralded state. Let $N_\mu$ be the total photon number of $m+1$ independent squeezed vacua with $\tanh r=\mu$, so that $\Prob(N_\mu=2k)=(1-\mu^2)^{(m+1)/2}\binom{k+(m-1)/2}{k}\mu^{2k}$. Then for every pattern with $N=\sum_jn_j$ photons and every integer $k_{\min}\ge0$,\vspace{-0.5em}
	\begin{equation}\label{eq:tailbound}
		\begin{aligned}
			\pS\sum_{k\ge k_{\min}}P_\psi(k) & \le\Prob(N_\mu\ge N+k_{\min})                                               \\
			                                 & =I_{\mu^2}\Big(\Big\lceil\frac{N+k_{\min}}{2}\Big\rceil,\frac{m+1}{2}\Big),
		\end{aligned}
	\end{equation}
	where $I_x(a,b)$ is the regularized incomplete beta function. If $N/(N+m+1)\ge\mu^2$, then also
	\begin{equation}\label{eq:expbound}
		\pS\le\Big[e(1-\mu^2)\Big(1+\frac{N}{m+1}\Big)\Big]^{(m+1)/2}\mu^N.
	\end{equation}
	With a displacement $\bs\beta$, for every $z$ with $1\le z<1/\mu$,
	\begin{equation}\label{eq:expbound-disp}
		\pS\le z^{-N}\Big(\frac{1-\mu^2}{1-\mu^2z^2}\Big)^{(m+1)/2}\exp\frac{z\norm{\bs\beta}^2}{1-z\mu}.
	\end{equation}
\end{proposition}
\begin{proof}
	The interferometer conserves the total photon number, so the total photon number $N_{\mathrm{tot}}$ of all modes is the sum of the photon numbers of the normal modes. The photon number of the $i$th normal mode is $2k$ with probability $(1-\sigma_i^2)^{1/2}\binom{2k}{k}(\sigma_i^2/4)^k$. The probability of at least $2k$ photons is its tail and increases with $\sigma_i$. Raising the squeezing of one normal mode to the bound therefore does not lower $\Prob(N_{\mathrm{tot}}\ge M)$, since for independent modes this probability is the sum over the photon numbers $r$ of the other modes of $\Prob(\text{others}=r)$ times the tail of the raised mode at $M-r$. Raising all modes in turn gives $\Prob(N_{\mathrm{tot}}\ge M)\le\Prob(N_\mu\ge M)$ for every $M$. The heralding modes register $\bs n$ and the signal mode holds at least $k_{\min}$ photons with probability $\pS\sum_{k\ge k_{\min}}P_\psi(k)$, and then $N_{\mathrm{tot}}\ge N+k_{\min}$. This gives the inequality~\eqref{eq:tailbound}.

	For the second form, take $k_{\min}=0$ and $1\le z<1/\mu$. Markov's inequality, $\Prob(Y\ge y)\le\Ex{Y}/y$ for nonnegative random variable $Y$, applied to $Y=z^{N_\mu}$ and $y=z^N$ gives
	\begin{equation}\label{eq:S-markov}
		\!\!\Prob(N_\mu\ge N)\le z^{-N}\Ex{z^{N_\mu}}=z^{-N}\bigg[\frac{1-\mu^2}{1-\mu^2z^2}\bigg]^{\frac{m+1}{2}}.
	\end{equation}
	The right-hand side is smallest at $\mu^2z^2=N/(N+m+1)$, which lies in the allowed range when $N/(N+m+1)\ge\mu^2$. There it equals\vspace{-0.25em}\vspace{-0.25em}
	\begin{equation}\label{eq:S-chernoff}
		(1-\mu^2)^{(m+1)/2}\,\mu^N\Big(1+\frac{m+1}{N}\Big)^{N/2}\Big(1+\frac{N}{m+1}\Big)^{(m+1)/2},\vspace{-0.25em}
	\end{equation}
	and $(1+\tfrac{m+1}{N})^{N/2}\le e^{(m+1)/2}$ gives Eq.~\eqref{eq:expbound}.

	With displacement, $\pS\le\Prob(N_{\mathrm{tot}}\ge N)\le z^{-N}\Ex{z^{N_{\mathrm{tot}}}}$ follows from the same inequality. Replacing every $\ad_j$ by $\sqrt z\,\ad_j$ in Eq.~\eqref{eq:state} multiplies the amplitude of every component with $M$ photons in total by $z^{M/2}$, so
	\begin{equation*}
		\Ex{z^{N_{\mathrm{tot}}}}=\frac{\langle\tilde G_{zA,\sqrt z\bs\beta}|\tilde G_{zA,\sqrt z\bs\beta}\rangle}{\langle\tilde G_{A,\bs\beta}|\tilde G_{A,\bs\beta}\rangle},
	\end{equation*}
	which Eq.~\eqref{eq:norm} gives for $z\smax(A)<1$. The exponent of Eq.~\eqref{eq:norm} in the denominator is nonnegative (App.~\ref{app:gauss}), and in the numerator it is at most $z\norm{\bs\beta}^2(1+z\smax)/(1-z^2\smax^2)\le z\norm{\bs\beta}^2/(1-z\mu)$. The determinants give $\prod_i[(1-\sigma_i^2)/(1-z^2\sigma_i^2)]^{1/2}$, and each factor increases with $\sigma_i$ for $z\ge1$, which gives Eq.~\eqref{eq:expbound-disp}.
\end{proof}

The heralded state does not depend on the damping, so Eq.~\eqref{eq:tailbound} also bounds $\pS^{\max}(\mu)$. For $k_{\min}=0$ it is the smallest bound that follows from $N_{\mathrm{tot}}\ge N$ alone, since $m+1$ normal modes at the bound have exactly the total photon number $N_\mu$. At a fixed number of heralding modes, both bounds decrease as $\mu^N$. Equation~\eqref{eq:expbound} requires the pattern to ask for at least as many photons as the $m+1$ squeezed vacua at the bound supply on average,\vspace{-0.25em}
\begin{equation*}
	N\ge \frac{(m+1)\mu^2}{1-\mu^2},
\end{equation*}
and Eq.~\eqref{eq:tailbound} with $k_{\min}=0$ likewise becomes weak when $m$ grows with $N$, e.g.\ for patterns with one photon per heralding mode.

For directly coupled sources (\ref{as:direct}), $\norm{\bs\beta}$ is bounded on $\Om_\mu$, because $\theta_j<-2\ln\abs{B_{0j}}$. For every $z<1/\mu$, the right-hand side of Eq.~\eqref{eq:expbound-disp} is then $z^{-N}$ times a factor that does not depend on $N$, so $\ln\pS^{\max}(\mu)$ decreases asymptotically by at least $\ln(1/\mu)$ per detected photon also with a displacement. The displacement can nevertheless raise $\pS^{\max}(\mu)$ considerably at moderate photon numbers. For one heralding mode with $s_1=0$ and the displacement $\bs b=(0,b_1)$, which heralds photon-added coherent states up to a displacement,\vspace{-0.25em}\vspace{-0.25em}
\begin{equation}\label{eq:pacs}
	\pS=(1-X)X^ne^{-X\abs{b_1}^2/(1-X)}\sum_{j=0}^n\binom nj\frac{\abs{b_1}^{2j}}{j!},\vspace{-0.25em}
\end{equation}
and at an active bound $X=\mu^2$. The sum is the Laguerre polynomial $L_n(-\abs{b_1}^2)$, which for large $n$ approaches $e^{2\abs{b_1}\sqrt n-\abs{b_1}^2/2}/(2\sqrt\pi\,\abs{b_1}^{1/2}n^{1/4})$, so the decrease of $2\ln(1/\mu)$ per photon of the Fock state is reached only slowly. At $10$\,dB and $n=80$, for example, $b_1=3$ raises $\ln\pS^{\max}$ from $-33.2$ to $-4.8$.

\vspace{-0.5em}\vspace{-0.25em}\vspace{-0.25em}
\section{Imperfect detection}\label{app:detection}
\vspace{-0.5em}\vspace{-0.25em}

This appendix supports Sec.~\ref{sec:detection}. Recall that the detectors register the pattern $\bs n$ with probability $g_{\bs n}(\bs k)$ when the photon numbers $\bs k$ arrive, so that $\pS(\bs\theta)=\sum_{\bs k}g_{\bs n}(\bs k)p_{\bs\theta}(\bs k)$ with the herald statistics $p_{\bs\theta}$ of Eq.~\eqref{eq:tilt}, and that $q(\bs k)=g_{\bs n}(\bs k)p_{\bs\theta}(\bs k)/\pS$ is the distribution of the arrived photon numbers given the registered pattern. Proposition~\ref{prop:projector} identifies the measurements for which the damping leaves the heralded state unchanged, which is property~\ref{pr:project} of Sec.~\ref{sec:detection}. The following subsections describe detector models beyond loss, among them dark counts, derive the gradient and the Hessian of Eq.~\eqref{eq:louis}, and prove the two uniqueness results, Thm.~\ref{thm:onemode} for one heralding mode and Prop.~\ref{prop:higheta} for several heralding modes with nearly ideal detectors.

\vspace{-0.5em}\vspace{-0.5em}
\subsection{Measurements preserved by damping}
\vspace{-0.5em}

A registered pattern of a real detector is described by a positive operator $\hat E$ on the heralding modes~\cite{lundeenTomographyQuantumDetectors2009}, and ideal PNR detection by the projector $\ketbra{\bs n}$.
\begin{proposition}\label{prop:projector}
	Let $\hat E$ be a nonzero positive operator on the heralding modes. The unnormalized heralded state found as $\trace_{1\dots m}[\hat E\,\Dop\rho\Dop]$ is independent of $\bs\theta$ up to normalization, for every state $\rho$ of the signal and heralding modes, if and only if $\hat E$ is proportional to the projector $\ketbra{\bs n}$ onto a single photon-number pattern.
\end{proposition}
\begin{proof}
	Write $\hat E=\sum_{\bs k,\bs k'}e_{\bs k\bs k'}\ketbra{\bs k}{\bs k'}$. Moving the damping onto the measurement gives
	\begin{equation}\label{eq:S-moveD}
		\trace_{1\dots m}\big[\hat E\,\Dop\rho\Dop\big]=\trace_{1\dots m}\big[(\Dop\hat E\Dop)\rho\big].
	\end{equation}
	For $\hat E\propto\ketbra{\bs n}$, $\Dop\hat E\Dop=e^{\bs n\cdot\bs\theta}\hat E$, so the heralded state is independent of $\bs\theta$ up to normalization, proving sufficiency.\newpage

	For necessity, take any two distinct patterns $\bs k\neq\bs k'$ and the state $\rho=\tfrac12(\ketbra{0}\otimes\ketbra{\bs k}+\ketbra{1}\otimes\ketbra{\bs k'})$. Since $\Dop\ket{\bs k}=e^{\bs k\cdot\bs\theta/2}\ket{\bs k}$, the unnormalized heralded state is $\tfrac12(e_{\bs k\bs k}e^{\bs k\cdot\bs\theta}\ketbra{0}+e_{\bs k'\bs k'}e^{\bs k'\cdot\bs\theta}\ketbra{1})$. If both $e_{\bs k\bs k}$ and $e_{\bs k'\bs k'}$ were nonzero, the ratio of the two weights would scale as $e^{(\bs k-\bs k')\cdot\bs\theta}$ and vary with $\bs\theta$. At most one diagonal element $e_{\bs n\bs n}$ can therefore be nonzero. Because $\hat E$ is positive, $\abs{e_{\bs k\bs k'}}^2\le e_{\bs k\bs k}e_{\bs k'\bs k'}=0$ for all $\bs k\neq\bs k'$, so all off-diagonal elements vanish. Since $\hat E\neq0$, $\hat E=e_{\bs n\bs n}\ketbra{\bs n}$.
\end{proof}
For loss, dark counts or threshold detectors, the operator $\hat E=\sum_{\bs k}g_{\bs n}(\bs k)\ketbra{\bs k}$ is diagonal in the photon numbers but not a projector onto a single pattern, so the heralded state depends on $\bs\theta$ for some sources.

\vspace{-0.5em}\vspace{-0.5em}\vspace{-0.25em}
\subsection{Detector models}
\vspace{-0.5em}\vspace{-0.25em}\vspace{-0.25em}

Accepting a set of patterns~\cite{ismailzadehMultioutcomeCircuitOptimization2026} makes $\pS$ a sum of log-concave functions of $\bs\theta$, which need not be log-concave, even when all accepted patterns herald the same state. For real $s_1=s_2=0.3$ and $\nu_{12}=0.2$, accepting the mirror-image patterns $(8,1)$ and $(1,8)$ gives two maxima, at $\bs\theta=(-2.412,-0.433)$ and its mirror image, where one pattern contributes almost all of $\pS=0.0109$. Accepting $(3,2)$ and $(2,3)$ gives a single maximum on the diagonal.

A detector of efficiency $\eta_j$ with Poissonian dark counts of mean $d_j$~\cite{semenovDetectionQuantumLight2008} registers $n_j$ with probability\vspace{-0.5em}
\begin{equation}\label{eq:S-gdef}
	g_{n_j}(k_j)=\Prob\big[Y(k_j,\eta_j)+Y'(d_j)=n_j\big]\vspace{-0.5em}
\end{equation}
when $k_j$ photons arrive, where $Y(k,\eta)$ is binomial with $k$ trials and $Y'(d)$ is Poissonian with mean $d$. The joint response is $g_{\bs n}(\bs k)=\prod_jg_{n_j}(k_j)$.

Loss with the same transmissivity $\eta_{\mathrm s}$ in all modes after the squeezers commutes with the interferometer. Moved behind it, the loss does not affect $\pS$ where it acts on the signal mode, and it acts as detectors of efficiency $\eta_{\mathrm s}\eta_j$ in the heralding modes. Losses between the interferometer and the detectors, such as propagation and coupling losses, act in the same way even when they differ between modes.
\vspace{-0.5em}\vspace{-0.25em}\vspace{-0.25em}\vspace{-0.25em}
\subsection{Gradient and Hessian for incomplete data}
\vspace{-0.5em}\vspace{-0.25em}

Substituting $p_{\bs\theta}(\bs k)=W_{\bs k}e^{\bs k\cdot\bs\theta-\Phi(\bs\theta)}$ from Eq.~\eqref{eq:tilt} into $\pS(\bs\theta)=\sum_{\bs k}g_{\bs n}(\bs k)p_{\bs\theta}(\bs k)$ gives
\begin{equation}\label{eq:S-lnpS-diff}
	\begin{aligned}
		\ln\pS(\bs\theta) & =\Phi_q(\bs\theta)-\Phi(\bs\theta),                                      \\
		\Phi_q(\bs\theta) & =\ln\sum_{\bs k}g_{\bs n}(\bs k)W_{\bs k}e^{\bs k\cdot\bs\theta}.
	\end{aligned}\vspace{-0.25em}\vspace{-0.25em}
\end{equation}

The distribution $q(\bs k)\!=\!g_{\bs n}(\bs k)W_{\bs k}e^{\bs k\cdot\bs\theta-\Phi_q(\bs\theta)}$ of Eq.~\eqref{eq:louis} is therefore an exponential family of the form~\eqref{eq:S-family}, with the carrier $g_{\bs n}(\bs k)W_{\bs k}$ and the log-partition function $\Phi_q$, which is finite on $\Om$ because $g_{\bs n}\le1$. As Eq.~\eqref{eq:derivs} gives the mean and the covariance matrix of $p_{\bs\theta}$ from $\Phi$, the first and second derivatives of $\Phi_q$ give those of $q$. The gradient of $\ln\pS$ is therefore the difference of the two means, and its Hessian the difference of the two covariance matrices, which is Eq.~\eqref{eq:louis}. In statistical terms, the registered pattern is \emph{incomplete data}, and Eqs.~\eqref{eq:louis} are the likelihood equation and the \emph{observed information}, the negative Hessian of the log-likelihood, for incomplete data from an exponential family~\cite{sundbergMaximumLikelihoodTheory1974, louisFindingObservedInformation1982}. Ideal detection has $q$ concentrated on $\bs n$ and recovers Eqs.~\eqref{eq:gradhess}. With loss and no dark counts, $q$ is supported on $\bs k\ge\bs n$, so every stationary point has $\Ex{\nhat_j}=\Ex{k_j}_q\ge n_j$.

\vspace{-1em}\vspace{-0.25em}
\subsection{One heralding mode}
\vspace{-0.5em}\vspace{-0.25em}\vspace{-0.25em}

The loss and dark-count models above satisfy the assumptions of Thm.~\ref{thm:onemode}.\vspace{-0.25em}

\begin{lemma}\label{lem:detector}
	For loss with efficiency $0<\eta\le1$ followed by Poissonian dark counts with mean $d\ge0$, $g_n$ is log-concave with an interval as support, satisfies $\inf_k g_n(k)=0<\sup_k g_n(k)$, and is therefore unimodal and nonconstant.
\end{lemma}
\begin{proof}
	When $k$ photons arrive, the registered count is $Y_k=Y(k,\eta)+Y'(d)$, the sum of $k$ independent Bernoulli variables with success probability $\eta$ and a Poissonian variable with mean $d$, and $g_n(k)=\Prob(Y_k=n)$. Detecting all $n$ arriving photons with zero dark counts gives $g_n(n)\ge\eta^ne^{-d}>0$, while $g_n(k)\le\Prob[Y(k,\eta)\le n]\to0$ as $k\to\infty$, so $\inf_k g_n(k)=0<\sup_k g_n(k)$. For $\eta=1$, $g_n(k)=\Prob[Y'(d)=n-k]$ is directly log-concave on its support ($\{n\}$ for $d=0$, $\{0,\dots,n\}$ for $d>0$). For $n=0$, $g_0(k)=e^{-d}(1-\eta)^k$ is log-linear on $k\ge0$. Take now $\eta<1$ and $n\ge1$; the support of $g_n$ is $k\ge n$ for $d=0$ and all $k\ge0$ for $d>0$. Every summand of $Y_k$ has a log-concave distribution on an interval, and so does their sum~\cite{hoggarChromaticPolynomialsLogarithmic1974} (extending to the Poissonian summand as a limit of binomial variables). One more arriving photon adds one Bernoulli variable,\vspace{-0.25em}
	\begin{equation}\label{eq:S-Yrec}
		\Prob(Y_{k+1}=j)=(1-\eta)\Prob(Y_k=j)+\eta\Prob(Y_k=j-1),\vspace{-0.25em}
	\end{equation}
	so wherever $g_n(k)>0$,\vspace{-0.25em}
	\begin{equation*}
		\frac{g_n(k+1)}{g_n(k)}=1-\eta+\eta\,\Gamma_k,\qquad \Gamma_k=\frac{\Prob(Y_k=n-1)}{\Prob(Y_k=n)}.\vspace{-0.25em}
	\end{equation*}
	By log-concavity of $Y_k$, $\Prob(Y_k=j-1)/\Prob(Y_k=j)$ is nondecreasing in $j$, $R(j)=\Prob(Y_{k+1}=j)/\Prob(Y_k=j)$ is non-decreasing by Eq.~\eqref{eq:S-Yrec}. Therefore, we have $\Gamma_{k+1}=\Gamma_k\,R(n-1)/R(n)\le\Gamma_k$, the ratio $g_n(k+1)/g_n(k)$ does not increase with $k$, and $g_n$ is log-concave. Finally, any log-concave sequence supported on an interval is unimodal, with the sets $\{k:g_n(k)>a\}$ forming intervals.
\end{proof}

\begin{proof}[Proof of Theorem~\ref{thm:onemode}]
	By~(\ref{as:pure}), for $m=1$ the heralding mode couples to the signal mode ($B_{01}\neq0$), so $W_k>0$ for every $k\ge0$ (App.~\ref{app:gauss}) and $\smax(A)$ grows without bound as $\theta\to\infty$; since $g_n\ge0$ is nonconstant, $\pS>0$ on $\Om$. Fix $a>0$ and write $X=e^\theta$. Since $\pS=\sum_kg_n(k)W_kX^ke^{-\Phi}$ and $e^{\Phi}=\sum_kW_kX^k$,\vspace{-0.25em}
	\begin{equation}\label{eq:S-series}
		e^{\Phi(\theta)}\big[\pS(\theta)-a\big]=\sum_{k\ge0}c_kX^k,\; c_k=\big[g_n(k)-a\big]W_k .\vspace{-0.25em}\vspace{-0.25em}
	\end{equation}
	The series converges on $\Om$, because $0\le g_n\le1$, and has at least one nonzero coefficient because $W_k>0$ for all $k$ and $\inf_k g_n(k)<\sup_k g_n(k)$.
	
	Since $g_n$ is unimodal, the indices with $g_n(k)>a$ form an interval with $c_k\le0$ outside of this interval. The nonzero coefficients therefore change sign at most twice, and if twice, in the order $-,+,-$. By Descartes' rule of signs for power series~\cite[Thm.~3.1]{jamesonCountingZerosGeneralised2006}, the series has at most as many positive zeros in its interval of convergence as its coefficients have sign changes, and near $X=0$ it has the sign of its first nonzero coefficient.
	
	With at most one sign change it has at most one positive zero, while with two it is negative near $X=0$ and has at most two; in neither case can $\pS-a$ take the signs $+,-,+$ at three increasing values of $\theta$ (which requires two zeros, or three if negative near $X=0$).

	Suppose that $\pS$ had a local maximum at $\theta_a$ and another point $\theta_b\neq\theta_a$ with $\pS(\theta_b)\ge\pS(\theta_a)$, and let $\theta'$ minimize $\pS$ on the closed interval between them. Because $\pS(\theta)=\pS(\theta_a)$ has at most two solutions on $\Om$, $\pS$ must drop below $\pS(\theta_a)$ near $\theta_a$ on the side of $\theta_b$, so $\pS(\theta')<\pS(\theta_a)$. For any $a\in(\pS(\theta'),\pS(\theta_a))$, $\pS-a$ takes the signs $+,-,+$ at $\theta_a$, $\theta'$ and $\theta_b$ (in either order of $\theta_a$ and $\theta_b$), which is excluded above. Hence a local maximum is the unique global maximum. For $m=1$ and $\mu>\abs{B_{00}}$ (otherwise $\Om_\mu$ is empty), $\Om_\mu$ is a half-line $\theta\le\theta_\mu$, because $\smax$ is convex in $\theta$, tends to $\abs{B_{00}}<\mu$ as $\theta\to-\infty$, and grows without bound as $\theta\to\infty$, so it does not decrease. The same argument applies on $\Om_\mu$, including a local maximum at $\theta_\mu$.\vspace{-0.25em}\vspace{-0.25em}\vspace{-0.25em}
\end{proof}

Theorem~\ref{thm:onemode} does not guarantee existence, but under pure loss ($\eta_j>0$) a maximum of $\pS$ exists on $\Om$ for any directly coupled source (\ref{as:direct}) with $n_j\ge1$. Since $p_{\bs\theta}(\bs k)\to0$ for all $\bs k\ge\bs n$ as $\bs\theta$ leaves compact subsets of $\Om$ (Cor.~\ref{cor:hill}) and $\sum_{\bs k}g_{\bs n}(\bs k)=\prod_j\eta_j^{-1}<\infty$, dominated convergence (with $p_{\bs\theta}\le1$) gives $\pS\to0$ there, so $\pS$ attains its maximum. With dark counts a maximum need not exist.
\vspace{-1em}
\subsection{Heralding with nearly ideal detectors}
\vspace{-0.75em}\vspace{-0.25em}
\begin{proposition}\label{prop:higheta}
	For a directly coupled source (\ref{as:direct}) and $n_j\ge1$ for every $j$ (\ref{as:photon}), let detector $j$ have efficiency $\eta_j$ and no dark counts. There is $\eta_0<1$, $\eta_0=\eta_0(B, \bs b, \bs n)$, such that for $\eta_j\in[\eta_0,1]$ the heralding probability has exactly one stationary point on $\Om$, its global maximum.\vspace{-0.25em}
\end{proposition}
\vspace{-0.25em}
\begin{proof}
	We first show that the stationary points lie in a compact set. Fix $\underline\eta\in(0,1)$ and let every $\eta_j\ge\underline\eta$. Model detector $j$ as a beam splitter of transmissivity $\eta_j$ followed by ideal detection on the transmitted mode, with reflected photons entering a \emph{loss mode} $l_j$ (creation operators $\hat t_j^{\dagger}$ and $\hat l_j^{\dagger}$). Given the registered pattern $\bs n$, the mean photon number of $l_j$ is $\Ex{k_j}_q-n_j$, since $k_j-n_j$ of the $k_j$ arriving photons are reflected. In Eq.~\eqref{eq:state}, the beam splitters replace each heralding operator $\ad_j$ by $\sqrt{\eta_j}\,\hat t_j^{\dagger}+\sqrt{1-\eta_j}\,\hat l_j^{\dagger}$. Projecting the transmitted modes onto $\ket{\bs n}$ then leaves the signal and loss modes in the state $h(\ad_0,\hat{\bs l}^{\dagger})|\tilde G_{LAL,L\bs\beta}\rangle$, where $L=\diag(1,\sqrt{1-\eta_1},\dots,\sqrt{1-\eta_m})$ and $h$ is a polynomial of degree at most $N=\sum_j n_j$, as in Eq.~\eqref{eq:S-heralded}.\newpage

	The loss modes are attenuated by $L$, this keeps them below infinite squeezing. For unit $\bs u,\bs v$, $\abs{\bs u^{\dagger}LAL\bs v}\le\norm{L\bs u}\norm{L\bs v}\smax(A)\le1$ for every $A$ in the closure of the admissible set. Equality would need $\norm{L\bs u}=\norm{L\bs v}=1$, that is, $\bs u$ and $\bs v$ supported on the signal mode, the only mode that $L$ does not attenuate, but there $\abs{\bs u^{\dagger}LAL\bs v}=\abs{A_{00}}=\abs{B_{00}}<1$. Since the admissible $A$ and the matrices $L$ range over sets with compact closure, $\smax(LAL)\le\bar\mu<1$ for all $\bs\theta\in\Om$ and all $\eta_j\ge\underline\eta$.

	It follows that $|\tilde G_{LAL,L\bs\beta}\rangle\propto\hat S\ket{\bs0}$ for a Gaussian unitary $\hat S$ with squeezing at most $\operatorname{argtanh}\bar\mu$ and displacement bounded on $\Om$ (where $\theta_j<-2\ln\abs{B_{0j}}$). Pulling $\hat S$ to the left writes the normalized signal--loss state as $\hat S\ket\phi$, with $\ket\phi\propto\hat S^{\dagger}h(\ad_0,\hat{\bs l}^{\dagger})\hat S\ket{\bs0}$. Since $\hat S$ transforms the mode operators affinely, $\ket\phi$ contains at most $N$ photons, and $\hat S^{\dagger}\hat l_j^{\dagger}\hat l_j\hat S$ is a quadratic polynomial in the mode operators whose coefficients are bounded in terms of $\bar\mu$, $B$ and $\bs b$. The mean photon number $\langle\phi|\hat S^{\dagger}\hat l_j^{\dagger}\hat l_j\hat S|\phi\rangle$ of every loss mode is therefore at most a constant $N_{\mathrm{loss}}$ depending only on $\bar\mu$, $m$, $N$, $B$ and $\bs b$. By Eq.~\eqref{eq:louis}, every stationary point has $\Ex{\nhat_j}=\Ex{k_j}_q$, and hence $n_j\le\Ex{\nhat_j}\le n_j+N_{\mathrm{loss}}$.

	By the mean-value parametrization (App.~\ref{app:expfam}), $\bs\theta\mapsto\Ex{\hat{\bs n}}$ maps $\Om$ one-to-one onto the open positive orthant, and its inverse is continuous because its Jacobian $\Sigma$ is positive definite. The preimage $\mathcal S$ of the box $\prod_j[n_j,n_j+N_{\mathrm{loss}}]$ is therefore a compact subset of $\Om$, and so is its convex hull, since $\Om$ is convex.

	On $\conv\mathcal S$, $\Sigma\succeq\varsigma I$ and $p_{\bs\theta}(\bs n)\ge p_0$ for some $\varsigma,p_0>0$, where $M\preceq M'$ means that $M'-M$ is positive semidefinite. With $\eta_{\min}=\min_j\eta_j$,\vspace{-0.5em}
	\begin{equation}\label{eq:S-qbound}
		q(\bs k)\le\frac{q(\bs k)}{q(\bs n)}\le\frac1{p_0}\prod_j\binom{k_j}{n_j}(1-\eta_{\min})^{k_j-n_j}
	\end{equation}
	for $\bs k\ge\bs n$, using $p_{\bs\theta}(\bs k)\le1$ and $g_{\bs n}(\bs k)/g_{\bs n}(\bs n)=\prod_j\binom{k_j}{n_j}(1-\eta_j)^{k_j-n_j}$. Hence\vspace{-0.5em}
	\begin{equation*}
		\begin{aligned}
			\Cov_q(\bs k) & \preceq\Big[\sum_{\bs k}\abs{\bs k-\bs n}^2q(\bs k)\Big]I                                                                    \\
			              & \preceq\frac{I}{p_0}\sum_{\bs k\ge\bs n}\abs{\bs k-\bs n}^2\prod_j\binom{k_j}{n_j}(1-\eta_{\min})^{k_j-n_j},\vspace{-0.75em}
		\end{aligned}\vspace{-0.75em}
	\end{equation*}
	which tends to zero as $\eta_{\min}\to1$. For $\eta_{\min}$ close enough to one, $\Cov_q(\bs k)\prec\varsigma I\preceq\Sigma$ on $\conv\mathcal S$, and by Eq.~\eqref{eq:louis} $\ln\pS$ is strictly concave there.

	A strictly concave function on the convex set $\conv\mathcal S$, which contains every stationary point, has at most one stationary point. A maximum exists, as shown above, and since it lies in the open set $\Om$, it is a stationary point.
\end{proof}
\vspace{-1em}
Dark counts keep the arrived photon numbers uncertain even at unit efficiency, so $\Cov_q(\bs k)$ does not vanish as $\eta_j\to1$, and the proof does not extend to them. Figure~\ref{fig:etac} in the Supplemental Material shows the largest efficiency $\eta_\mathrm{c}$ with several maxima for symmetric sources.

\clearpage
\bibliographystyle{apsrev4-2}
\bibliography{paper}

\clearpage
\onecolumngrid
\renewcommand{\topfraction}{0.9}
\renewcommand{\bottomfraction}{0.8}
\renewcommand{\textfraction}{0.07}
\renewcommand{\floatpagefraction}{0.75}
\setcounter{equation}{0}
\setcounter{table}{0}
\setcounter{figure}{0}
\renewcommand{\theequation}{S\arabic{equation}}
\renewcommand{\thetable}{S\arabic{table}}
\renewcommand{\thefigure}{S\arabic{figure}}
\renewcommand{\theHequation}{S\arabic{equation}}
\renewcommand{\theHtable}{S\arabic{table}}
\renewcommand{\theHfigure}{S\arabic{figure}}
\setcounter{proposition}{0}
\renewcommand{\theproposition}{S\arabic{proposition}}
\renewcommand{\theHproposition}{S\arabic{proposition}}
\newcounter{suppsec}
\renewcommand{\thesuppsec}{S\arabic{suppsec}}
\renewcommand{\theHsuppsec}{S\arabic{suppsec}}
\newcommand{\suppsection}[2]{\refstepcounter{suppsec}\label{#2}\section*{\texorpdfstring{\thesuppsec.\quad #1}{\thesuppsec. #1}}}

\begin{center}
	{\large\bfseries Supplemental Material for ``\papertitle'' \par}
	\vspace{1.0em}
	Vojt\v{e}ch Kucha\v{r}\,$^{1,2}$ and Jarom\'{\i}r Fiur\'a\v{s}ek\,$^{1}$\par
	\vspace{0.5em}
	{\small\itshape $^{1}$\affUPOL\par}
	\vspace{0.15em}
	{\small\itshape $^{2}$\affDTU\par}
	\vspace{0.5em}
	{\small(Dated: \today)\par}
\end{center}

\vspace{0.5em}

We use the notation of the main text: the source~\eqref{eq:state} with one signal mode, labeled $0$ (Sec.~\ref{supp:signal} treats several), and $m$ heralding modes, the damping $\theta_j=\ln X_j$, which turns the reference source with $B$ and $\bs b$ into $A=\Lambda B\Lambda$ and $\bs\beta=\Lambda\bs b$, and the normal modes with $\sigma_i=\tanh r_i$ and the squeezing $10\log_{10}e^{2r_i}$\,dB. Directly coupled sources are written in the form~\eqref{eq:B}, and all examples have $s_0=0$.

\vspace{-0.5em}
\suppsection{Scaling of the heralding probability}{supp:onemode}\vspace{-0.5em}

Section~\ref{sec:trials-free} estimates $\pS^{\max}$ by the Gaussian estimate~\eqref{eq:clt}, $\pS^{\max}\approx[(2\pi)^m\det\Sigma^\star]^{-1/2}$, which decays as a power of the number of detected photons $N=\sum_jn_j$, and Sec.~\ref{sec:trials-bound} shows that a squeezing bound makes the decay exponential, so that $\ln\pS^{\max}$ decreases by at least $\ln(1/\mu)$ per detected photon (Prop.~\ref{prop:expbound}). We test both statements on sources with one heralding mode, which can be solved exactly and show why the Gaussian estimate works and when it fails.

\vspace{-0.5em}
\subsection*{One heralding mode}\vspace{-0.25em}\vspace{-0.25em}\vspace{-0.25em}

\begin{figure}[t]
	\centering
	\includegraphics[width=0.9\textwidth]{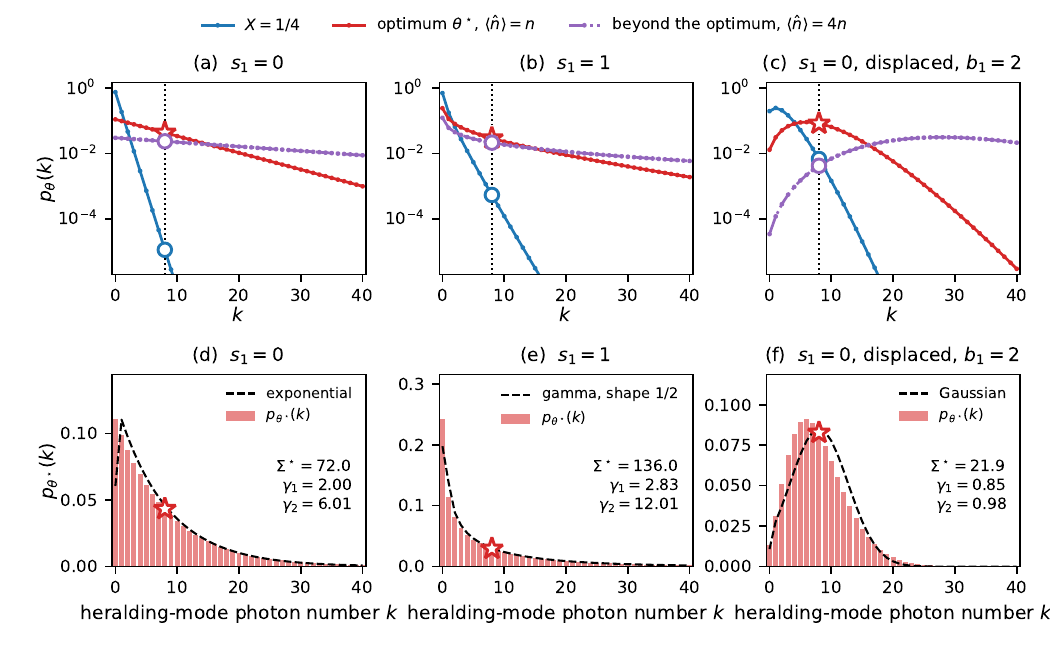}\vspace{-0.25em}\vspace{-0.25em}
	\caption{Damping as exponential tilting, as in Fig.~\ref{fig:tilting}, for one heralding mode and the pattern $n=8$. \textbf{(a)} $s_1=0$, whose heralding mode is thermal. \textbf{(b)} $s_1=1$, Eq.~\eqref{eq:S-squeezed-stats}. \textbf{(c)} $s_1=0$ with the displacement $\bs b=(0,2)$ in the heralding mode (App.~\ref{app:trials}). The top row shows the herald distributions $p_\theta(k)$ at $X=1/4$, at the optimum $\theta^\star$, where $\Ex{\nhat}=n$, and beyond it, where $\Ex{\nhat}=4n$, and the markers give $p_\theta(n)=\pS(\theta)$. \textbf{(d)}--\textbf{(f)} The optimal distributions $p_{\theta^\star}(k)$ (bars) with their continuum limits for large $n$, integrated over $[k-\tfrac12,k+\tfrac12]$ (dashed). These are the exponential distribution with mean $n$, the gamma distribution with shape $1/2$ and mean $n$, and the Gaussian distribution with mean $n$ and variance $\Sigma^\star$. The panels give the variance $\Sigma^\star$, the skewness $\gamma_1$ and the excess kurtosis $\gamma_2$ of $p_{\theta^\star}$, computed from the derivatives of $\Phi$ at $\theta^\star$, which are the cumulants of $p_{\theta}$.\vspace{-0.25em}\vspace{-0.25em}}
	\label{fig:shapes}\vspace{-0.5em}
\end{figure}

For one heralding mode, the form~\eqref{eq:B} with $s_0=0$ has $B_{01}=1$ and $B_{11}=s_1$, and the admissible set is $X<1/(1+\abs{s_1})$ with $X=e^\theta$. Without displacement, mean matching is a quadratic equation for $X$, whose admissible root is the closed form of Ref.~\PRA. Figure~\ref{fig:shapes} shows three families. For $s_1=0$, a two-mode squeezed vacuum that heralds the Fock state $\ket n$, the herald statistics are thermal, $p_\theta(k)=(1-X)X^k$, and mean matching gives $X^\star=n/(n+1)$ and $\pS^{\max}=n^n/(n+1)^{n+1}$~\PRA. For $s_1=1$, they are\vspace{-0.25em}\vspace{-0.25em}
\begin{equation}\label{eq:S-squeezed-stats}
	p_\theta(k)=(1-2X)^{1/2}\binom{2k}{k}\Big(\frac X2\Big)^k,\vspace{-0.25em}\vspace{-0.25em}
\end{equation}
half the photon number of a squeezed vacuum with $\tanh^2r=2X$, and $X^\star=n/(2n+1)$. In each case $\pS^{\max}=p_{\theta^\star}(n)$ is the probability that the optimal distribution assigns to its own mean, which the Gaussian estimate replaces by the height of a Gaussian with the same variance. For large $n$, the optimal distributions for $s_1=0$ and $s_1=1$ approach an exponential distribution and a gamma distribution with shape $1/2$, whose heights at the mean are smaller than the estimate by the constant factors $e/\sqrt{2\pi}=1.08$ and $\sqrt{e/2}=1.17$. The estimate thus has the right power of $n$.

The displaced source heralds photon-added coherent states~\PRA. Its heralding mode is a displaced thermal state with the thermal photon number $n_{\mathrm{th}}=X/(1-X)$ and the coherent photon number $n_{\mathrm{coh}}=X\abs{b_1}^2/(1-X)^2$, whose photon-number variance is $\Sigma=n_{\mathrm{th}}(n_{\mathrm{th}}+1)+n_{\mathrm{coh}}(2n_{\mathrm{th}}+1)$. For $n\gg\abs{b_1}^2$ and $n\gg1/\abs{b_1}^2$, mean matching gives $1-X^\star\approx\abs{b_1}/\sqrt n$. The optimum then holds about $\sqrt n/\abs{b_1}$ thermal photons and nearly $n$ coherent ones, $\Sigma^\star\approx2n^{3/2}/\abs{b_1}$, and the Gaussian estimate gives $\pS^{\max}\approx\abs{b_1}^{1/2}/(2\sqrt\pi\,n^{3/4})$, the $n^{-3/4}$ law of Ref.~\PRA\ including its prefactor. Here the estimate becomes exact for large $n$, because the optimal distribution approaches a Gaussian. For $b_1=1$, the ratio of $\pS^{\max}$ to the estimate is $0.947$ at $n=8$ and $0.998$ at $n=10^4$.

\vspace{-0.25em}\vspace{-0.25em}
\subsection*{Limits of the Gaussian estimate}\vspace{-0.25em}\vspace{-0.25em}

The Gaussian estimate would follow from the local central limit theorem~\cite[Ch.~VII]{petrovSumsIndependentRandom1975} for sums of many independent contributions. Here the photons come from only $m+1$ interfered normal modes, a number that stays fixed while the pattern grows, so the theorem does not apply, and Eq.~\eqref{eq:clt} is a heuristic. It needs the optimal distribution to vary smoothly on the scale of one photon near its mean, and it fails in two situations.

First, the optimal herald statistics can have a dip at the registered pattern. For the two-mode source $s_1=5.6179$, $s_2=3.3716$, $\nu_{12}=-3.1118$ with the pattern $(4,3)$, the optimal statistics assign the probability $1.4\times10^{-5}$ to $(4,3)$ but about $10^{-2}$ to each of its neighbors $(5,3)$ and $(4,4)$, and $\pS^{\max}$ is $1.8\times10^{-3}$ times the estimate.

Second, the estimate assumes a fixed number of heralding modes. If the photons that a two-mode squeezed vacuum sends towards the detectors are split over $m=N$ heralding modes that each register one photon ($s_j=\nu_{jk}=0$, $n_j=1$), the optimal damping parameters are equal by symmetry, $\pS=N!\,X^N(1-NX)$, and $\pS^{\max}=N!/(N+1)^{N+1}\approx\sqrt{2\pi N}\,e^{-N-1}/N$ decays exponentially even without a bound.

\subsection*{Decay under a bound}\vspace{-0.25em}\vspace{-0.25em}

Under a bound $\mu$, Prop.~\ref{prop:expbound} shows that $\ln\pS^{\max}$ decreases by at least $\ln(1/\mu)$ per detected photon. For $s_1=0$, $\smax(A)=\sqrt X$, so the bound is active for $\mu^2<n/(n+1)$, and then\vspace{-0.25em}
\begin{equation}\label{eq:S-fock-bounded}
	\pS^{\max}(\mu)=(1-\mu^2)\,\mu^{2n},\vspace{-0.25em}
\end{equation}
which decreases by $2\ln(1/\mu)$ per photon. For $s_1=1$, $\smax(A)=[X+(X^2+4X)^{1/2}]/2$, so the bound reads $X\le X_\mu=\mu^2/(1+\mu)$, which is active for all large $n$, and then\vspace{-0.25em}
\begin{equation}\label{eq:S-squeezed-bounded}
	\pS^{\max}(\mu)=(1-2X_\mu)^{1/2}\binom{2n}{n}\Big(\frac{X_\mu}2\Big)^n,\vspace{-0.25em}
\end{equation}
which decreases by $\ln[(1+\mu)/(2\mu^2)]$ per photon, which is $1.52$ to $1.56$ times $\ln(1/\mu)$ for bounds of $6$ to $12$\,dB. For $N$ heralding modes with one photon each, $\pS^{\max}(\mu)=N!\,(\mu^2/N)^N(1-\mu^2)$ for $\mu^2<N/(N+1)$, which decreases by $1+2\ln(1/\mu)$ per photon.

For random sources with two and three heralding modes, chosen as in Table~\ref{tab:newton}, we computed the exact bounded optima for \emph{scaled patterns} $c\,\bs n$, a base pattern with every entry multiplied by $c=1,2,\dots$, up to $N\approx200$. Fitting $-\ln\pS^{\max}=a_0+a_1N+a_2\ln N$ over the upper half of the range of $c$ (to extract the asymptotic decrease $a_1$ per photon) gives $a_1$ between $1.40$ and $1.97$ times $\ln(1/\mu)$ at bounds of $6$ to $12$\,dB, and $a_1$ close to zero without a bound. For these sources, the universal lower bound $\ln(1/\mu)$ of Prop.~\ref{prop:expbound} is therefore not tight.

The fitted $a_1$ exceed $\ln(1/\mu)$ because the heralded state carries photons as well, which the bound~\eqref{eq:tailbound} with $k_{\min}>0$ takes into account. For the Fock state $\ket n$, $k_{\min}=n$ gives $\pS\le\mu^{2n}$, which exceeds the exact value~\eqref{eq:S-fock-bounded} only by the factor $1/(1-\mu^2)$ and decreases as fast. For the random sources, minimizing the right-hand side of Eq.~\eqref{eq:tailbound} over $k_{\min}$ gives values within $0.07\ln(1/\mu)$ of the fitted $a_1$. The bound is weak only when the number of heralding modes grows with $N$. For $N$ heralding modes with one photon each, it tends to one above $10$\,dB, although $\ln\pS^{\max}$ decreases by $1+2\ln(1/\mu)$ per photon.

\vspace{-0.25em}
\suppsection{Newton's method}{supp:newton}\vspace{-0.25em}\vspace{-0.25em}\vspace{-0.25em}

Newton's method maximizes $\ln\pS$ by solving the mean-matching condition $\Ex{\hat{\bs n}}=\bs n$, so its result is easy to check: for every instance of Table~\ref{tab:newton}, the mean photon numbers of the damped source agreed with the pattern to better than $10^{-12}$ at the end of the run.

Near a point $\bs\theta$, Newton's method replaces $\ln\pS$ by its second-order expansion and moves to the maximum of this quadratic, which gives the step $\bs\Delta=\Sigma^{-1}(\bs n-\Ex{\hat{\bs n}})$. The quadratic model predicts that the step increases $\ln\pS$ by $\lambda^2/2$, where $\lambda^2=(\bs n-\Ex{\hat{\bs n}})^{\mathsf T}\bs\Delta$ is called the \emph{squared Newton decrement}. Near the optimum, $\lambda^2/2$ approximates the remaining gain $\ln\pS^{\max}-\ln\pS$ [Eq.~\eqref{eq:S-gain}], which makes it a natural stopping criterion. Far from the optimum, a full step can leave $\Om$ or decrease $\pS$. The step is then shortened by \emph{backtracking}, that is, the points $\bs\theta+t\bs\Delta$ are tested for $t=1,\tfrac12,\tfrac14,\dots$ until one lies in $\Om$ and increases $\ln\pS$ by at least a small fixed fraction of $t\lambda^2$.

The convergence guarantee of Sec.~\ref{sec:newton}~\cite[Sec.~9.5.3]{boydConvexOptimization2004} gives no usable bound on the number of steps, because it depends on curvature constants of $\ln\pS$ that are not known in advance. Steps become short when an iterate lands very close to the boundary of $\Om$ while the optimum also lies close to it, which happens for patterns with many photons per mode or for many heralding modes.

Our implementation~\cite{heraldingDampingRepository} starts on the diagonal $\theta_1=\dots=\theta_m$ at the point where $\smax(A)=1/2$ (for $s_0=0$) and stops when $\lambda^2/2<10^{-12}$. Table~\ref{tab:newton} gives the number of Newton steps for random sources with up to nine photons per heralding mode. All instances converged.

\begin{table}[t]
	\centering
	\caption{Newton steps until $\lambda^2/2<10^{-12}$ for random sources: $s_j$ and $\nu_{jk}$ independent circularly symmetric complex Gaussian variables with $\langle\abs{s_j}^2\rangle=\langle\abs{\nu_{jk}}^2\rangle=1$, and $n_j$ drawn uniformly from $\{1,\dots,9\}$. There are $1000$ instances for each $m$ ($500$ for $m=30$ and $300$ for $m=50$). The random seeds are in Ref.~\cite{heraldingDampingRepository}. The last row gives the median squeezing of the most squeezed normal mode at the optimum.}
	\label{tab:newton}
	\small
	\begin{tabular*}{\textwidth}{@{\extracolsep{\fill}}lccccccccc@{}}
		\toprule
		$m$            & 1 & 2 & 3 & 5 & 8 & 12 & 20 & 30 & 50 \\
		\midrule
		median         & 6 & 6 & 7 & 8 & 8 & 9 & 10 & 9 & 10 \\
		99th percentile & 9 & 10 & 11 & 12 & 13 & 14 & 15 & 17 & 19 \\
		largest        & 14 & 15 & 14 & 14 & 15 & 18 & 21 & 23 & 35 \\
		squeezing (dB) & 15.0 & 17.8 & 19.5 & 21.7 & 23.9 & 25.6 & 27.9 & 29.7 & 31.9 \\
		\bottomrule
	\end{tabular*}\vspace{-0.25em}\vspace{-0.25em}
\end{table}

One optimization, including the computation of the starting point, took a median of $1.8$\,ms for $m=1$, $23$\,ms for $m=50$ and $122$\,ms for $m=100$ on one core of an Intel Xeon processor at $2.1$\,GHz.

A \emph{continuation in the pattern} avoids the short steps and is the default in our implementation. It first solves the problem for the pattern $\bs n$ divided by the smallest power of $4$ for which no entry exceeds $16$, and then for $4$ times larger patterns in turn, each time starting from the previous optimum, until it reaches $\bs n$. The intermediate patterns need not be integers, because the maximizer of $\bs n\cdot\bs\theta-\Phi(\bs\theta)$ exists for all real $n_j>0$ (remark after Cor.~\ref{cor:hill}). With this continuation, at most $51$ steps sufficed for patterns with up to $9\times10^4$ photons per mode. Without the continuation, patterns $10^3$ to $10^4$ times larger than those of Table~\ref{tab:newton} needed up to several hundred steps for $m\ge8$, and some instances with $m=20$ and $50$ did not converge within $1000$ steps.

\vspace{-0.25em}\vspace{-0.25em}
\suppsection{Bounded squeezing}{supp:bounded}\vspace{-0.25em}\vspace{-0.25em}\vspace{-0.25em}

With a bound $\smax(A)\le\mu$, the optimum lies on the boundary of $\Om_\mu$ whenever the bound is active. Mean matching is then replaced by the condition~\eqref{eq:kkt}, $\Ex{\nhat_j}=n_j-\xi\abs{U_{j1}}^2$. The source delivers fewer photons than the pattern prescribes, mostly in the heralding modes that make up the most squeezed normal mode, and $\xi=\d\ln\pS^{\max}/\d\ln\mu$ measures what the bound costs.\vspace{-0.25em}\vspace{-0.25em}

\vspace{-0.25em}\vspace{-0.25em}\vspace{-0.25em}\vspace{-0.25em}
\subsection*{Barrier method}\vspace{-0.25em}\vspace{-0.25em}

The bounded optimum lies on the boundary of $\Om_\mu$ whenever the bound is active, so Newton's method of Sec.~\ref{sec:newton} cannot be used directly. We use a logarithmic barrier in the sense of Ref.~\cite[Sec.~11.2]{boydConvexOptimization2004}, which replaces the constraint by a term that diverges at the boundary. On $\Om_\mu$,\vspace{-0.5em}
\begin{equation}\label{eq:S-barrier}
	\Barr(\bs\theta)\equiv\ln\det(\mu^2I-AA^{\dagger})=2(m+1)\ln\mu-2\Phi_{B/\mu}(\bs\theta),
\end{equation}
with $\Phi_{B/\mu}$ the log-partition function of the reference source $|\tilde G_{B/\mu,\bs0}\rangle$. The gradient and the Hessian of $\Barr$ are therefore respectively $-2\Ex{\hat{\bs n}}_{A/\mu}$ and $-2\Sigma_{A/\mu}$, the moments of $|\tilde G_{A/\mu,\bs0}\rangle$, so $\Barr$ is strictly concave and tends to $-\infty$ at the boundary of $\Om_\mu$.

The method maximizes $\tau\ln\pS+\Barr$ by Newton's method with backtracking for $\tau=1,10,100,\dots$, each time starting from the previous maximizer. Large $\tau$ weights the objective more, so the maximizer approaches the bounded optimum. The following proposition says how fast, and certifies the optimum.\newpage

\begin{proposition}\label{prop:barrier}
	For directly coupled sources (\ref{as:direct}), let $\tau>0$ and let $\bs\theta_\tau$ maximize $\tau\ln\pS+\Barr$ over $\Om_\mu$. Then
	\begin{equation}\label{eq:S-gap}
		0\le\ln\pS^{\max}(\mu)-\ln\pS(\bs\theta_\tau)\le\frac{m+1}{\tau}.
	\end{equation}
\end{proposition}
\begin{proof}
	The lower bound holds because $\bs\theta_\tau\in\Om_\mu$. The upper bound follows the textbook argument for barrier methods, in which the stationarity condition at $\bs\theta_\tau$ bounds $\ln\pS$ on all of $\Om_\mu$ from above. In the textbook case, the log-det barrier certifies the gap $(m+1)/\tau$, one unit per eigenvalue, for a matrix inequality $\mu^2I-M\succeq0$ in which $M$ is a convex function of the variables in the matrix sense, for instance an affine one~\cite[Sec.~11.6]{boydConvexOptimization2004}. Here $M=AA^{\dagger}$ is not of this kind, and the convexity of the functions $\ell_k$ of Prop.~\ref{prop:logconvex} takes over its role. Let\vspace{-0.5em}
	\begin{equation}\label{eq:S-zeta}
		\zeta_i=\frac{2\sigma_i^2}{\mu^2-\sigma_i^2}\quad\text{at }\bs\theta_\tau,\vspace{-0.3em}
	\end{equation}
	which does not increase with $i$ because $\sigma_1\ge\sigma_2\ge\dots$, and set $\zeta_{m+2}=0$. Terms with $\sigma_i=0$, which occur only if $B$ is singular, have $\zeta_i=0$ and are omitted below. Since $\Barr=\sum_i\ln(\mu^2-\sigma_i^2)$, its gradient is $-\sum_i\zeta_i\nabla\ln\sigma_i$ where the $\sigma_i$ are distinct. Writing $\ln\sigma_i=\ell_i-\ell_{i-1}$ with $\ell_0=0$ and collecting the terms of each $\ell_k$ gives the second form of Eq.~\eqref{eq:S-stat} below, in which $\ell_k$ has the coefficient $\zeta_k-\zeta_{k+1}\ge0$. This form also holds where singular values coincide. By first-order degenerate perturbation theory, a group of equal singular values $\sigma_{p+1}=\dots=\sigma_{p+r}$ enters the gradient of $\Barr$ only through the sum of their logarithms, which is $\ell_{p+r}-\ell_p$, and the coefficients of the $\ell_k$ inside the group vanish. The function $\ell_k$ is differentiable wherever $\sigma_k>\sigma_{k+1}$, so only differentiable $\ell_k$ appear. Setting the gradient of $\tau\ln\pS+\Barr$ to zero gives\vspace{-0.25em}\vspace{-0.25em}\vspace{-0.25em}
	\begin{equation}\label{eq:S-stat}
		\nabla\ln\pS=\frac1\tau\sum_i\zeta_i\nabla\ln\sigma_i=\frac1\tau\sum_k(\zeta_k-\zeta_{k+1})\nabla\ell_k .\vspace{-0.25em}
	\end{equation}
	Let $\bs\theta'\in\Om_\mu$ and $\bs d=\bs\theta'-\bs\theta_\tau$. A concave function lies below its tangent plane and a convex function above it, so $\ln\pS(\bs\theta')-\ln\pS(\bs\theta_\tau)\le\nabla\ln\pS\cdot\bs d$ and $\nabla\ell_k\cdot\bs d\le\ell_k(\bs\theta')-\ell_k(\bs\theta_\tau)$. With Eq.~\eqref{eq:S-stat}, whose coefficients $\zeta_k-\zeta_{k+1}$ are nonnegative, and with $\ell_k(\bs\theta')\le k\ln\mu$ on $\Om_\mu$,
	\begin{equation}\label{eq:S-gapchain}
		\begin{aligned}
			 & \ln\pS(\bs\theta')-\ln\pS(\bs\theta_\tau)                                                        \\
			 & \quad\le\frac1\tau\sum_k(\zeta_k-\zeta_{k+1})\big[\ell_k(\bs\theta')-\ell_k(\bs\theta_\tau)\big] \\
			 & \quad\le\frac1\tau\sum_k(\zeta_k-\zeta_{k+1})\sum_{i\le k}\ln\frac\mu{\sigma_i}                  \\
			 & \quad=\frac1\tau\sum_i\zeta_i\ln\frac\mu{\sigma_i}\le\frac{m+1}\tau .
		\end{aligned}
	\end{equation}
	The last step uses $\zeta_i\ln(\mu/\sigma_i)=-w\ln w/(1-w)\le1$ with $w=\sigma_i^2/\mu^2$, which is $\ln x\le x-1$ for $x=1/w$, and the fact that there are $m+1$ singular values. The supremum over $\bs\theta'\in\Om_\mu$ then gives Eq.~\eqref{eq:S-gap}.
\end{proof}

\vspace{-0.25em}\vspace{-0.25em}\vspace{-0.25em}
\subsection*{Effect of the bound on the heralding probability}\vspace{-0.25em}\vspace{-0.25em}

We follow the bounded optimum from $1$ to $30$\,dB, each time starting from the optimum of the next tighter bound. For the sources with two heralding modes of Fig.~\ref{fig:bounded}, a bound of $15$\,dB keeps at least $99\,\%$ of $\pS^{\max}$, $10$\,dB between $42$ and $73\,\%$, and $6$\,dB between $2$ and $13\,\%$. For the source of Fig.~\ref{fig:bounded} that heralds $(\ket0+\ket2+\ket4+\ket6)/2$ with the pattern $(3,3)$, the best of them, a bound of $10$\,dB lowers $\pS^{\max}=0.0132$ to $0.0079$, and the heralding modes then deliver $1.03$ photons each on average instead of $3$.

\begin{figure}[t]
	\centering
	\includegraphics[width=0.75\textwidth]{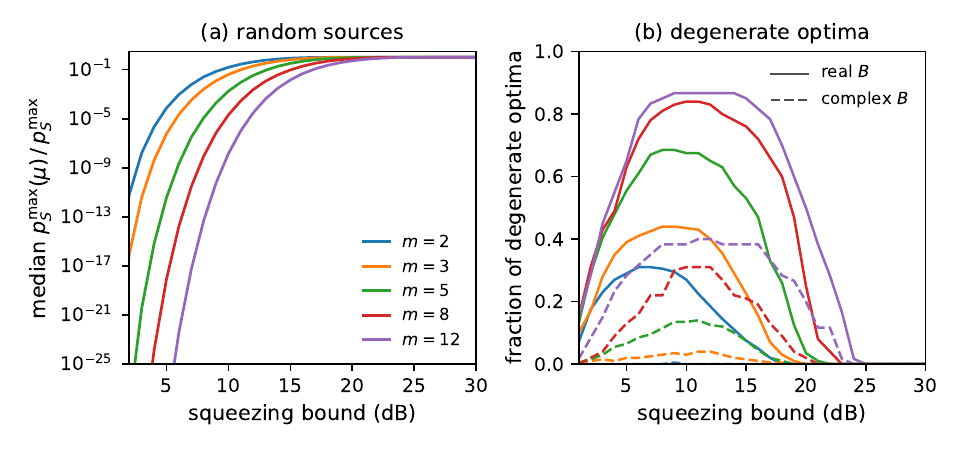}
	\caption{Effect of a squeezing bound with ideal detection for random sources. \textbf{(a)} Random sources with complex control parameters chosen as in Table~\ref{tab:newton}: median of $\pS^{\max}(\mu)/\pS^{\max}$ for $m$ heralding modes, with $200$ instances for $m=2$, $3$ and $5$, $100$ for $m=8$ and $60$ for $m=12$. \textbf{(b)} Fraction of degenerate optima, $\sigma_1=\sigma_2=\mu$, for the complex sources of (a) (dashed) and for sources with real control parameters (solid) for each $m$ (colors as in (a)).\vspace{-0.25em}\vspace{-0.25em}}
	\label{fig:bounded-random}
\end{figure}

Figure~\ref{fig:bounded-random} shows the same for an independent sample of random sources drawn from the same distribution as in Table~\ref{tab:newton}. These are benchmarks for the method rather than designs for particular targets, and their unbounded optima need more squeezing, with medians of $18.0$\,dB for $m=2$ and $25.3$\,dB for $m=12$ in this sample. A bound of $10$\,dB is therefore active in essentially every instance, and both the surviving fraction of $\pS^{\max}$ and the delivered photons fall steeply with $m$. At this bound, the heralding modes deliver a median of $16\,\%$ of the photons of the pattern for $m=2$ and $4\,\%$ for $m=12$.

\vspace{-0.25em}\vspace{-0.25em}\vspace{-0.25em}
\subsection*{Equality bound}\vspace{-0.25em}\vspace{-0.25em}\vspace{-0.25em}

Fixing $\smax=\mu$, as in Ref.~\cite[Fig.~5]{fiurasekHeraldingProbabilityOptimization2026}, restricts $\bs\theta$ to the boundary of $\Om_\mu$, which is not convex, and $\pS$ can have several local maxima on it. For example, the source $s_1=-0.3427$, $s_2=0.5129$, $\nu_{12}=-1.5773$ with the pattern $(1,3)$ has two maxima on this boundary for $\mu=0.9639$ ($17.4$\,dB), whose heights differ by $5.6\,\%$, although its unbounded optimum needs only $12.1$\,dB. Such local maxima do not arise if the bound is imposed as the inequality~\eqref{eq:bound}, whose unique maximum the barrier method finds from any admissible starting point.\vspace{-0.25em}\vspace{-0.25em}

\subsection*{Optimality condition at degenerate optima}

Where the largest singular value is degenerate, $\smax$ need not be differentiable, and Eq.~\eqref{eq:kkt}, derived in App.~\ref{app:capped} for a nondegenerate one, has to be generalized. Write the largest singular value as a maximum,
\begin{equation}\label{eq:S-smaxvar}
	\smax(A)=\max\big\{\mathrm{Re}\,(\bs u^{\dagger}A\bs v):\ \bs u,\bs v\in\C^{m+1}\ \text{unit}\big\},
\end{equation}
and let $\smax=\sigma_1=\dots=\sigma_\kappa$ with the Takagi vectors $U_D=(\bs U_1,\dots,\bs U_\kappa)$. The maximizing pairs in Eq.~\eqref{eq:S-smaxvar} are $(U_D\bs a,U_D^*\bs a)$ with unit $\bs a\in\C^\kappa$, since we have $AA^{\dagger}U_D=\smax^2U_D$ and $A^{\dagger}U_D=\smax U_D^*$. Danskin's theorem~\cite{danskinTheoryMaxMin1967} states that the directional derivative of a maximum such as Eq.~\eqref{eq:S-smaxvar} is the largest directional derivative among the maximizing pairs. With $\partial_{\theta_j}A=\tfrac12(\Pi_jA+A\Pi_j)$ it gives the one-sided directional derivative\vspace{-0.25em}
\begin{equation}\label{eq:S-danskin}
	\lim_{t\to0^+}\!\frac{\ln\smax(\bs\theta+t\bs\delta)-\ln\smax(\bs\theta)}{t} = \max_Z\sum_j\delta_j\trace(R_jZ),\vspace{-0.5em}
\end{equation}
with\vspace{-0.5em}
\begin{equation}\label{eq:S-Rj}
	(R_j)_{ab}=\mathrm{Re}\big(U_{ja}^*U_{jb}\big).
\end{equation}
For the maximizing pair with $\bs a=\bs x+i\bs y$, where $\bs x$ and $\bs y$ are real, the derivative of $\mathrm{Re}(\bs u^{\dagger}A\bs v)$ along $\bs\delta$ is $\smax\sum_j\delta_j(\bs x^{\mathsf T}R_j\bs x+\bs y^{\mathsf T}R_j\bs y)=\smax\sum_j\delta_j\trace(R_jZ)$ with $Z=\bs x\bs x^{\mathsf T}+\bs y\bs y^{\mathsf T}$. The maximum in Eq.~\eqref{eq:S-danskin} can therefore be taken over the real symmetric positive semidefinite $\kappa\times\kappa$ matrices $Z$ of unit trace, which are the convex combinations of $\bs x\bs x^{\mathsf T}$ with real unit vectors $\bs x$, because a linear function of $Z$ is largest at one of these extreme points. The largest value of $\trace(MZ)$ over such $Z$ is the largest eigenvalue of $M$, so the right-hand side of Eq.~\eqref{eq:S-danskin} is the largest eigenvalue of the real symmetric $\kappa\times\kappa$ matrix $\sum_j\delta_jR_j$. This is first-order degenerate perturbation theory.

Along $\bs\delta$, the $\kappa$-fold eigenvalue $\smax^2$ of $AA^{\dagger}$ splits according to the eigenvalues of $2\smax^2\sum_j\delta_jR_j$, which is the derivative of $AA^{\dagger}$ projected onto the span of $U_D$, and $\smax$ follows the largest of them. Only a one-sided limit appears because for $\kappa>1$ the left- and right-hand derivatives, the smallest and the largest eigenvalue of $\sum_j\delta_jR_j$, generally differ. For $\kappa=1$ there is one matrix $Z=1$ and Eq.~\eqref{eq:S-danskin} reduces to the ordinary derivative $\partial\ln\smax/\partial\theta_j=\abs{U_{j1}}^2$.

It is convenient to collect the right-hand side of Eq.~\eqref{eq:S-danskin} into a set of vectors,\vspace{-0.5em}
\begin{equation}\label{eq:S-omegaW}
	\begin{aligned}
		\bs\omega(Z) & =\big(\trace R_1Z,\dots,\trace R_mZ\big),                \\
		\mathcal N   & =\{\bs\omega(Z):Z\succeq0,\ \trace Z=1\},\vspace{-0.5em}
	\end{aligned}
\end{equation}
so that the directional derivative is $\max_{\bs\omega\in\mathcal N}\bs\omega\cdot\bs\delta$. The set $\mathcal N$ is compact and convex. For $\kappa=1$ it is the single vector $(\abs{U_{j1}}^2)_j$, the gradient of $\ln\smax$ and thus the outward normal of the boundary of $\Om_\mu$. For $\kappa>1$ the boundary generally has an edge at $\bs\theta$, and the vectors in $\mathcal N$ play the role of the outward normals there.

The next proposition is the familiar statement about Lagrange multipliers for optima under constraints.

\begin{proposition}\label{prop:kkt}
	For directly coupled sources (\ref{as:direct}), let the bound be active at the bounded optimum $\bs\theta^\star$, with $\smax=\sigma_1=\dots=\sigma_\kappa$ there. Then\vspace{-1em}\vspace{-0.25em}
	\begin{equation}\label{eq:S-kkt-general}
		\bs n-\Ex{\hat{\bs n}}=\bs\omega(\Xi)\vspace{-0.5em}
	\end{equation}
	for a real symmetric positive semidefinite $\kappa\times\kappa$ matrix~$\Xi$. For $\kappa=1$ this is Eq.~\eqref{eq:kkt} with $\xi=\Xi$, and for $\kappa=2$ it reads\vspace{-0.5em}\vspace{-0.25em}\vspace{-0.25em}\vspace{-0.25em}
	\begin{equation}\label{eq:kkt2}
		\Ex{\nhat_j}=n_j-\sum_{a,b=1}^{2}\Xi_{ab}\,\mathrm{Re}\big(U_{ja}^*U_{jb}\big).\vspace{-0.5em}
	\end{equation}
\end{proposition}
\begin{proof}
	Every $\bs\omega\in\mathcal N$ has $\sum_j\omega_j>0$. Since $U$ is unitary, $\sum_{j\ge1}\omega_j(\bs x\bs x^{\mathsf T})=\sum_{j\ge1}\abs{(U_D\bs x)_j}^2$ is the weight of the unit vector $U_D\bs x$ on the heralding modes, and $\sum_j\omega_j(Z)$ is an average of such weights. If it vanished, some vector $U_D\bs x$ with real $\bs x$ would have no weight on the heralding modes, hence equal $e^{i\phi}\bs e_0$ with $\bs e_0$ the basis vector of the signal mode, and satisfy $A\bs e_0=\smax e^{2i\phi}\bs e_0$, which contradicts $A_{j0}=A_{0j}\neq0$. Hence shifting any direction by $-\varepsilon(1,\dots,1)$ with $\varepsilon>0$ strictly lowers $\max_{\mathcal N}\bs\omega\cdot\bs\delta$, which we use below.

	A direction $\bs\delta$ with $\max_{\mathcal N}\bs\omega\cdot\bs\delta<0$ keeps $\bs\theta^\star+t\bs\delta$ in $\Om_\mu$ for small $t>0$, by Eq.~\eqref{eq:S-danskin}. Since $\bs\theta^\star$ is optimal,\vspace{-0.25em}
	\begin{equation}\label{eq:S-nodescent}
		\big(\bs n-\Ex{\hat{\bs n}}\big)\cdot\bs\delta\le0\quad\text{whenever}\quad\max_{\mathcal N}\bs\omega\cdot\bs\delta<0 .\vspace{-0.5em}
	\end{equation}
	Suppose $\bs n-\Ex{\hat{\bs n}}$ were outside the convex cone $\{t\bs\omega:t\ge0,\ \bs\omega\in\mathcal N\}$, which is closed because $\mathcal N$ is compact and does not contain $\bs0$. By the separating hyperplane theorem~\cite[Sec.~2.5.1]{boydConvexOptimization2004}, a plane then separates the point from the cone, and since the cone is invariant under scaling, the plane can be taken through the origin, and the normal of this plane is a direction $\bs\delta$ with $(\bs n-\Ex{\hat{\bs n}})\cdot\bs\delta>0\ge\max_{\mathcal N}\bs\omega\cdot\bs\delta$. Shifting $\bs\delta$ by $-\varepsilon(1,\dots,1)$ with $\varepsilon$ small makes the second inequality strict while keeping the first, which contradicts Eq.~\eqref{eq:S-nodescent}. Hence $\bs n-\Ex{\hat{\bs n}}$ lies in that cone, which is Eq.~\eqref{eq:S-kkt-general} after absorbing the scale of the cone into $\Xi$.
\end{proof}
By the sensitivity interpretation of multipliers~\cite[Sec.~5.6]{boydConvexOptimization2004}, $\trace\Xi=\d\ln\pS^{\max}(\mu)/\d\ln\mu$ wherever this derivative exists, generalizing Eq.~\eqref{eq:kkt}.

\vspace{-0.25em}\vspace{-0.25em}\vspace{-0.25em}\vspace{-0.25em}\vspace{-0.25em}
\subsection*{Degenerate optima}\vspace{-0.25em}\vspace{-0.25em}\vspace{-0.25em}

An optimum is degenerate if the two largest singular values both equal $\mu$, so that two normal modes are squeezed up to the bound. Such optima are common [Fig.~\ref{fig:bounded-random}(b)]: at $10$\,dB they occur in $27$ to $87\,\%$ of the random instances with real control parameters, from $m=2$ to $m=12$, and with complex control parameters in up to $38\,\%$ for $m=12$ but essentially never for $m=2$. They fill whole regions of patterns, because the edge $\sigma_1=\sigma_2=\mu$ of $\Om_\mu$ generically has dimension $m-2$ for real and $m-3$ for complex control parameters, and the multiplier $\Xi$ of Eq.~\eqref{eq:kkt2} adds two or three dimensions, so that edge points and multipliers together cover a set of patterns of dimension $m$. Numerically they cause no difficulty: the barrier $\Barr=\ln\det(\mu^2I-AA^{\dagger})$ is smooth also where singular values coincide, so the barrier method approaches an edge like any other part of the boundary, and the certificate~\eqref{eq:S-gap} holds unchanged.

\vspace{-0.25em}\vspace{-0.25em}\vspace{-0.25em}
\suppsection{Imperfect detection}{supp:detection}\vspace{-0.25em}\vspace{-0.25em}

With imperfect detection, the detectors register the pattern $\bs n$ with probability $g_{\bs n}(\bs k)$ when the photon numbers $\bs k$ arrive, and $\pS(\bs\theta)=\sum_{\bs k}g_{\bs n}(\bs k)p_{\bs\theta}(\bs k)$. For loss with the efficiencies $\eta_j$, $g_{\bs n}(\bs k)=\prod_j\binom{k_j}{n_j}\eta_j^{n_j}(1-\eta_j)^{k_j-n_j}$, and Poissonian dark counts with the means $d_j$ can be added (App.~\ref{app:detection}). Several local maxima of $\pS$ are common for sources that are symmetric under exchange of two heralding modes, rare for random sources and absent for the sources of Ref.~\QST, and the optimum for ideal detection with the pattern $(n_j-d_j)/\eta_j$ comes within about $1\,\%$ of the best $\pS$ for random sources.\vspace{-0.25em}\vspace{-0.25em}

\vspace{-0.25em}\vspace{-0.25em}\vspace{-0.25em}\vspace{-0.25em}
\subsection*{Evaluation of the heralding probability}\vspace{-0.25em}\vspace{-0.25em}\vspace{-0.25em}

For a source without displacement, let $P(\bs l)$ be the probability to register the pattern $\bs l$, so that $\pS=P(\bs n)$, and write $\bs y^{\bs l}=\prod_jy_j^{l_j}$. With the matrix $J(\bs X)=I-BKB^{\dagger}K$, $K=\diag(1,X_1,\dots,X_m)$, for which $\det J(\bs X)=\det(I-AA^{\dagger})$ (App.~\ref{app:gauss}), the generating function of the registered photon numbers is
\begin{equation}\label{eq:S-pgf}
	\sum_{\bs l}P(\bs l)\,\bs y^{\bs l}=\bigg[\frac{\det J(\bs X)}{T(\bs y)}\bigg]^{1/2},\qquad T(\bs y)=\det J\big(\bs X\circ(\bs1-\bs\eta+\bs\eta\circ\bs y)\big),
\end{equation}
where $\circ$ is the entrywise product~\cite{dodonovMultidimensionalHermitePolynomials1994}. For ideal detectors, $\bs\eta=\bs1$, this follows from Eq.~\eqref{eq:tilt}. Shifting $\bs\theta$ by $\ln\bs y$ replaces $X_j$ by $X_jy_j$, so $\sum_{\bs k}p_{\bs\theta}(\bs k)\bs y^{\bs k}=\exp[\Phi(\bs\theta+\ln\bs y)-\Phi(\bs\theta)]$ with $\Phi=-\tfrac12\ln\det J$, first for $0<y_j\le1$ and then, since both sides are analytic near $\bs y=\bs0$, as an identity of power series. Loss registers each photon independently with probability $\eta_j$, which replaces $y_j$ by $1-\eta_j+\eta_jy_j$. Dark counts multiply the generating function by $\prod_je^{d_j(y_j-1)}$, which convolves the $P(\bs l)$ with Poisson distributions.

Without dark counts, the probabilities $P(\bs l)$ are therefore $[\det J(\bs X)]^{1/2}$ times the Taylor coefficients $f_{\bs l}$ of $T^{-1/2}$. Let $T(\bs y)=\sum_{\bs l}t_{\bs l}\bs y^{\bs l}$, which has at most $3^m$ coefficients because $T$ has degree at most two in each $y_j$, and let $\abs{\bs l}=\sum_jl_j$. The operator $\sum_jy_j\partial_{y_j}$ multiplies the coefficient of $\bs y^{\bs l}$ by $\abs{\bs l}$, and applied to $T^{-1/2}$ it gives $-\tfrac12T^{-1/2}\sum_jy_j\partial_{y_j}T/T$. Multiplying by $T$ and comparing the coefficients of $\bs y^{\bs k}$ gives the recursion\vspace{-0.25em}
\begin{equation}\label{eq:S-recursion}
	f_{\bs0}=t_{\bs0}^{-1/2},\qquad f_{\bs k}=-\frac{1}{t_{\bs0}\abs{\bs k}}\sum_{\bs0\neq\bs l\le\bs k}t_{\bs l}\big(\abs{\bs k}-\tfrac12\abs{\bs l}\big)f_{\bs k-\bs l},\vspace{-0.25em}
\end{equation}
where $\bs l\le\bs k$ holds entrywise and $t_{\bs0}=T(\bs0)>0$.

The gradient and the Hessian of $\ln\pS$ follow from Eq.~\eqref{eq:louis}, with $\Ex{\hat{\bs n}}$ and $\Sigma$ from Eq.~\eqref{eq:cov} and the moments of $q$ from $P$ at neighboring patterns. For pure loss, $(k-n)\binom kn=(n+1)\binom k{n+1}$ gives $(k_j-n_j)\,g_{\bs n}(\bs k)=(n_j+1)\,r_j\,g_{\bs n+\bs e_j}(\bs k)$ with $r_j=(1-\eta_j)/\eta_j$ and the unit vector $\bs e_j$ of heralding mode $j$. Summing over $\bs k$ with the weights $p_{\bs\theta}(\bs k)$, and applying the identity twice, gives\vspace{-0.25em}\vspace{-0.25em}
\begin{equation}\label{eq:S-kq}
	\begin{aligned}
		\Ex{k_j-n_j}_q                  & =(n_j+1)\,r_j\,\frac{P(\bs n+\bs e_j)}{P(\bs n)},                                                        \\
		\Ex{(k_j-n_j)(k_l-n_l)}_q       & =(n_j+1)(n_l+1)\,r_jr_l\,\frac{P(\bs n+\bs e_j+\bs e_l)}{P(\bs n)}\quad(j\neq l), \\
		\Ex{(k_j-n_j)(k_j-n_j-1)}_q     & =(n_j+1)(n_j+2)\,r_j^2\,\frac{P(\bs n+2\bs e_j)}{P(\bs n)},
	\end{aligned}
\end{equation}
from which $\Ex{\bs k}_q$ and $\Cov_q(\bs k)$ follow. Evaluating $\pS$, its gradient and its Hessian therefore needs $f_{\bs k}$ for $\bs k\le\bs n+2\cdot\bs1$.

\vspace{-0.5em}
\subsection*{Search for local maxima}
\vspace{-0.5em}

To locate all local maxima of $\pS$ for $m=2$ and $m=3$, we evaluate $\ln\pS$ on a grid over $\Om$ (and along the boundary of $\Om_\mu$ when a squeezing bound is imposed) and refine every discrete local maximum by Newton's method in the interior and one-dimensional maximization along the boundary. For symmetric sources we also test whether the best point on the symmetry axis is a saddle point, which indicates mirror-image maxima, and locate the threshold efficiency $\eta_c$ by bisection. Grid parameters and numerical checks are detailed in Ref.~\cite{heraldingDampingRepository}; as with any grid search, maxima narrower than the grid resolution cannot be excluded.

\vspace{-0.25em}\vspace{-0.25em}\vspace{-0.25em}
\subsection*{Results of the searches}\vspace{-0.25em}\vspace{-0.25em}\vspace{-0.25em}

\begin{table}[t]
	\centering
	\caption{Searches for several local maxima of $\pS$ with imperfect detection. Sources: number of source--pattern combinations, with two heralding modes unless stated. Detectors: number of settings, either efficiencies $\eta$ or mean dark counts $d$. Bounds: number of bound settings, including no bound. Last column: cases with several maxima.}
	\label{tab:scans}
	\small
	\setlength{\tabcolsep}{3pt}
	\begin{tabular*}{\textwidth}{@{\extracolsep{\fill}}llcrr@{}}
		\toprule
		sources & detectors & bounds & cases & several \\
		\midrule
		symmetric (incl.\ Fig.~\ref{fig:etac}), $24$ & $21$ $\eta$ & $31$ & $15\,624$ & Fig.~\ref{fig:etac} \\
		symmetric, $540$ & $13$ $\eta$ & $31$ & $217\,620$ & $11\,710$ \\
		perturbed, $56$ & $7$ $\eta$ & $11$ & $4312$ & $1606$ \\
		random, $300$ & $5$ $\eta$ & $31$ & $46\,500$ & $73$ \\
		random, $300$ & $4$ $d$, $\eta=1$ & $31$ & $37\,200$ & $0$ \\
		Ref.~\QST, $24$ & $22$ $\eta$ & $31$ & $16\,368$ & $0$ \\
		Ref.~\QST, $24$ & $3$ $\eta$, $4$ $d$ & $31$ & $8928$ & $0$ \\
		Ref.~\QST, $m=3$, $9$ & $5$ $\eta$ & $7$ & $315$ & $0$ \\
		random, $m=3$, $60$ & $3$ $\eta$ & $4$ & $720$ & $1$ \\
		heuristic, $200$ & $1$ $\eta$, $1$ $d$ & $5$ & $1000$ & $0$ \\
		\bottomrule
	\end{tabular*}\vspace{-0.25em}\vspace{-0.25em}
\end{table}

Table~\ref{tab:scans} summarizes the searches. Several maxima are common for sources that are symmetric under exchange of the two heralding modes, $s_1=s_2$, and persist when this symmetry is slightly broken; they are rare for random sources.

For the source of Fig.~\ref{fig:detection}, $s=-0.5$, $\nu_{12}=5$ with the pattern $(4,4)$ and a bound of $16$\,dB, all local maxima lie on the boundary of $\Om_\mu$. Up to $\eta=0.279$, the two mirror-image maxima sit at the corners of $\Om_\mu$, where also $\sigma_2=\mu$. There the heralding modes carry $14.5$ photons on average, against $8.9$ at the best point on the symmetry axis, which is why the corners win at low efficiency. A third maximum appears on the axis at $\eta=0.355$, all three are equally high at $\eta=0.3654$ (just above the $\eta=0.365$ shown in Fig.~\ref{fig:detection}(c), where the central maximum is still slightly lower), and above $\eta_c=0.367$ the maximum on the axis is the only one. Without the bound, several maxima exist up to $\eta_c=0.32$, and with ideal detection the source has a single maximum, on the axis.

\begin{figure}[t]
	\centering
	\includegraphics[width=0.8\textwidth]{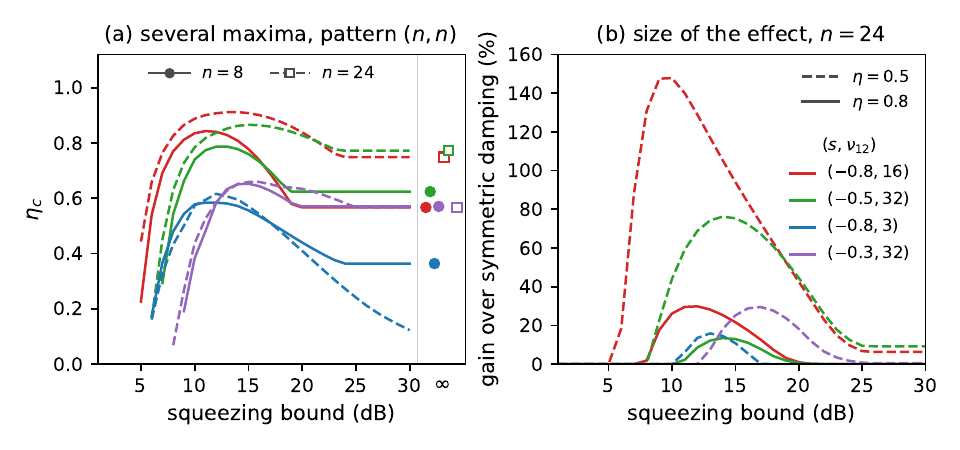}\vspace{-0.25em}\vspace{-0.25em}
	\caption{Several maxima under loss for four symmetric sources, $s_1=s_2=s$ with $(s,\nu_{12})$ given in the legend of (b) and the pattern $(n,n)$. \textbf{(a)} Largest efficiency $\eta_c$ at which $\pS$ has several local maxima, against the squeezing bound, for $n=8$ (solid) and $n=24$ (dashed). The markers at $\infty$ give $\eta_c$ without a bound (filled for $n=8$, open for $n=24$). \textbf{(b)} The same sources with $n=24$: relative increase of $\pS$ at the best local maximum over the best point on the symmetry axis, at $\eta=0.8$ (solid) and $\eta=0.5$ (dashed).\vspace{-0.25em}\vspace{-0.25em}\vspace{-0.25em}\vspace{-0.25em}\vspace{-0.25em}\vspace{-0.25em}\vspace{-0.25em}\vspace{-0.25em}}
	\label{fig:etac}
\end{figure}

Figure~\ref{fig:etac}(a) shows the threshold $\eta_c$ as a function of the bound for eight of the $24$ symmetric source--pattern combinations in the first row of Table~\ref{tab:scans} (four sources, each with $n=8$ and $n=24$). It rises with the bound, peaks at bounds between $9$ and $17$\,dB, where it exceeds its value without a bound, and falls back once the bound no longer constrains the maxima. Among these $24$ combinations, the highest threshold is $\eta_c=0.91$, for $s=-0.8$, $\nu_{12}=16$ and the pattern $(24,24)$, where $\eta_c$ exceeds $0.90$ at bounds from $11$ to $16$\,dB. The gain of the best maximum over the best point on the symmetry axis [Fig.~\ref{fig:etac}(b)] can be large, but only where $\pS$ is small, for example $30\,\%$ at $\eta=0.8$ with $\pS=6.9\times10^{-6}$. Close to the threshold it is at most about $10\,\%$, and the fidelity of the heralded state is lower at the maxima away from the axis than at the best point on the axis. A broader scan of $540$ symmetric sources found several maxima for $189$ of them, all with $s\le0$, at efficiencies up to $0.95$ (at a bound of $3$\,dB), but at efficiencies of $0.9$ and above the best maximum exceeded the best point on the axis by less than $1\,\%$.

Several maxima persist when the symmetry is broken by unequal $s_j$ or asymmetric patterns, or when $s_1=s_2$ acquires a common complex phase (Table~\ref{tab:scans}). The random sources of Table~\ref{tab:scans} have, with equal probability, real Gaussian $s_j$ and $\nu_{12}$ with standard deviations drawn from $\{0.3,1,2\}$, or circularly symmetric complex Gaussian ones with one common standard deviation from the same set, and patterns with $1\le n_j\le5$. For them, several maxima are rare: $73$ of $46\,500$ cases with two heralding modes, all from $8$ sources with real control parameters, and one of $720$ cases with three heralding modes. For the $24$ sources of Ref.~\QST\ with two heralding modes, the search never found more than one maximum, and at efficiencies from $0.8$ to $0.99$ the damping that maximizes $\pS$ under loss raised $\pS$ by at most $5.7\,\%$ over the damping that is optimal for ideal detection. With dark counts, the maximum can fail to exist (App.~\ref{app:detection}). In our scans this happened only with bounds of at most $3$\,dB and mean dark counts of at least $0.01$, where $\pS$ increases towards the limit in which a heralding mode is in the vacuum and its count is supplied by dark counts.
\vspace{-0.25em}\vspace{-0.25em}\vspace{-0.25em}\vspace{-0.25em}\vspace{-0.25em}
\subsection*{Starting point of the search}\vspace{-0.25em}\vspace{-0.25em}\vspace{-0.25em}

Without a criterion for the global maximum, the search needs good starting points. A detector of efficiency $\eta_j$ with the mean dark count $d_j$ registers on average $\eta_jk_j+d_j$ photons when $k_j$ arrive. Matching the registered pattern on average suggests the damping that is optimal for ideal detection with the non-integer pattern $(n_j-d_j)/\eta_j$, which reduces to $n_j/\eta_j$ without dark counts. Because the mean-matching condition has a unique solution for every pattern with positive real entries (remark after Cor.~\ref{cor:hill}), the starting point is found by Newton's method (or, with a squeezing bound, by the barrier method), provided $n_j>d_j$.

\begin{figure}[t]
	\centering
	\includegraphics[width=0.8\textwidth]{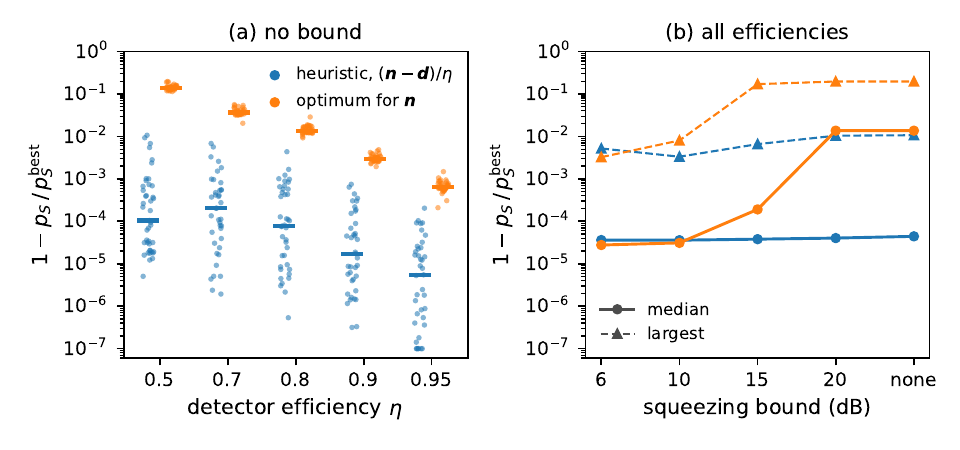}
	\caption{Starting point under imperfect detection, for $200$ random sources with two heralding modes (text): relative loss $1-\pS/\pS^{\mathrm{best}}$ (lower is better) with respect to the largest $\pS$ found by the search, comparing the ideal-detection optimum for the rescaled pattern $(\bs n-\bs d)/\eta$ (heuristic, blue) against the naive ideal-detection optimum for the registered pattern $\bs n$ (orange). \textbf{(a)} Without a bound, one marker per source and bars at the medians. Losses below $10^{-7}$ are drawn at $10^{-7}$. \textbf{(b)} Median (solid) and largest (dashed) loss over all sources against the squeezing bound.}
	\label{fig:heuristic}
\end{figure}

Figure~\ref{fig:heuristic} compares this starting point with the largest $\pS$ found by the search, for $200$ random sources with two heralding modes, complex control parameters drawn as in Table~\ref{tab:newton}, $1\le n_j\le5$, one efficiency $\eta\in\{0.5,0.7,0.8,0.9,0.95\}$ and one mean dark count $d\in\{0,10^{-3},10^{-2}\}$ per source. Without a bound, it came within a median of $4\times10^{-5}$ and at most $1.1\,\%$ of the best $\pS$, while the optimum for the registered pattern $\bs n$ itself lost a median of $1.4\,\%$ and up to $20\,\%$, most at low efficiency [Fig.~\ref{fig:heuristic}(a)]. With a tight bound, both lie on the boundary of $\Om_\mu$ close to the best point, and the difference disappears [Fig.~\ref{fig:heuristic}(b)]. None of these sources had several maxima. For symmetric sources, the starting point lies on the symmetry axis and misses maxima away from it, so it is only a starting point for the search.

\subsection*{Heralding probability and fidelity}

Constrained numerical maximization of $\pS$ under loss subject to $F\ge F_0$ and $\smax(A)<1$~\cite{heraldingDampingRepository} confirms that the trade-off optimum for Fig.~\ref{fig:fidelity} lies on the line of uniform damping ($\theta_1=\theta_2$). For this source at $\eta=0.99$, $F=0.958$ at the maximum of $\pS$, and requiring $F\ge0.98$ or $F\ge0.99$ lowers $\pS$ by factors of $26$ and $1390$.

\vspace{-0.5em}
\suppsection{Several signal modes}{supp:signal}
\vspace{-0.5em}

Some targets are states of several signal modes, for example the two-mode states of Ref.~\QST. All results of Secs.~\ref{sec:ideal} to~\ref{sec:detection} carry over to $S$ signal modes. The assumptions on the source then refer to some signal mode instead of the signal mode, and the signal--signal block of $B$ takes the place of $B_{00}$. Most proofs apply word for word, as listed in Table~\ref{tab:signal}. Four steps need a new argument: the choice of the reference source, the positivity of the weights, the feasibility of a squeezing bound, and the optimality condition at an active bound.\vspace{-0.25em}\vspace{-0.25em}

\subsection*{Setting}

The source has the form~\eqref{eq:state} with $S$ signal modes and $m$ heralding modes, so $A$ and $B$ are complex symmetric $(S+m)\times(S+m)$ matrices, and it has $S+m$ normal modes. We label the signal modes $0_1,\dots,0_S$ and use the index $i$ for them and $j$ for the heralding modes. Damping acts on the heralding modes only, $\Lambda=\diag(1,\dots,1,e^{\theta_1/2},\dots,e^{\theta_m/2})$ with $S$ entries equal to one. It therefore does not change the signal--signal block $B_{\mathrm{ss}}$ of $B$, which is also the signal--signal block of $A$. A pattern $\bs n$ heralds a state of all $S$ signal modes, and the weights $W_{\bs k}=\sum_{\bs k_{\mathrm s}}\bigl|\langle\bs k_{\mathrm s},\bs k|\tilde G_{B,\bs b}\rangle\bigr|^2$ sum over the photon numbers $\bs k_{\mathrm s}$ of all signal modes.

Three conditions take the place of those for one signal mode. Assumption~\ref{as:pure}, that no set of heralding modes is in a pure state, keeps its form. Assumption~\ref{as:direct} becomes that every heralding mode $j$ has $B_{ij}\neq0$ for some signal mode $i$, and we fix one such signal mode $i(j)$ for every $j$. The third condition is $\smax(B_{\mathrm{ss}})<1$. Since $B_{\mathrm{ss}}$ is a submatrix of $A$, $\smax(A)\ge\smax(B_{\mathrm{ss}})$ for every $\bs\theta$, and $\smax(A)\to\smax(B_{\mathrm{ss}})$ as all $\theta_j\to-\infty$. The admissible set $\Om$ is therefore nonempty if and only if $\smax(B_{\mathrm{ss}})<1$, which for one signal mode reads $\abs{B_{00}}<1$.\newpage

\begin{table}[t]
	\centering
	\caption{Results for one signal mode that hold for $S$ signal modes, and the changes in their proofs.}
	\label{tab:signal}
	\footnotesize
	\setlength{\tabcolsep}{5pt}
	\renewcommand{\arraystretch}{1.15}
	\begin{tabularx}{\textwidth}{@{}ll>{\raggedright\arraybackslash}X@{}}
		\toprule
		Result & Ref. & Change in proof \\
		\midrule
		\multicolumn{3}{@{}l}{\emph{Ideal detection}} \\
		\addlinespace[1.5pt]
		Heralded state, exponential family & Eqs.~\eqref{eq:amp}--\eqref{eq:derivs} & None ($e^{\bs k\cdot\bs\theta/2}$ depends only on herald counts). \\
		Convex $\Om$, (strict) concave $\ln\pS$ & Thm.~\ref{thm:main} & H\"older step unchanged; set $v_i=0$ on all signal modes so $\mathcal P$ contains only heralding modes. \\
		Moments, Newton, continuation & Eqs.~\eqref{eq:cov}, \eqref{eq:S-displaced-cov} & $(S+m)\times(S+m)$ matrices; Newton step takes $\mathcal O((S+m)^3)$ ops. \\
		Existence, uniqueness, mean matching & Cor.~\ref{cor:hill} & Signal column $i(j)$ of $A$ bounds $\theta_j$ from above; $W_{\bs k}>0$ shown below. \\
		Supremum without direct coupling & Sec.~\ref{supp:nodirect} & None (no step assumes $S=1$). \\
		Removing modes with $n_j=0$ & App.~\ref{app:gauss} & None (vacuum projection still yields a submatrix of $A$). \\
		\addlinespace[5pt]
		\multicolumn{3}{@{}l}{\emph{Bounded squeezing and scaling}} \\
		\addlinespace[1.5pt]
		Convexity of $\Om_\mu$ & Prop.~\ref{prop:logconvex} & None (three-lines proof is independent of $S$). \\
		Active-bound optimality, multiplier & Prop.~\ref{prop:kkt} & Positivity of $\sum_j\omega_j$ shown below. \\
		Barrier method and certificate & Prop.~\ref{prop:barrier} & $\Barr=2(S+m)\ln\mu-2\Phi_{B/\mu}$; gap bound becomes $(S+m)/\tau$. \\
		Exponential decay under a bound & Prop.~\ref{prop:expbound} & $S+m$ normal modes replace $m+1$; $P_\psi$ tracks total signal photons. \\
		\addlinespace[5pt]
		\multicolumn{3}{@{}l}{\emph{Imperfect detection}} \\
		\addlinespace[1.5pt]
		Gradient, Hessian and evaluation & Eqs.~\eqref{eq:louis}, \eqref{eq:S-pgf}--\eqref{eq:S-kq} & $K=\diag(1,\dots,1,X_1,\dots,X_m)$ with $S$ ones. \\
		At most one maximum ($m=1$) & Thm.~\ref{thm:onemode} & For $m=1$, (\ref{as:pure}) gives $B_{i1}\neq0$ for some signal mode $i$, so $W_k>0$ (shown below) and $\Om_\mu$ stays a half-line as $\smax(A)\to\smax(B_{\mathrm{ss}})<\mu$. \\
		Unique maximum near $\eta_j=1$ & Prop.~\ref{prop:higheta} & Equality in $\abs{\bs u^{\dagger}LAL\bs v}\le1$ needs signal support, where $\abs{\bs u^{\dagger}B_{\mathrm{ss}}\bs v}\le\smax(B_{\mathrm{ss}})<1$. \\
		Measurements preserved by damping & Prop.~\ref{prop:projector} & None (proof holds for arbitrary states $\rho$). \\
		\bottomrule
	\end{tabularx}
\end{table}

\subsection*{Modified proofs}

With several signal modes, only one coupling of every heralding mode can be set to one. For a directly coupled source we choose the reference with $B_{i(j)\,j}=1$, by a phase rotation of heralding mode $j$ and the choice of the origin of $\theta_j$, as in Sec.~\ref{sec:Dop}. The couplings $B_{ij}$ of heralding mode $j$ to the other signal modes, $i\neq i(j)$, are then control parameters, and the block $B_{\mathrm{ss}}$ takes the place of $s_0$. Every admissible damping of this reference still has $X_j<1$. The column of $A$ that belongs to signal mode $i$ contains the entries $A_{ji}=\sqrt{X_j}$ of the heralding modes with $i(j)=i$, so its squared norm is at least $\sum_{j:\,i(j)=i}X_j$, and it cannot exceed $\smax(A)^2<1$. Relative to this reference, the damping therefore again always attenuates.

Every pattern lies in the support of the weights. In the Bargmann representation (App.~\ref{app:gauss}), the signal variables $z_i$ enter the exponent only through $\tfrac12\sum_{i,i'}A_{ii'}z_iz_{i'}+\sum_iz_i(\beta_i+\sum_jA_{ij}z_j)$, where $i$ and $i'$ run over the signal modes and $j$ over the heralding modes. The Bargmann function of the state heralded by the pattern $\bs k$ is therefore $e^{\frac12\sum_{i,i'}A_{ii'}z_iz_{i'}+\sum_i\beta_iz_i}$ times a polynomial of degree $\sum_jk_j$, whose part of the highest degree is proportional to $\prod_j(\sum_iA_{ij}z_i)^{k_j}$. Each factor $\sum_iA_{ij}z_i$ is a nonzero linear function, because $A_{i(j)\,j}\neq0$, so their product and the heralded state are nonzero. Hence $W_{\bs k}>0$ for every $\bs k$, and the closed convex hull of the support is again the closed nonnegative orthant (App.~\ref{app:expfam}).

Under a squeezing bound, $\Om_\mu$ contains points with $\smax(A)<\mu$ exactly when $\mu>\smax(B_{\mathrm{ss}})$, by the two limits above, and we assume this. It replaces the condition $\mu>\abs{B_{00}}$ of Sec.~\ref{sec:bounded}. The strict inequality $\smax(A)>\abs{B_{00}}$ of Sec.~\ref{sec:bounded} has no counterpart, because a leading singular vector of $B_{\mathrm{ss}}$ need not couple to any heralding mode. For example, let $B_{\mathrm{ss}}$ have the diagonal entries $0.3$ and the off-diagonal entry $0.2$, and let one heralding mode couple to the two signal modes with $B_{0_1,1}=1$, $B_{0_2,1}=-1$, and $B_{11}=0$. The signal vector $(1,1)/\sqrt2$ has the singular value $\smax(B_{\mathrm{ss}})=0.5$ and does not couple to the heralding mode, and $\smax(A)=0.5$ for all $X_1\le0.1$.

The proof of Prop.~\ref{prop:kkt} uses $A_{0j}\neq0$ to show that every $\bs\omega\in\mathcal N$ has $\sum_j\omega_j>0$. With several signal modes, suppose that a unit vector $\bs u=U_D\bs x$ with real $\bs x$ had no weight on the heralding modes, so that $\bs u=(\bs w,\bs0)$ with a unit vector $\bs w$ of the signal modes. Since $\bs x$ is real, $A\bs u^*=\smax\bs u$, and the signal rows of this equation give $B_{\mathrm{ss}}\bs w^*=\mu\bs w$. Then $\smax(B_{\mathrm{ss}})\ge\mu$, which contradicts $\mu>\smax(B_{\mathrm{ss}})$. The rest of the proof is unchanged.

Statements that are specific to one signal mode do not carry over. These are the form~\eqref{eq:B} with a single entry $s_0$ and the heralded Fock states of Eq.~\eqref{eq:fockm} and Sec.~\ref{supp:onemode}. Numerical checks for $S=2$ and $3$ (including confirmation that $\pS^{\max}$ retains the $n^{-(m+1)/2}$ scaling for $S=2$, $m=2$) are provided in Ref.~\cite{heraldingDampingRepository}.

\suppsection{Sources without direct coupling}{supp:nodirect}

Without direct coupling (\ref{as:direct}), the first step of the proof of Cor.~\ref{cor:hill}, which bounds $\theta_j$ from above by $A_{0j}\neq0$, fails, and the maximum need not exist. Let $\mathcal H = \overline{\conv}\supp W$. For a minimal family, the maximum exists iff $\bs n$ lies in the interior of $\mathcal H$~\cite[Cor.~9.6]{barndorffnielsenInformationExponentialFamilies1978}. For directly coupled sources, $\mathcal H$ is the closed nonnegative orthant, whose interior contains every pattern with $n_j\ge1$. Otherwise $\bs n$ can lie on the boundary of $\mathcal H$, and then a smaller scheme is better.

\begin{proposition}\label{prop:supremum}
	Let no set of heralding modes be in a pure state (\ref{as:pure}), and let $\bs n\in\supp W$ lie on the boundary of $\mathcal H$. Then there is a vector $\bs\ell\neq\bs0$ with $\bs\ell\cdot\bs n=0$ and $\bs\ell\cdot\bs k\ge0$ for all $\bs k\in\supp W$, and for every $\bs\theta\in\Om$
	\begin{equation}\label{eq:S-face}
		\lim_{t\to\infty}\pS(\bs\theta-t\bs\ell)=\frac{W_{\bs n}e^{\bs n\cdot\bs\theta}}{\sum_{\bs k:\,\bs\ell\cdot\bs k=0}W_{\bs k}e^{\bs k\cdot\bs\theta}}>\pS(\bs\theta).
	\end{equation}
	The limit is the heralding probability of a physical source in which the heralding modes with $\ell_j\neq0$ form a set in a pure state. Leaving these modes out gives a scheme with fewer heralding modes that heralds the same state with a higher probability at the remaining damping parameters than $\pS(\bs\theta)$.
\end{proposition}
\begin{proof}
	We first show that $\mathcal H$ is a cone. Let $\bs v\cdot\bs k\le h$ for all $\bs k\in\supp W$. Then
	\begin{equation*}
		\Phi(\bs\theta+t\bs v)\le\Phi(\bs\theta)+th,\qquad t>0,
	\end{equation*}
	so $\forall t, \bs\theta+t\bs v$ stays in $\Om$. Along it, all entries of $A$ and $\bs\beta$ equal their value at $t=0$ times an exponential of $t$.
	
	Entries of $A$ cannot grow, because $\abs{A_{jk}}\le\smax<1$ on $\Om$, and entries of $\bs\beta$ cannot grow, because $\norm{\bs\beta}^2/2\le\Phi$ grows at most linearly [Eq.~\eqref{eq:S-expbound}]. All entries therefore converge, and the limit has $\smax<1$, because $\ln\smax$ is convex along the half-line (Prop.~\ref{prop:logconvex}) and negative, so it does not increase. Hence $\Phi$ stays bounded along the half-line, which requires
	\begin{equation*}
		\bs v\cdot\bs k\le0\quad\text{for all}\quad\bs k\in\supp W .
	\end{equation*}
	Every half-space that contains $\supp W$ therefore still contains it when its boundary is moved through the origin, so $\mathcal H$ is a cone, and the boundary point $\bs n$ has a supporting plane through the origin with a normal $\bs\ell$ as stated.

	Next, along $\bs\theta-t\bs\ell$, $W_{\bs n}e^{\bs n\cdot(\bs\theta-t\bs\ell)}=W_{\bs n}e^{\bs n\cdot\bs\theta}$ of $\pS$ is constant, and the denominator decreases term by term,
	\begin{equation*}
		e^{\Phi(\bs\theta-t\bs\ell)}=\sum_{\bs k}W_{\bs k}e^{\bs k\cdot\bs\theta}e^{-t\,\bs\ell\cdot\bs k}\ \xrightarrow{\ t\to\infty\ }\ \sum_{\bs k:\,\bs\ell\cdot\bs k=0}W_{\bs k}e^{\bs k\cdot\bs\theta}.
	\end{equation*}
	This gives Eq.~\eqref{eq:S-face}~\cite{geyerLikelihoodInferenceExponential2009}. The inequality is strict, because some $\bs k\in\supp W$ has $\bs\ell\cdot\bs k>0$, as otherwise $\bs\ell\in\mathcal V$, which assumption~\ref{as:pure} excludes.

	Finally, along $\bs\theta-t\bs\ell$ the term $\ad_jA_{jk}\ad_k$ of the exponent in Eq.~\eqref{eq:state} is multiplied by $e^{-t(\ell_j+\ell_k)/2}$ and the term $\beta_j\ad_j$ by $e^{-t\ell_j/2}$, with $\ell_0=0$ for the signal mode. By the first step with $\bs v=-\bs\ell$, none of them grows, so each either tends to zero or stays fixed. Let $B^{\bs\ell}$ and $\bs b^{\bs\ell}$ be $B$ and $\bs b$ with the entries of the vanishing terms set to zero. Damped by $\bs\theta$, this source has the Fock amplitudes of the original source for the patterns with $\bs\ell\cdot\bs k=0$ and zero amplitudes for all others. It heralds the same state, with the probability~\eqref{eq:S-face}, and it is physical, although no damping of $B$ prepares it. Since it produces only patterns with $\bs\ell\cdot\bs k=0$, $\Var(\bs\ell\cdot\hat{\bs n})=0$, and by the proof of Thm.~\ref{thm:main} its heralding modes with $\ell_j\neq0$ form a set in a pure state. This set does not affect the heralded state, and leaving it out raises the heralding probability further.
\end{proof}
Removing all decoupled pure-state heralding modes and repeating the reduction (at most $m$ times) if $\bs n$ still lies on a boundary face yields a smaller scheme that attains its maximum and outperforms every $\bs\theta\in\Om$. In numerical tests~\cite{heraldingDampingRepository}, Newton's method indicates this case by iterates that diverge along $-\bs\ell$ while $\pS$ approaches the supremum.

\end{document}